\documentclass[12pt, letterpaper]{report}

\usepackage{amssymb, amsmath, amsthm, amsfonts, latexsym}
\usepackage{graphicx}
\usepackage{epstopdf}
\usepackage{float}
\usepackage{color}
\usepackage{xcolor}
\usepackage{longtable}
\usepackage[figuresright]{rotating}
\usepackage{listings}
\usepackage{etoolbox}
\usepackage{sectsty}
\usepackage{xspace}
\usepackage{fancyhdr}
\usepackage{setspace}
\usepackage{ifthen,xkeyval,xfor,amsgen}
\usepackage[acronym,toc,nogroupskip,nonumberlist]{glossaries}
\usepackage[lmargin=1.5in, rmargin=1in, vmargin=1in, headsep=0.083334in]{geometry}
\usepackage{booktabs}
\usepackage{lscape}
\usepackage{gensymb}
\usepackage{textcomp}
\usepackage{svg}
\usepackage{caption}
\usepackage{subcaption}
\usepackage{multirow}
\usepackage{tabularx}
\usepackage[version=3]{mhchem} 
\usepackage{enumitem}
\usepackage{soul}
\usepackage{times}
\usepackage[pagewise]{lineno}
\usepackage{algorithm}
\usepackage{algpseudocode}
\usepackage[numbers]{natbib}
\usepackage[hidelinks]{hyperref}

\hypersetup{
    colorlinks=true,
    linkcolor=blue,
    filecolor=magenta,
    urlcolor=cyan,
    pdftitle={PhD Thesis - Mohamed Sy},
    pdfpagemode=FullScreen,
}
\chapterfont{\fontsize{14}{15}\selectfont}
\sectionfont{\fontsize{14}{15}\selectfont}
\subsectionfont{\fontsize{14}{15}\selectfont}
\subsubsectionfont{\fontsize{14}{15}\selectfont}
\fancyheadoffset[L]{0.25in}
\makeatletter
\patchcmd{\@makechapterhead}{50\p@}{20pt}{}{}
\patchcmd{\@makeschapterhead}{50\p@}{20pt}{}{}
\makeatother

\fancypagestyle{plain}{
\fancyhf{}
\fancyhead[C]{\thepage}

}

\renewcommand{\thechapter}{\arabic{chapter}}
\renewcommand\bibname{\centering BIBLIOGRAPHY}

\newcommand{\thesis}{Machine Learning-Enhanced Laser Spectroscopy for Multi-Species Gas Detection in Complex Environments\xspace}
\newcommand{\longname}{Mohamed Sy \xspace}
\newcommand{\name}{\longname}
\newcommand{\specialization}{}
\newcommand{\monthyear}{August, 2025}

\newglossary[slg]{symbols}{syi}{sbl}{List of Abbreviations and Symbols}

\newglossaryentry{symb:UC}{
name=$U_{C}$, type=symbols,
description=Coflow velocity (ms$^{-1}$),
sort=symboluc
}

\makeglossaries

\begin{document}
    \doublespacing

\vspace{2pt}
\thispagestyle{empty} 
\addvspace{10mm} 

\begin{center}
    {\large\textbf{Machine Learning Enhanced Laser Spectroscopy for Multi-Species Gas Detection in Complex Environments}}\vfill 
    
    {\fontsize{12}{15} \selectfont
        Dissertation by\\
        Mohamed Sy\vfill 
        
        In Partial Fulfillment of the Requirements\\
        For the Degree of \\
        Doctor of Philosophy\\
        \specialization\vfill 
        
        King Abdullah University of Science and Technology\\
        Thuwal, Kingdom of Saudi Arabia\vfill
        
        \copyright \monthyear\\ 
        All Rights Reserved\\
        
        \includegraphics[height=12pt]{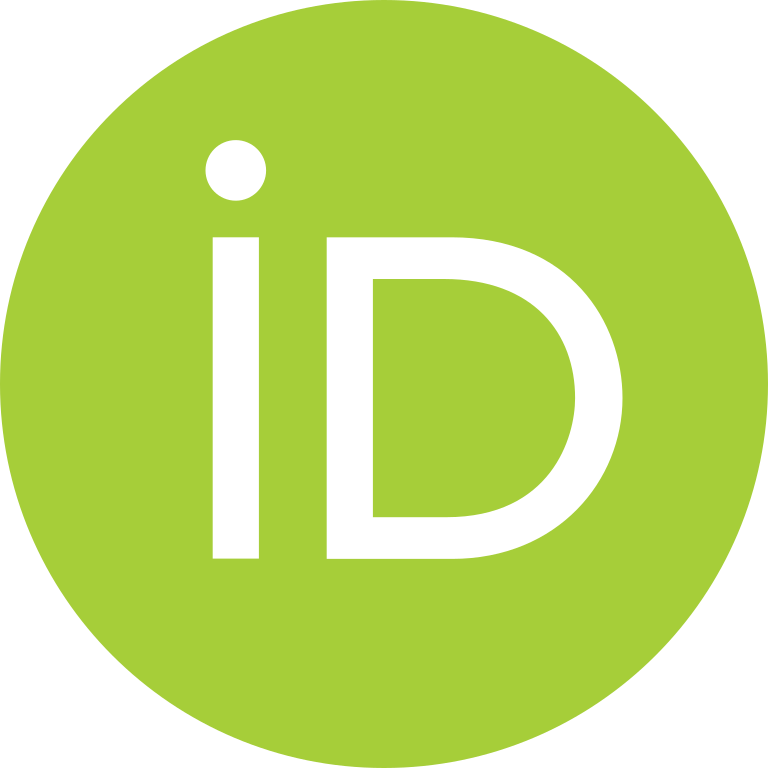} 
        \hspace{3pt} 
        \href{https://orcid.org/0000-0003-3600-8437}{https://orcid.org/0000-0003-3600-8437} 
    }
\end{center}

\begin{center}
\end{center}
\begin{center}
    {{\bf\fontsize{14pt}{14.5pt}\selectfont \uppercase{ABSTRACT}}}
\end{center}
\doublespacing
\addcontentsline{toc}{chapter}{Abstract}
\begin{center}
    {{\fontsize{14pt}{14.5pt}\selectfont {\thesis\\
    \name}}}
\end{center}

Laser absorption spectroscopy (LAS) is a well-established technique for non-intrusive measurement of gas species concentrations in combustion and atmospheric environments. However, conventional LAS methods often struggle when applied to multi-species gas mixtures in dynamically changing or interference-laden conditions. The presence of overlapping spectral features, instrumental noise, and incomplete reference data limits the reliability of standard analysis approaches, especially when unknown species or low-absorbing compounds are present.

This dissertation investigates the development of advanced diagnostics that combine laser spectroscopy with machine learning (ML) to address these limitations. Several classes of ML models are explored for denoising, spectral decomposition, and multi-species quantification. Deep denoising autoencoders (DDAEs) are developed and applied to spectroscopic measurements in a shock tube to enhance signal fidelity during high-speed pyrolysis of hydrocarbon mixtures. The DDAEs demonstrate improved detection limits for trace species.

A second focus of the work is the development of a structured unsupervised learning framework, HT-SIMNet, capable of mitigating spectral interference from unknown species without requiring full calibration datasets. The method employs spectral augmentation strategies and a Noise2Noise-inspired training scheme to isolate individual species in reactive systems. In cases where reference spectra are unavailable or incomplete, an autoencoder-based blind source separation method—termed UnblindMix is proposed. This framework reconstructs both concentration profiles and spectral signatures directly from mixture data and is validated on complex synthetic and experimental mixtures containing up to eight components.

To enhance the detectability of weakly absorbing species that are often masked by broader absorbers, spectral feature engineering method is applied using first derivatives and convolution operations. The resulting processed spectra enable selective highlighting of minor species with low signal contributions. Finally, the reliability and certifiability of ML-based classification is addressed through the development of a model termed VOC-certifire. This framework employs randomized smoothing and Voigt-based spectral perturbation to provide robust classification of volatile organic compounds under varying conditions.

The techniques developed in this dissertation are experimentally validated and benchmarked against existing methods. The integration of spectroscopic hardware with ML architectures presented herein offers a practical path toward real-time, interference-resilient, and reference-free gas detection in environments where conventional techniques often fail. The results provide a foundation for next-generation gas sensing technologies applicable to combustion science, environmental monitoring, and industrial safety.

\begin{center}
\end{center}
\begin{center}
    {\bf\fontsize{14pt}{14.5pt}\selectfont \uppercase{Acknowledgements}}\\\vspace{1cm}
\end{center}
\addcontentsline{toc}{chapter}{Acknowledgements}
In the name of Allah, the Most Gracious, the Most Merciful. All praise and thanks be to Allah for granting me the strength, patience, and perseverance to complete this journey.

I would like to express my deepest gratitude to my advisor, Professor Aamir Farooq, for his unwavering support, mentorship, and belief in my potential. Coming from a relatively unknown university, I was well aware of the challenges ahead, yet Professor Farooq took a chance on me; a leap of faith that transformed my life. His insistence on scientific rigor, attention to detail, and clarity of thought has shaped not only my research but also the way I approach problems beyond the lab. His guidance has been the most defining experience of my academic career. I am also thankful to Professor Mohamed-Slim Alouini, whose kind recommendation made it possible for me to meet and join Professor Farooq’s group.

I am also grateful to my PhD committee members, Professor Greg Rieker (external examiner), Professor Xin Gao (committee chairperson), and Professor Mani Sarathy for their willingness to dedicate time and effort to evaluate my dissertation and participate in my defense.

A heartfelt thank you goes to Mhanna Mhanna, whose mentorship, friendship, and generosity have been invaluable. I also thank Emad Al-Ibrahim for his mentorship in machine learning, Ali Elkhazraji, Mohammed Al-Momtan, Mohamad Abou Daher and Khalil for their constant support, friendship and camaraderie. To all members of the FASTER group, past and present, alumni and juniors alike, thank you for making the research environment both productive and enjoyable.

Finally, I extend my deepest love and gratitude to my family for their endless sacrifices, prayers, and unconditional belief in me. And to my friends, both near and far, thank you for your unwavering support and presence throughout this journey.
    
    \begin{onehalfspacing}
    \renewcommand{\contentsname}{\centerline{\textbf{{\large TABLE OF CONTENTS}}}}
    \tableofcontents
    \cleardoublepage
    
    \addcontentsline{toc}{chapter}{\listfigurename} 
    \renewcommand*\listfigurename{\centerline{LIST OF FIGURES}} 
    \listoffigures
    \cleardoublepage
    
    \addcontentsline{toc}{chapter}{\listtablename}
    \renewcommand*\listtablename{\centerline{LIST OF TABLES}} 
    \listoftables
    
    \glsaddall
\end{onehalfspacing}

    \chapter{Introduction}\label{ch1:intro}
\section{Motivation and Significance of Multi-Species Gas Detection}

Accurate detection of multiple gases is critical in a wide range of applications where real-time, selective, and interference-resistant sensing is required. Monitoring complex mixtures, rather than isolated components, enables significant progress in environmental regulation, public safety, industrial operations, and healthcare. The following examples highlight key areas where multi-species gas sensing is essential and motivate the developments presented in this work.

\subsection*{Environmental Monitoring}

Air quality management depends on tracking mixtures of pollutants such as NO$_x$, SO$_2$, CO, O$_3$, and volatile organic compounds (VOCs), whose combined effects influence smog formation, human health, and climate~\cite{zhang2019drivers}. Regulatory agencies including the WHO~\cite{who2021aqg, hoffmann2021air} and EPA~\cite{epa2021naaqs} mandate strict emission thresholds, often requiring concurrent monitoring of several gases. For instance, BTEX compounds must be measured together due to their shared sources and cumulative toxicity~\cite{chauhan2014recent}. Satellite missions such as GOSAT monitor CO$_2$ and CH$_4$ simultaneously to assess greenhouse gas emissions~\cite{yokota2009global}. In all cases, multi-species sensing provides a more accurate and comprehensive picture than isolated measurements.

\subsection*{Safety and Leak Detection}

Industrial safety protocols rely on the ability to detect hazardous gas mixtures. Facilities handling hydrocarbons, ammonia, or hydrogen sulfide must monitor species such as CH$_4$, H$_2$S, CO, and O$_2$ in parallel, as leaks typically involve multi-gas emissions~\cite{panda2024comprehensive}. OSHA standards and commercial multi-gas detectors reflect this requirement, emphasizing rapid and accurate identification during emergency events~\cite{osha2021chemicals, osha2021confined}. Sensors capable of distinguishing between gases in complex environments improve safety, reduce false alarms, and support early intervention.

\subsection*{Indoor Air Quality}

Indoor environments contain a mix of gases including CO$_2$, CO, formaldehyde, and various VOCs, originating from occupants, building materials, and cooking~\cite{saini2020indoor}. High CO$_2$ levels signal inadequate ventilation, while formaldehyde can exceed WHO limits of 0.08 ppm in enclosed spaces. Since many pollutants coexist, multi-gas sensing is required to accurately assess air quality and guide ventilation control~\cite{dai2023achieving}. Smart building systems increasingly use such sensors to improve occupant health and energy efficiency~\cite{baqer2023development}.

\subsection*{Medical Diagnostics}

Exhaled breath contains a mixture of volatile organic compounds that serve as potential biomarkers for disease. Elevated acetone is associated with diabetes, while isoprene and ammonia provide insight into lipid metabolism and renal function~\cite{owen2014calibration}. Breath analysis systems must detect multiple compounds simultaneously to account for biological variability and improve diagnostic accuracy~\cite{pavlou2000sniffing, dragonieri2017electronic}. Recent efforts focus on developing compact analyzers that combine multi-species detection with statistical classification techniques for early disease identification~\cite{broza2013nanomaterial}.

\subsection*{Combustion Diagnostics}

Combustion processes produce a range of gaseous species, including CH$_4$, CO, CO$_2$, H$_2$O, and various reactive intermediates. Accurate quantification of these components is required to validate reaction models and study fuel oxidation pathways~\cite{kuenning2024multiplexed}. High-speed sensing systems are needed to capture transient species profiles in shock tubes, engines, and burners~\cite{tancin2021ultrafast}. Multi-wavelength laser absorption and signal deconvolution methods are used to resolve overlapping features and enable simultaneous detection of multiple species under dynamic conditions.

\subsection*{Forensics and Security}

Detection of hazardous chemicals in forensic or security scenarios often requires identifying mixtures of compounds. Illicit drug production, for example, may release ammonia, ether, and acetone, while chemical warfare agents are detected through both parent compounds and degradation products~\cite{ahrens2022detection}. Multi-species analyzers deployed in the field improve detection specificity and reduce false positives~\cite{bednar2011field, evans2021fieldable}. These systems play a key role in first response, counter-terrorism, and chemical threat mitigation.

\begin{figure}[H]
    \centering
    \includegraphics[width=\linewidth]{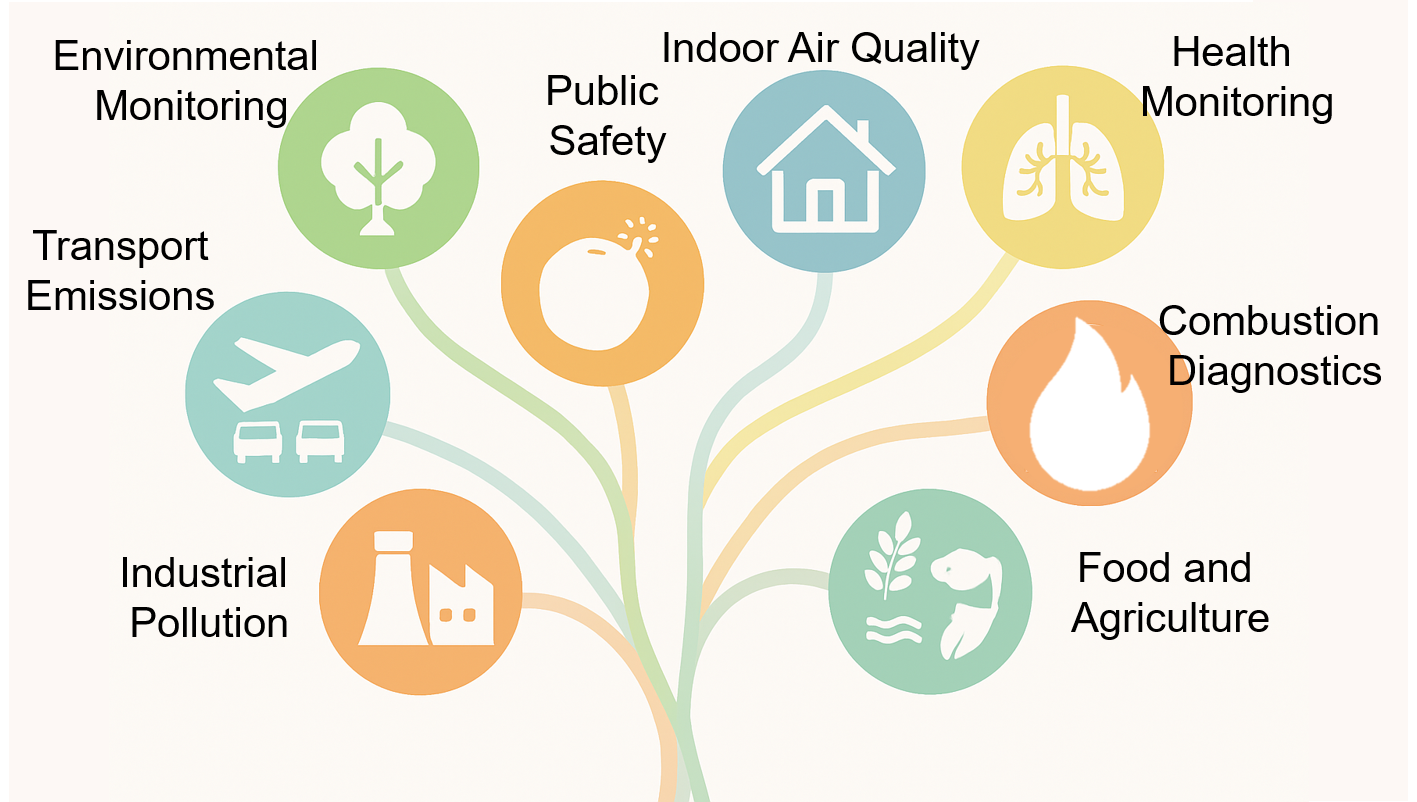}
    \caption{Major application areas for gas sensing technologies.}
    \label{fig:gas_sensing_applications}
\end{figure}

\vspace{0.5em}
\noindent These examples illustrate the broad relevance of multi-species gas sensing across sectors. As laser-based sensors and analytical tools continue to evolve, new applications will emerge in fields such as agriculture, aerospace, and smart city infrastructure. Figure~\ref{fig:gas_sensing_applications} provides a summary of major application areas for multi-species sensing technologies.

\section{Limitations of Conventional Multi-Species Detection Techniques}

Despite the increasing demand for gas sensing across many application areas, the accurate and simultaneous detection of multiple gas species remains a significant technical challenge. Although each gas exhibits characteristic spectral features in the infrared region, these features frequently overlap when multiple species are present. In practical settings, factors such as fluctuating temperature, pressure variations, and background noise further complicate measurements by altering line shapes and introducing nonlinear signal behavior~\cite{tian2023gas}. While the theoretical framework for selective detection is well established, real-time implementation under dynamic conditions remains difficult.

Several established techniques have been applied to the problem of multi-species gas detection, including gas chromatography (GC), mass spectrometry (MS), Fourier-transform infrared (FTIR) spectroscopy, and laser absorption spectroscopy (LAS). Each method offers specific advantages but also exhibits limitations in terms of speed, portability, and scalability.

Gas chromatography is widely used for species separation based on physical properties~\cite{franceschelli2023real}. It provides high sensitivity and selectivity but is inherently a slow, batch-mode technique. Each measurement typically requires several minutes and offline sample processing. The serial detection process restricts its use in real-time or field-deployable applications, particularly in safety-critical or rapidly evolving environments.

Mass spectrometry enables compound identification based on mass-to-charge analysis following ionization~\cite{sparkman2011gas}. When coupled with GC, it achieves high resolution and broader coverage. However, MS systems require vacuum conditions and complex sample handling~\cite{snyder2016miniature}, which limits their use in portable sensors. Isobaric species such as CO and C$_2$H$_4$, both with molecular weight 28, present additional challenges unless high-resolution or tandem mass spectrometers are used~\cite{marshall2008high}. Time-of-flight (TOF) MS systems offer improved temporal resolution but remain costly and complex~\cite{cotter2004time}.

FTIR spectroscopy captures broadband infrared absorption spectra and can detect many gases simultaneously~\cite{griffiths1983fourier}. While FTIR is well suited to multi-species sensing, it struggles with overlapping spectral features and reduced sensitivity at trace concentrations. Quantitative analysis requires access to accurate spectral libraries and expert calibration, particularly when interferents are unknown or poorly characterized.

Laser absorption spectroscopy, including tunable diode laser absorption spectroscopy (TDLAS), has gained popularity due to its high sensitivity and rapid time response~\cite{goldenstein2017infrared, farooq2022laser, liu2019laser}. However, traditional laser-based sensors typically monitor a single species by tuning to one absorption line. Extending these systems to multiple species often requires several lasers or multiplexed configurations, increasing cost and system complexity~\cite{liu2019laser}. Experimental setups such as those by Nicolas et al.~\cite{pinkowski2019multi} and Cassady et al.~\cite{cassady2020thermal} utilize multiple lasers and wavelengths to track several species. While effective, these systems are expensive, alignment-sensitive, and difficult to scale.

\begin{figure}[H]
    \centering
    \includegraphics[width=0.85\linewidth]{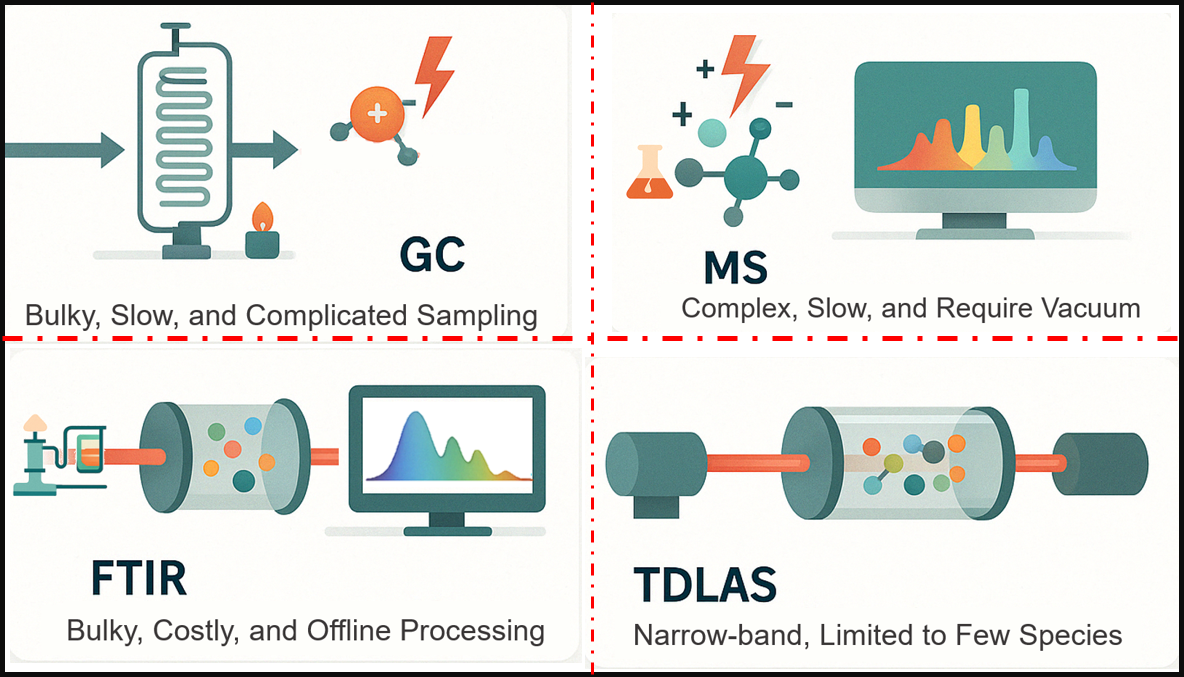}
    \caption{Conventional techniques for multi-species gas sensing and their limitations.}
    \label{fig:conventional_methods}
\end{figure}

Figure~\ref{fig:conventional_methods} summarizes key conventional techniques—GC, MS, FTIR, and TDLAS—and the limitations associated with each. These include large instrument size, long response times, the need for vacuum systems, and limited scalability to multiple species. These challenges underscore the need for alternative sensing strategies that can handle overlapping spectra, correct for nonlinear effects, and scale efficiently without requiring extensive hardware or prior knowledge of the sample composition. This motivates the data-driven approaches developed in this work.

In addition, common analysis methods such as multi-dimensional linear regression (MLR) assume linear superposition of spectral features. This assumption is often invalid in the presence of overlapping features, strong absorbers, or pressure-broadened lines~\cite{bernath2020spectra}. In practical multi-species environments, absorption spectra are rarely isolated; instead, they consist of a superposition of many partially overlapping rovibrational transitions. When multiple gases absorb in the same spectral window, the measured signal no longer reflects the contribution of a single species in a straightforward, additive manner. This leads to ambiguities and errors in concentration retrieval, particularly when weaker absorbers are masked by stronger ones or when nonlinear interactions distort the apparent line shapes~\cite{barik2022selectivity, tennyson2012exomol}. 

\begin{figure}[H]
    \centering
    \includegraphics[width=0.85\linewidth]{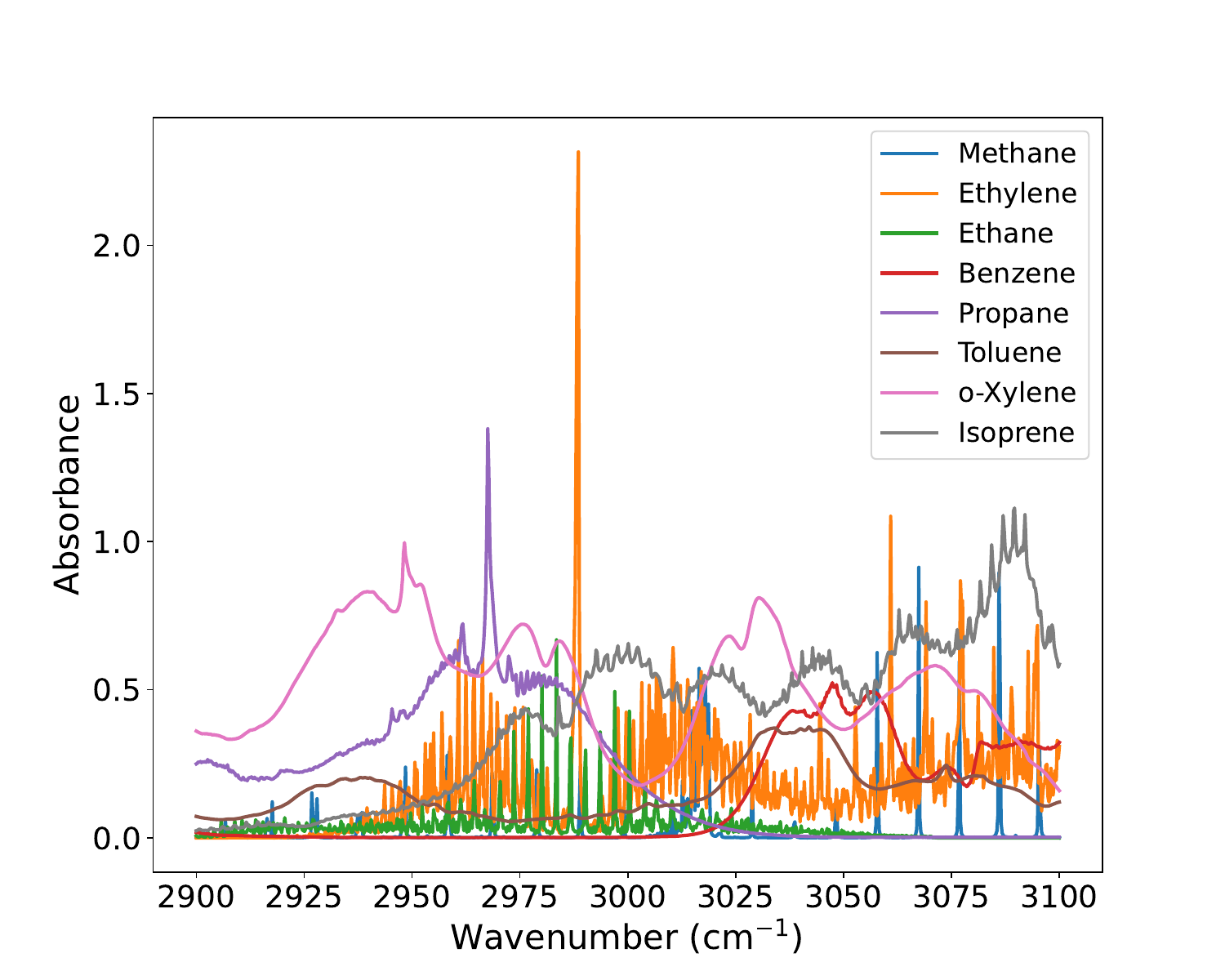}
    \caption{Gas-phase infrared spectra of target species obtained from the PNNL database \cite{sharpe2004gas}, at conditions of 298 K, 1 atm, and 1 ppm-meter.}
     \label{fig:overlap}
\end{figure}

Figure~\ref{fig:overlap} illustrates this challenge by showing the infrared absorption spectra of several hydrocarbon gases—methane, ethylene, ethane, benzene, propane, toluene, o-xylene, and isoprene—within the C–H stretch region (2900–3100~cm$^{-1}$). The significant overlap across these spectra demonstrates how difficult it is to distinguish and quantify individual species when relying solely on conventional line-by-line analysis. Such overlap is especially problematic under atmospheric pressure conditions, where collisional broadening further increases the spectral congestion. These limitations motivate the use of advanced, data-driven methods capable of disentangling complex absorption features and enabling robust multi-species quantification, as explored in this dissertation.

\section{Machine Learning for Multi-Species Detection: Capabilities and Challenges}
\subsection{The Potential of Machine Learning for Multi-Species Detection}
Machine learning methods have been introduced in gas sensing to overcome several challenges faced by traditional techniques. These models can learn nonlinear relationships between spectral measurements and gas composition, which allows them to handle overlapping absorption features, spectral interference, and sensor imperfections~\cite{chowdhury2025artificial, nicolle2024mixtures}. Applications in combustion science have expanded considerably~\cite{ihme2022combustion, ihme2024artificial}, with machine learning used to predict reference spectra~\cite{mcgill2021predicting}, estimate reaction rates, and classify fuel properties~\cite{al2019octane, heid2023chemprop, al2021prediction}.

Supervised regression models have been used to estimate the concentrations of multiple gas species from spectral or sensor data. Techniques such as partial least squares, support vector machines, random forests, and neural networks have been trained on datasets that pair measured spectra with known gas concentrations~\cite{chowdhury2025artificial, yaqoob2021chemical}. Once trained, these models can analyze new data and estimate the concentration of each species based on patterns in the spectra.

Neural networks, particularly deep models, have proven effective at resolving overlapping spectral signals and capturing nonlinear behavior. Deep neural networks (DNNs) have been used to distinguish between gases with similar absorbance profiles~\cite{mhanna2021selective, chowdhury2024deep, elkhazraji2024selective}. One example is the work by Mhanna et al., who used a tunable laser and a DNN to detect benzene, toluene, ethylbenzene, and xylene (BTEX)~\cite{mhanna2022deep, mhanna2022laser, mhanna2024multi}. These compounds have broad, overlapping mid-infrared spectra, yet the model could quantify each species using a single laser. A similar approach allowed real-time detection of combustion intermediates in a shock tube, even during rapid changes in concentration~\cite{mhanna2024multi}.

Unsupervised learning methods have also been applied to gas sensing. Blind source separation (BSS) algorithms recover individual spectral components from mixed measurements without requiring reference concentrations~\cite{Liu2021, Badeau2021}. These models extract basis spectra and their time evolution directly from mixture data, which is useful when isolated reference spectra are unavailable.

Machine learning methods are also used to reduce measurement noise and correct for non-ideal conditions. Neural denoisers and filter-based architectures can suppress signal fluctuations, baseline drift, and laser instability~\cite{steed2023human}. In one case, a deep residual network improved methane detection by correcting nonlinear baselines, resulting in enhanced sensitivity under fluctuating experimental conditions~\cite{xu2024leveraging}.

\begin{figure}[H]
    \centering
    \includegraphics[width=0.75\linewidth]{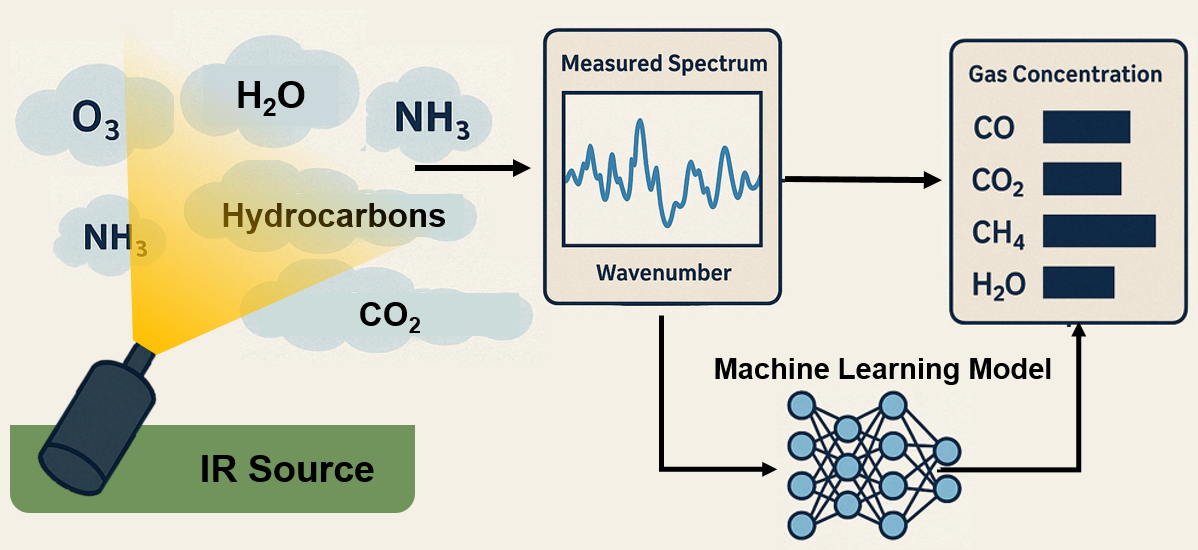}
    \caption{Typical machine learning framework for gas sensing applications.}
    \label{fig:ml_framework_gas_sensing}
\end{figure}

Figure~\ref{fig:ml_framework_gas_sensing} presents an example of a data-driven gas sensing framework. Spectral inputs are processed through algorithms that perform denoising, classification, and quantification. Combined with modern optical hardware, these tools enable fast, selective, and reliable detection of multiple gases, even in complex environments. The methods described here form the basis for the approach developed in this work.

\subsection{Gaps in Current Knowledge}

Although machine learning tools have been increasingly applied to gas sensing, several important limitations remain when these methods are used in practical multi-species detection scenarios:

\begin{itemize}

    \item Experimental setups that rely on multi-laser configurations or broadband light sources with multiplexed detection enable tracking of several species but involve high system complexity. Optical alignment is often sensitive, and the overall footprint and cost make these systems difficult to scale for widespread use~\cite{pinkowski2019multi, cassady2020thermal}.
    
    \item Supervised regression models, including neural networks and ensemble techniques, require large training datasets with known concentrations. These datasets must span a wide range of conditions, including temperature, pressure, and composition. In most cases, collecting such data is experimentally expensive and time-intensive~\cite{chowdhury2025artificial, yaqoob2021chemical}.
    
    \item Many approaches assume that high-quality reference spectra are available for all species present in the mixture. When an unknown or uncalibrated gas is introduced, the model cannot provide accurate estimates unless new training data is acquired and the model is retrained~\cite{mcgill2021predicting}.
   
    \item Traditional source separation algorithms, exhibit scaling ambiguity and cannot resolve nonlinear distortions or random noise present in measured spectra. These shortcomings limit their reliability when applied to mixtures with complex absorption behavior~\cite{Liu2021, Badeau2021}..
    
    \item Baseline correction and denoising routines are often used to improve detection limits, but these methods typically assume linear baseline behavior. Under field conditions, nonlinear drift, abrupt transients, or fluctuations in laser power can lead to errors in quantification~\cite{steed2023human, xu2024leveraging}.
        
    \item Most machine learning models developed for gas sensing have not been certified through formal procedures. Model certification is essential for use in safety-critical environments and ensures consistency across instruments and operating conditions.
\end{itemize}

These limitations motivate the development of alternative approaches presented in this dissertation. The methods introduced here are designed to improve noise handling, reduce reliance on complete reference data, adapt to unknown species, and support robust, real-time multi-species detection using compact optical setups and data-driven models.

\section{Research Objectives}

The aim of this research is to develop data-driven methods for gas sensing that address the key limitations of conventional and existing laser-based approaches. The objectives are defined as follows:

\begin{itemize}
    \item \textbf{Enhance signal quality through denoising.} Develop machine learning models that improve the signal-to-noise ratio of measured spectra. These models are intended to reduce baseline drift, suppress instrumental noise, and improve the reliability of trace gas detection in real-time measurements.

    \item \textbf{Address interference from known species.} Design analysis techniques that account for spectral overlap between target gases and known background components. These methods aim to improve the accuracy of gas quantification in mixtures where absorption features are not well separated.

    \item \textbf{Adapt to unknown species and incomplete references.} Implement algorithms capable of detecting and compensating for uncalibrated or unknown gases. This includes blind source separation and adaptive modeling strategies that operate without requiring a complete reference library.

    \item \textbf{Establish model robustness and interpretability.} Develop tools to assess the reliability of model predictions under different conditions. This includes calibration transfer, error quantification, and techniques for interpreting model outputs to ensure consistent performance in laboratory and field applications.
\end{itemize}

Each of these objectives targets a specific limitation in current gas sensing methods. The first improves measurement quality by filtering out noise and fluctuations. The second allows quantification in the presence of overlapping absorption features. The third extends model utility to cases with limited prior information. The fourth ensures that the results can be trusted and explained, which is necessary for broader adoption in applied systems.

\section{Dissertation Structure}

This dissertation is organized into eight chapters as outlined below:

\begin{itemize}
    \item \textbf{Chapter 1: Introduction.} Outlines the motivation for multi-species gas detection, discusses limitations of conventional sensing methods, and introduces the proposed data-driven framework. The chapter also presents the research objectives and the overall structure of the dissertation.

    \item \textbf{Chapter 2: Background and Literature Review.} Provides a review of spectroscopic principles and relevant machine learning methods. It covers infrared absorption fundamentals, the Beer--Lambert law, and prior work in traditional and data-driven gas sensing techniques.

    \item \textbf{Chapter 3: Machine Learning for Noise Reduction.} Develops and evaluates models for improving signal quality through denoising. The chapter focuses on reducing baseline drift and measurement noise to enhance detection limits for trace species.

    \item \textbf{Chapter 4: Spectral Interference Mitigation.} Focuses on mitigating interference from unknown species. Techniques are introduced to isolate target signals in the presence of unexpected or unmodeled components using model-based and data-driven approaches.

    \item \textbf{Chapter 5: Blind Source Separation and Reference-Free Detection.} Introduces algorithms designed to recover individual species signals in cases where reference spectra are incomplete or unavailable. Blind source separation frameworks are presented and validated experimentally.

    \item \textbf{Chapter 6: Spectral Feature Engineering.} Proposes methods to enhance spectral contrast for low-absorbing species. These methods are applied to cases where dominant absorbers obscure weaker signals, improving selectivity through feature enhancement.

    \item \textbf{Chapter 7: Model Certification and Reliability.} Presents strategies for validating machine learning model predictions. Calibration transfer, uncertainty quantification, and interpretability tools are used to assess prediction robustness across different operating conditions.

    \item \textbf{Chapter 8: Conclusions and Future Work.} Summarizes the main contributions of the dissertation and revisits the research objectives. Limitations of the current work are discussed, and directions for future research are outlined.
\end{itemize}

Each chapter addresses a specific technical challenge and contributes to a unified sensing framework that integrates spectroscopy with machine learning. The developed methods are tested under realistic conditions to evaluate their ability to improve detection accuracy, interference handling, and reliability.

\section*{Summary of Motivation and Approach}

Multi-species gas detection is essential in applications such as combustion diagnostics, environmental monitoring, industrial safety, and medical analysis. Conventional sensing methods face challenges when spectral features from multiple gases overlap, or when measurement conditions vary with time. These limitations are compounded by the presence of species for which reference spectra are incomplete or unavailable, making accurate quantification difficult in practical environments.

This dissertation presents a series of methods that integrate infrared absorption measurements with data-driven models. Techniques are developed to reduce measurement noise, separate overlapping signals, identify species without prior reference data, and highlight features of weak absorbers masked by stronger background signals. Model reliability is addressed through validation under varying conditions and by applying methods for calibration transfer and error estimation.

Each chapter addresses a specific technical objective. Signal quality is improved through denoising methods. Spectral interference is mitigated through modeling approaches that isolate unknown contributions. Blind source separation is applied to mixtures containing species lacking full spectral information. Feature engineering is used to enhance the detectability of low-absorbing gases. Finally, model robustness is assessed to ensure applicability across a range of scenarios. Together, these contributions form a sensing framework capable of operating under conditions that limit the performance of conventional techniques.

    \chapter{Background and Literature Review}\label{chapter2}

\section{Fundamentals of Laser-Based Gas Sensing}

\section*{Absorption Principles}

Laser absorption spectroscopy for gas sensing is based on the Beer--Lambert law, which describes the attenuation of light as it passes through an absorbing medium. The absorbance $A(\nu)$ at a given frequency $\nu$ is defined as the logarithmic ratio of incident to transmitted intensity and is expressed for a gas mixture as~\cite{swinehart1962beer}:

\begin{equation}
A(\nu) = \sum_{i=1}^{N} \epsilon_i(\nu)\, c_i\, L \,,
\label{eq:beer_lambert}
\end{equation}

where $\epsilon_i(\nu)$ is the absorption cross-section of species $i$ at frequency $\nu$, $c_i$ is its concentration, $L$ is the optical path length, and $N$ is the total number of absorbing species. Each species contributes independently to the total absorbance through its unique absorption cross-section. The cross-section, which has units of area per molecule or molar absorptivity depending on the formulation, quantifies the strength of absorption at each frequency and is determined by the line strength and line shape of the transition.

\begin{figure}[H]
    \centering
    \includegraphics[width=0.85\linewidth]{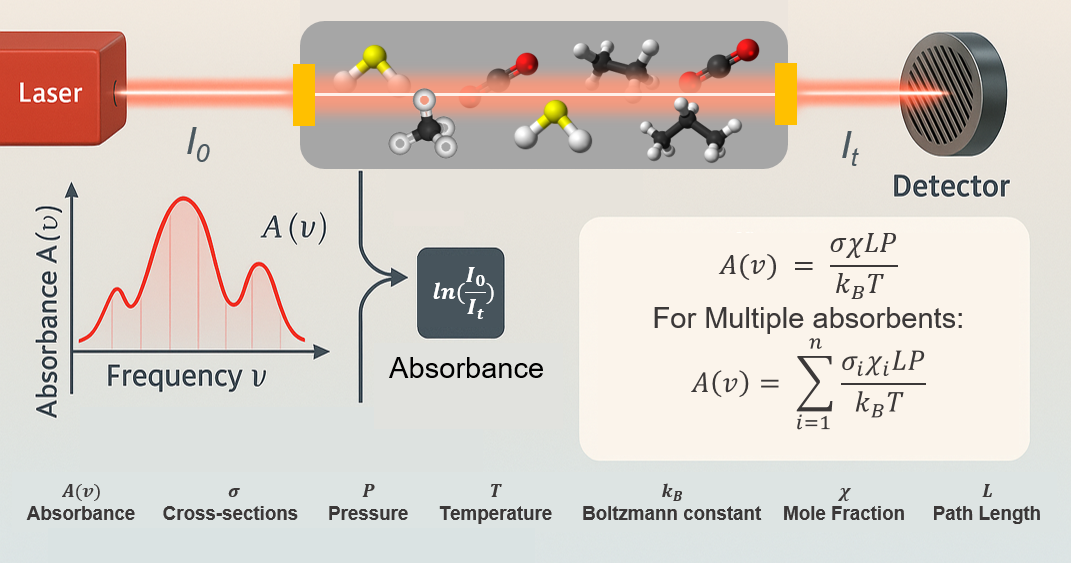}
    \caption{Schematic illustration of the Beer--Lambert law in gas sensing.}
    \label{fig:beer_lambert_schematic}
\end{figure}

According to Eq. \ref{eq:beer_lambert}, the absorbance increases linearly with either the species concentration $c_i$ or the path length $L$. The largest contributions occur at frequencies where $\epsilon_i(\nu)$ is largest, typically near the center of absorption lines. This relationship forms the basis for quantitative gas measurements. Given a measured absorbance spectrum and knowledge of the absorption cross-sections from reference databases such as HITRAN, the concentrations of species in the mixture can be inferred. Figure~\ref{fig:beer_lambert_schematic} illustrates the principle of Beer--Lambert absorption, showing the reduction in laser intensity as light propagates through an absorbing gas sample.

\section*{Infrared Spectroscopy of Gases}

Infrared (IR) laser absorption spectroscopy has emerged as a leading method for gas sensing due to the strong and selective absorption features exhibited by many molecular species in the IR region. In particular, the mid-infrared region, spanning from approximately 2.5 to 25~$\mu$m (4000 to 400~cm$^{-1}$), contains fundamental vibrational transitions that are highly sensitive to gas composition. These transitions, when combined with rotational fine structure, result in dense absorption spectra unique to each molecule. Such spectra serve as diagnostic fingerprints for gas identification.

Hydrocarbon species, for example, exhibit strong C--H stretching vibrations near 3.3~$\mu$m ($\sim$3000~cm$^{-1}$), along with bending modes distributed across the mid-IR range. Only vibrational modes that induce a change in molecular dipole moment are infrared active. As a result, homonuclear diatomics such as N$_2$ and O$_2$ do not exhibit fundamental IR absorption, while heteronuclear and polyatomic molecules like CO, CO$_2$, CH$_4$, and NO$_x$ produce distinct and measurable absorption lines~\cite{goldenstein2017infrared, farooq2022laser}.

In the near-infrared (NIR) and visible regions (approximately 0.7--2.5~$\mu$m), gas absorption arises from overtone and combination bands. These transitions have significantly lower line strengths than the fundamental bands but can still be detected using sensitive techniques. The availability of compact laser sources in this region, such as distributed-feedback diode lasers and interband cascade lasers (ICLs), enables practical implementation. Quantum cascade lasers (QCLs) operating between 4 and 12~$\mu$m are well suited for mid-IR sensing, while terahertz sources access pure rotational transitions of polar molecules in the far-infrared and THz ranges.

Wavelength selection plays a central role in laser-based diagnostics. Absorption lines must be well isolated from interference and must exhibit sufficient line strength under expected pressure and temperature conditions. The Beer--Lambert law relates absorbance to concentration and path length at each wavelength. Practical implementations often involve scanning the laser across a set of absorption lines or using broadband sources such as frequency combs or Fourier-transform infrared (FTIR) spectrometers to acquire wide spectral coverage. The recorded spectra are then analyzed to retrieve gas concentrations using curve-fitting methods or precomputed absorption models.

Laser absorption techniques provide several advantages. These include high selectivity due to the uniqueness of each absorption spectrum, fast time response enabled by direct optical interrogation, and compatibility with in-situ and non-intrusive measurements. However, accurate quantification becomes more difficult when absorption features from multiple gases overlap. Addressing this challenge requires careful wavelength selection and, in some cases, the use of advanced signal processing methods.

\section{Machine Learning Methods for Spectroscopic Analysis}

Machine learning models provide a practical solution for analyzing spectral data in cases where traditional rule-based techniques are limited. These models are trained to recognize patterns in measured spectra and can be used to reduce noise, resolve overlapping signals, and estimate gas concentrations under varying measurement conditions~\cite{mishra2022deep, alajaji2025scoping}. At a high level, ML models can be divided into supervised and unsupervised learning approaches:

\begin{itemize}
    \item Supervised learning requires labeled data where inputs (spectra) are paired with known outputs (concentrations or species identities). The model learns a mapping between inputs and outputs during training and can then make predictions on unseen data. Techniques such as linear regression, support vector machines, and deep neural networks (DNNs) fall into this category~\cite{cunningham2008supervised, contreras2025explainable}.
    \item Unsupervised learning is used when labeled data are not available. These methods find patterns or latent structures within the data, such as separating mixture spectra into constituent components. Examples include principal component analysis (PCA), non-negative matrix factorization (NMF), and autoencoder-based blind source separation frameworks such as UnblindMix~\cite{ghahramani2004unsupervised, naik2025blind, puleio2023calibration}.
\end{itemize}

Deep learning is a subset of ML that uses multi-layered neural networks to capture complex, nonlinear relationships. In a neural network, layers of interconnected nodes (neurons) process the input data through weighted connections, activation functions, and iterative optimization. Training typically involves adjusting these weights to minimize a loss function that quantifies the difference between predicted and actual outputs. Models with many layers are referred to as “deep,” and these architectures have shown remarkable success in a variety of applications, ranging from noise reduction and quantitative analysis to blind source separation and molecular classification, due to their ability to model high-dimensional data and nonlinear absorption phenomena~\cite{lecun2015deep, mishra2022deep}.

\subsection{Regression Models for Concentration Estimation}

A primary application is the prediction of gas concentrations from measured absorbance spectra. Classical approaches include linear regression and partial least squares (PLS) regression, which have been used extensively in chemometrics. For example, Su et al.\ applied PLS to near-infrared spectra to quantify mixture components in complex samples~\cite{casian2021testing}.

Machine learning extends these approaches by introducing nonlinear models that can capture subtle spectral variations. Neural networks, such as multilayer perceptrons, have been trained on labeled spectra to predict species concentrations from high-dimensional inputs. These models are capable of learning complex relationships that may not be captured by linear methods~\cite{lakhmi2024linear}.

Recent developments include the use of convolutional neural networks (CNNs) and physics-informed networks to further improve prediction accuracy. For example, Zhang et al.\ used a CNN architecture to extract relevant features from absorption spectra for gas mixture analysis~\cite{zhang2022tdacnn}. In another study, Wang et al.\ developed a co-training neural network for an NDIR sensor array~\cite{wang2022co}. Their method used two models trained on different features from the same signal to improve accuracy while reducing the need for extensive calibration. The system combined labeled and unlabeled data in a semi-supervised training scheme, allowing for more efficient use of experimental measurements.

Regression models remain central to quantitative gas sensing. By learning the relationship between spectra and concentrations, these models are able to resolve overlapping features and account for nonlinearity in the sensor response. When trained on well-characterized datasets, they provide a reliable approach for retrieving gas concentrations in complex mixtures.

\subsection{Classification for Gas Identification}

In many applications, determining which gases are present is as important as quantifying their concentrations. Classification models are used to assign labels to spectral measurements, indicating the presence or absence of target species. Since several gases may coexist in a mixture, this task is typically treated as a multi-label classification problem.

Conventional qualitative methods often rely on detecting peak absorbance at known wavelengths for specific gases. While effective under ideal conditions, such approaches are limited when spectral features overlap or when background interference is present. Machine learning models can analyze the full spectral profile, allowing for more robust identification under variable conditions.

A range of classifiers have been applied to spectroscopic data. These include support vector machines (SVMs), random forests, and fully connected neural networks. More recent work has focused on convolutional neural networks (CNNs), which are well suited to high-dimensional inputs and can extract local spectral features automatically.

Early studies using SVMs demonstrated reliable classification of species based on infrared absorption spectra by learning distinctive patterns in the data~\cite{chowdhury2024deep}. Chowdhury et al.\ later proposed a one-dimensional CNN model (VOC-Net) for classifying volatile organic compounds (VOCs) using terahertz rotational spectra~\cite{chowdhury2022vocnet}. The model was trained on both simulated and experimental data covering the 220–330~GHz range and achieved accurate classification of twelve VOCs based on their spectral fingerprints.

Further work has extended classification to more complex environments. Tian et al.\ combined a deep neural network with broadband dual-comb spectroscopy to identify methane, acetone, and water vapor from overlapping mid-infrared spectra~\cite{tian2023dual}. Zhang et al.\ developed a domain-adapted CNN model that compensated for sensor drift and achieved stable classification across varying measurement conditions~\cite{zhang2021tdacnn}. 

\subsection{Denoising and Spectral Preprocessing}

Spectroscopic measurements are often affected by noise sources such as detector fluctuations, interference fringes, and baseline drift. These artifacts degrade signal quality and can reduce the accuracy of gas identification and quantification. Data-driven denoising methods have been developed to improve signal fidelity and enhance the performance of downstream models.

Autoencoders are neural networks trained to reconstruct their input through a compressed internal representation. When trained on clean reference spectra, they can remove noise from distorted inputs by learning dominant spectral features. For example, Han et al.\ applied a convolutional autoencoder to Raman spectra, demonstrating effective noise removal while preserving peak structure~\cite{han2024cae}.

Deep convolutional denoisers have also been used in combustion diagnostics and time-resolved spectroscopy. Yoon et al.\ designed a CNN architecture incorporating reversible down-up sampling and a proper orthogonal decomposition (POD)–based loss function to denoise fast-gated flame emission spectra measured under variable conditions~\cite{yoon2022flame}.

Other approaches focus on balancing noise suppression with the preservation of fine spectral features. CNN models have been trained with custom loss functions to outperform traditional filters such as Savitzky–Golay and wavelet denoisers, particularly in low-light Raman measurements~\cite{barton2021}. Liu et al.\ combined guided filtering with CNNs for speckle reduction in SAR imaging, achieving noise suppression while retaining edge definition~\cite{liu2019sar}. Similar principles have been applied in spectroscopy to protect narrowband absorption features.

Ren et al.\ developed a dual-CNN framework that performs both denoising and gas concentration estimation in a single model. Their system showed improved accuracy in detecting methane and N$_2$O mixtures~\cite{ren2023dual}. Another strategy involves adding synthetic noise to training spectra, which forces models to learn robust features that remain stable under varying conditions. This approach improves generalization and allows the model to distinguish true absorption patterns from random fluctuations.

\subsection{Blind Source Separation and Unsupervised Learning}

One of the most difficult problems in multi-species gas sensing is the recovery of individual spectral components from a measured mixture without access to known reference spectra or calibration data. This task falls under blind source separation (BSS), where both the spectral profiles and their relative contributions are unknown.

Linear methods such as principal component analysis (PCA)~\cite{jolliffe2016principal} and independent component analysis (ICA)~\cite{hyvarinen2000independent} have been applied to spectral data to extract basis vectors. However, the results often include components that are mathematically valid but lack physical correspondence to actual gas spectra.

Non-negative matrix factorization (NMF) introduces constraints on the basis and mixing matrices to improve interpretability~\cite{lee1999learning}. This aligns better with absorption spectra, which are inherently non-negative. However, when spectral features of different species overlap, NMF often fails to produce a clean separation of signals.

To improve performance in these scenarios, unsupervised learning methods have been developed. Autoencoders and related deep models can be trained to represent mixtures as combinations of latent features that approximate individual gas contributions~\cite{banko2021deep}. Both convolutional and recurrent autoencoder structures have been used, along with deep attractor networks, to map spectral mixtures into low-dimensional representations that preserve species-specific information~\cite{chen2017deep}.

Recent work has explored the inclusion of physical constraints within these models. Physics-informed architectures enforce conditions such as non-negativity, conservation laws, or spectral consistency through custom loss functions or network design. One approach applies a reference-free autoencoder-BSS framework to infrared spectra, recovering both unknown absorbance profiles and their corresponding concentrations directly from mixture measurements~\cite{sy2024reference}.

Unsupervised methods offer a path forward in cases where reference data are incomplete or unavailable. They enable spectral decomposition without requiring prior knowledge of the system and are particularly useful in field settings, where gas compositions can vary and calibration data may be limited.

\section{Summary of Recent Works}

Machine learning has been applied to a range of gas sensing problems, including controlled laboratory studies and field-based measurements. Table~\ref{tab:supervised_spectroscopy} (adapted from \cite{al2023augmentations} summarizes representative studies, highlighting the methods used, spectral domains, target species, and reported performance. These examples illustrate how different modeling approaches are adapted to specific sensing objectives.

\begin{table}[H]
\centering
\tiny
\caption{Literature review of supervised learning on spectroscopic data.}
\begin{tabular}{p{2cm} p{3.1 cm} p{2cm} p{3cm} p{4cm}}

\hline
\textbf{Reference} & \textbf{Task} & \textbf{Method} & \textbf{Data} & \textbf{Application} \\
\hline
\cite{blazhko2021} & Regression & CNN & Visible/NIR spectroscopy & Corn, tablets, wheat, and soil \\

\cite{huang2021} & Classification & CNN & FTIR/NIR spectroscopy & Corn, tablets, yeast, and mould \\

\cite{huang_csi2021} & Detection/Classification & CNN on CWT & Hand-held Raman spectrometer & Gasoline forensics \\

\cite{fan2019} & Detection/Classification & CNN & Raman spectroscopy & Liquid and powder mixture \\

\cite{ibrahim2020} & Regression & PCA-ANN / CNN & FTIR & Fuel property \\

\cite{dalmiya2022} & Regression & CNN & FTIR & Fuel property \\

\cite{brandt2021} & Denoising & Autoencoder & FTIR/Raman spectroscopy & Microplastic analysis \\

\cite{wahl2020} & Denoising & CNN & Raman spectroscopy & Polyethylene, paraffin and ethanol \\

\cite{horgan2020} & Denoising & ResUNet & Raman spectroscopy & Molecular imaging \\

\cite{zhang2020} & Denoising & Autoencoder & Near-IR hyperspectral images & Milk powder, rice flour and soybean flour \\

\cite{pan2020} & Denoising & CNN & Raman spectroscopy & Hazardous chemicals \\

\cite{barton2021} & Denoising & CNN & Raman spectroscopy & Bio/Polymers \\

\cite{fan2021} & Denoising & CDAE & Raman spectroscopy & Collection of spectra \\

\cite{wu2020} & Denoising & GANs & Multi-object spectroscopy & Astronomy \\

\cite{depaoli2019} & Regression & CNN & Retinal oximetry & Eye hemodynamics \\

\cite{gan2019} & Detection/Classification & FNN-OT & Infrared spectroscopy & Gas sensing \\

\cite{ouyang2019} & Regression & ELM & Infrared spectroscopy & Gas sensing – NOx \\

\cite{alibrahim2020dnn} & Regression & DNN & Infrared spectroscopy & Gas sensing – BTEX \\

\cite{wang2021} & Detection/Classification & CNN & Optical emission spectroscopy of plasma & Gas sensing – VOCs \\

\cite{mozaffari2021} & Regression/Classification & CNN & Raman spectroscopy & Gas sensing – VOCs \\

\cite{fufurin2020machine,fufurin2021numerical,hesham2021deep} & Detection/Classification & CNN & Infrared spectroscopy & Gas sensing – Breath analysis \\
\hline
\end{tabular}
\label{tab:supervised_spectroscopy}
\end{table}

The studies reviewed in this chapter demonstrate the versatility of machine learning in handling tasks such as concentration estimation, classification, spectral denoising, and source separation. However, several limitations remain. Many models rely on fully labeled datasets and assume prior knowledge of all interfering species. Generalization to variable operating conditions, such as changing temperature, pressure, or noise levels, is often limited. In addition, most implementations address one objective—either denoising, quantification, or spectral decomposition—rather than integrating multiple capabilities into a unified model.

This work builds upon the existing literature by developing an end-to-end machine learning framework tailored for multi-species gas detection using laser-based spectroscopy. Both supervised and unsupervised models are employed to address the key challenges of noise, known and unknown interference, unknown species spectra, and model reliability. The methods presented aim to improve performance across a wide range of scenarios, including those where reference data are incomplete or measurement conditions vary.

The following chapters detail the development and evaluation of these models through a series of experimental studies and algorithmic contributions.
    \chapter{Multi-Species Detection in Shock Tube Experiments Using a Single Laser and Deep Neural Networks}\label{chapter3}

This chapter is adapted from Sy et al.~\cite{sy2023multi}, titled ``Machine learning-enhanced laser absorption spectroscopy for multi-species detection in shock tubes using a single laser,'' published in \textit{Combustion and Flame}.

\section{Introduction}

Chemical kinetic experiments involving the oxidation or pyrolysis of fuels often produce a dynamic mixture of intermediate species that must be measured simultaneously to resolve reaction pathways. Traditional laser-based diagnostics typically rely on single-species detection and require multiple lasers or repeated experiments to track different gases. For instance, Pinkowski et al.~\cite{pinkowski2019multi} and Cassady et al.~\cite{cassady2020ethane, cassady2020propane} demonstrated multi-laser strategies for speciation, but these methods entail complex optical setups and increase experimental uncertainty due to the need for repeated shock events. Furthermore, multidimensional linear regression (MLR) has been applied in scanned-wavelength diagnostics~\cite{mhanna2020benzene}, but it remains limited when species exhibit broad, overlapping absorption features.

This chapter introduces a compact mid-infrared diagnostic for selective and simultaneous detection of multiple species in shock tube experiments using a single interband cascade laser (ICL) tuned over the 3038–3039.6~cm$^{-1}$ region. The diagnostic leverages a deep denoising autoencoder (DDAE) to clean measured absorbance spectra, which are subsequently analyzed using MLR to determine the contributions of methane, ethane, ethylene, propane, and propyne. This integrated approach allows time-resolved multi-species detection without the need for multiple lasers, broadband sources, or repeated shots, and represents, to the best of our knowledge, the first such implementation using a narrow-scanning laser in shock tube environments.

\section{Experimental Details}
\subsection{Optical Setup}
\begin{figure}[H]
    \centering
    \includegraphics[width=0.75\linewidth]{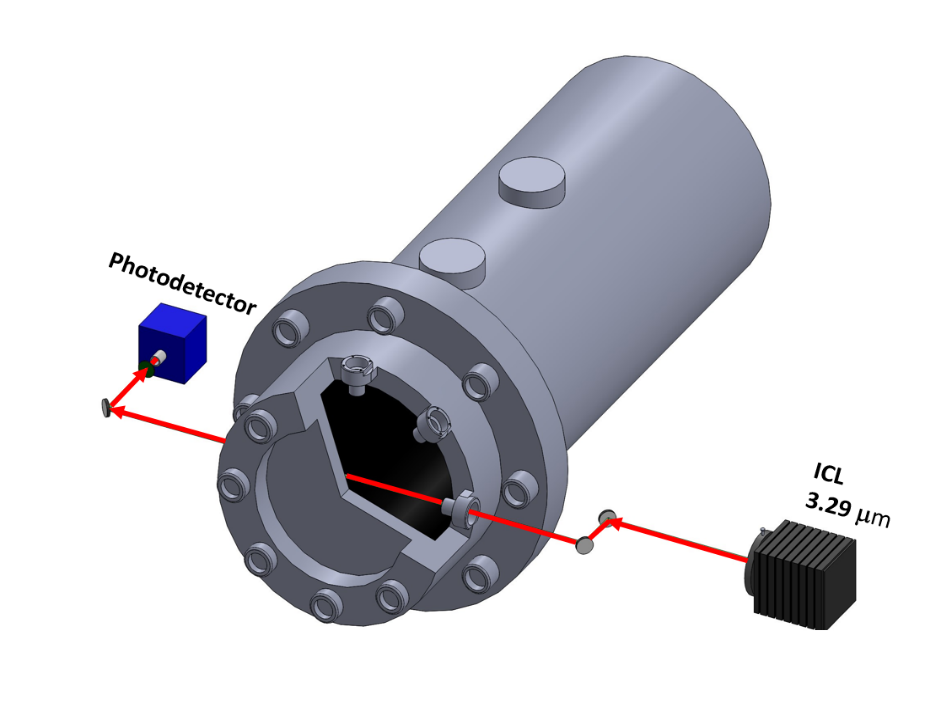}
    \caption{Optical schematic of the sensor through the shock tube cross-section. Red arrow indicates the direction of the propagation of the laser.}
    \label{fig:ST_cnF}
\end{figure}

The simple optical setup employed in this work is shown in Fig. \ref{fig:ST_cnF}. A distributed feedback (DFB) inter-band cascade laser (ICL, Nanoplus GmbH) centered at 3.29 m with an output power of 1 mW was used in this work. The shock tube was equipped with two ZnSe windows (Thorlabs) to transmit the laser beam. A 7.62 cm germanium etalon with a free-spectral range (FSR) of 0.0164 $cm^{-1}$ was used to convert the scan time to wavenumbers. A band-pass filter (FB-3300-300, Thorlabs) was used to minimize thermal emission reaching the detector. A ramp injection current was applied to scan the laser at a repetition frequency of 10 kHz, which enabled tuning range of 3038 – 3039.6 $cm^{-1}$ in every laser scan. A DC-coupled, four-stage thermoelectrically-cooled PV (photovoltaic) detector was used to collect the laser signal (bandwidth of 1.5 MHz, Vigo Systems).
\subsection{Wavelength Selection}
\begin{figure}[H]
    \centering
    \includegraphics[width=\linewidth]{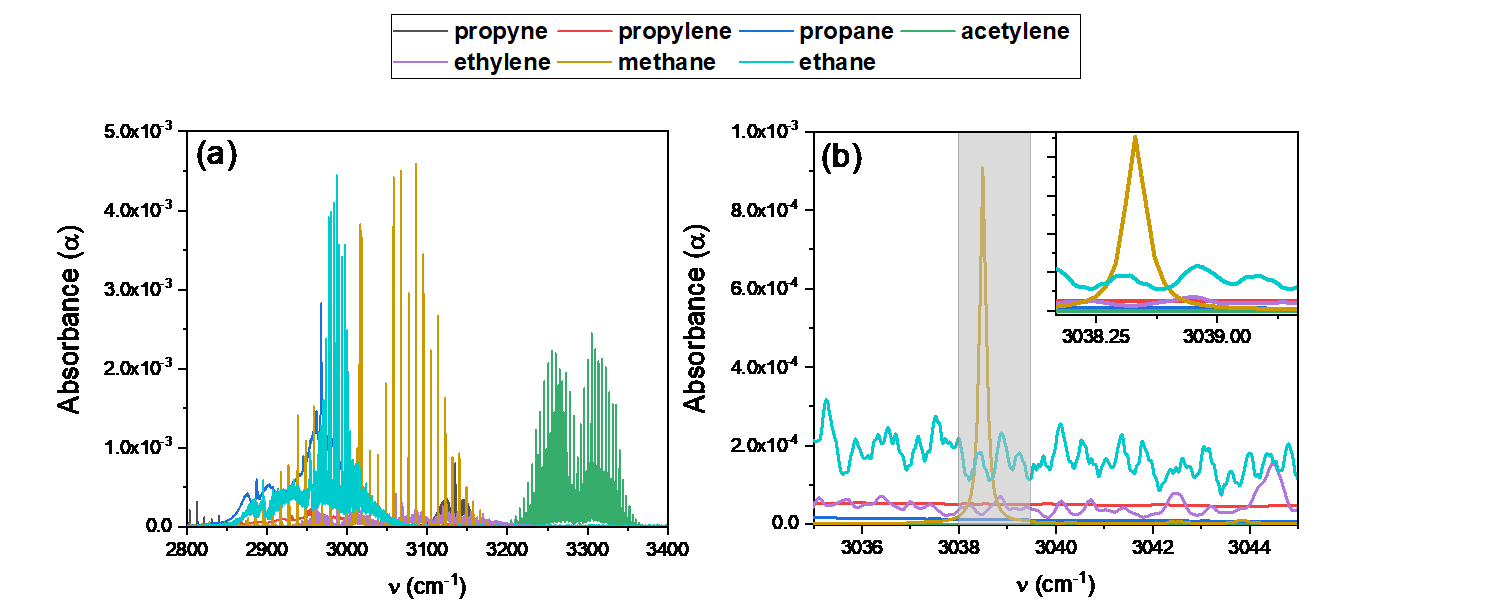}
    \caption{(a) Spectra of methane, ethane, ethylene, acetylene, propane, propene, and propyne at T = 298 K, P = 1 atm, ppm, and L = 1 m in the C-H stretch range of 2800 – 3400 $cm^{-1}$. The spectral data were obtained from PNNL database [40]. (b) Spectra over 3035 - 3045 $cm^{-1}$ which is the scan range of the ICL. The inset presents a zoomed-in view of the selected region for this work.}
    \label{fig:WS_CnF}
\end{figure}

In order to demonstrate our multispeciation diagnostic methodology, we targeted the pyrolysis of ethane and propane as representation cases. These two hydrocarbons are important constituents of natural gas and have been the subject of numerous experimental and chemical kinetic modeling studies. We conducted pyrolysis simulations using Chemkin-Pro software and AramcoMech3.0 model \cite{zhou2018butadiene} to predict the key evolving species during ethane / propane pyrolysis. AramcoMech3.0 has been shown to predict ignition delay times, flame speeds and species concentrations that are in good agreement with experimental measurements, as demonstrated by the previous work of Cassady et al. \cite{cassady2020ethane, cassady2020propane}. To accurately simulate temperature and pressure changes during pyrolysis, it is necessary to use an appropriate gas dynamic model. In our case, we chose to use the constant volume (CV) model, which has been found to be suitable for our shock tube facility \cite{sajid2015qcl}. 
During the pyrolysis of ethane, the major evolving hydrocarbons are methane, ethane, ethylene, and acetylene (see Fig. S1). In the case of propane, the major species are methane, ethane, ethylene, propane, propylene, and propyne (see Fig. S2). Most hydrocarbons absorb in the infrared region near 3000 $cm^{-1}$ due to the C–H stretch vibration. Fig. \ref{fig:WS_CnF} (a) plots  the absorption cross-sections of the evolving species within the C-H stretch region of 2800-3400 $cm^{-1}$ \cite{sharpe2004database}. In order to detect the target species, we used an interband cascade laser (ICL) capable of emitting near 3035-3045 $cm^{-1}$, where most of the evolving/targeted hydrocarbons have a measurable absorbance, as shown in Fig. \ref{fig:WS_CnF} (b). By scanning the laser at a repetition frequency of 10 kHz, we were able to cover a maximum of 2 $cm^{-1}$ in each scan and targeted the wavelength region of 3038.0 – 3039.6 $cm^{-1}$. In this region, we were able to detect three major hydrocarbons (methane, ethane, and ethylene) during ethane pyrolysis, and five major hydrocarbons (methane, ethane, ethylene, propene, and propyne) during the pyrolysis of propane.

\subsection{High temperature Absorption Cross-sections}
\begin{figure}[H]
    \centering
    \includegraphics[width=0.65\linewidth]{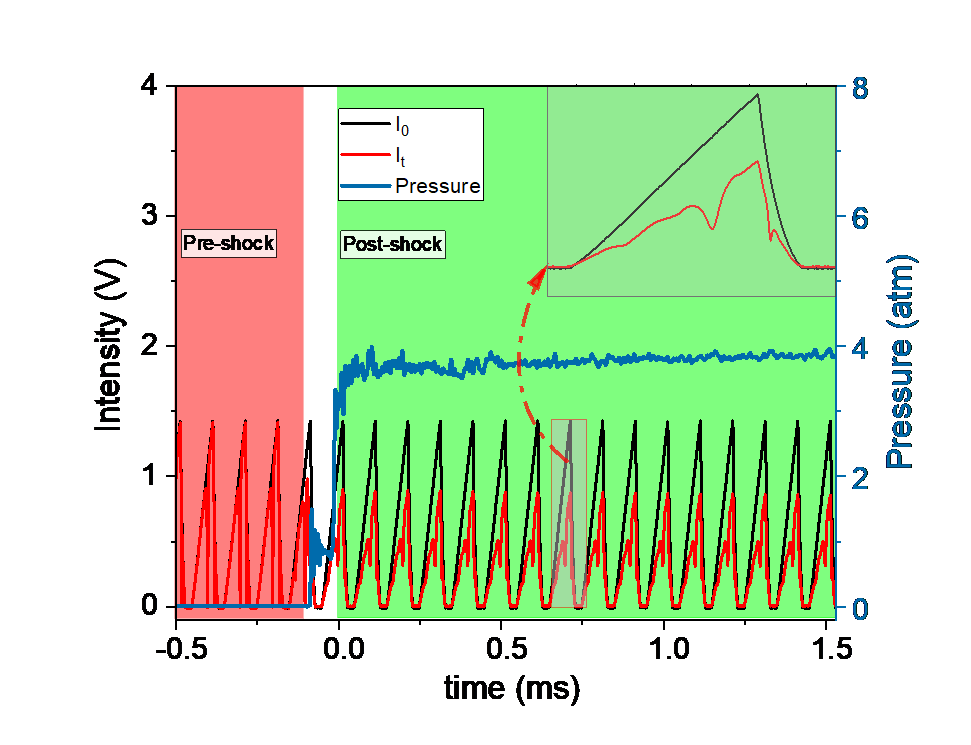}
    \caption{A representative shock experiment of CH4/Ar. The red region indicates pre-shock conditions at T = 298 K and P = 0.105 atm, while the green region indicates post-reflected-shock conditions at T = 1300 K and P = 3.78 atm. Left vertical axis corresponds to laser intensities, and the right vertical axis corresponds to the pressure trace. The inset shows a post-shock laser scan.}
    \label{fig:rep_shock}
\end{figure}

\begin{figure}[H]
    \centering
    \includegraphics[width=\linewidth]{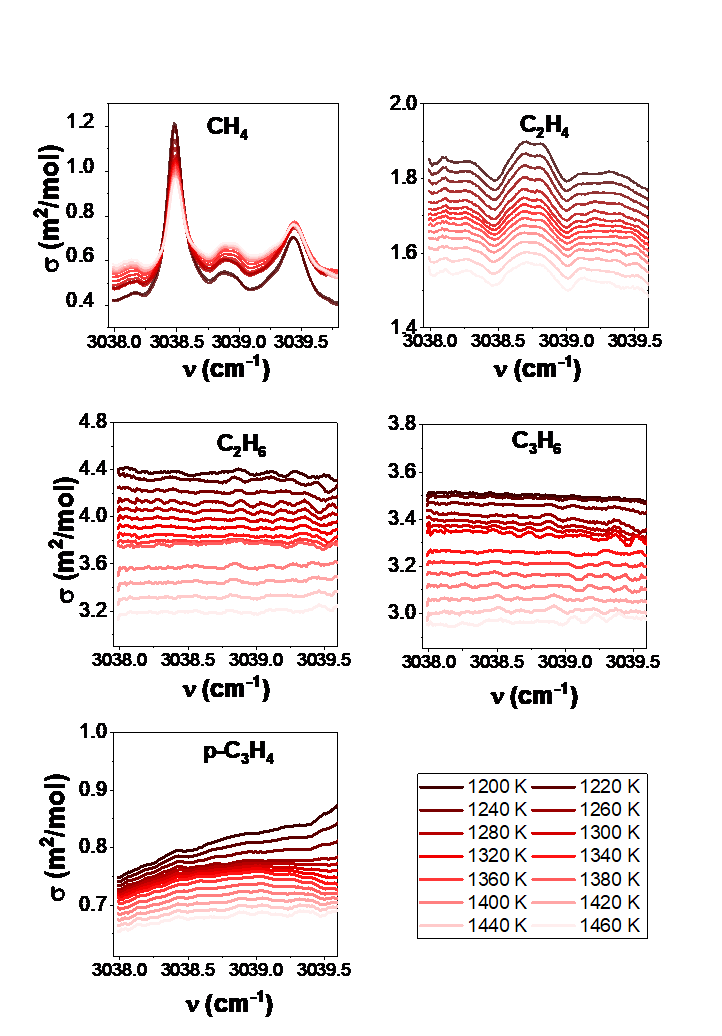}
    \caption{Measured temperature-dependent absorption cross-sections of methane, ethane, ethylene, propylene, and propyne over T = 1200 – 1460 K and P = 3.5 – 4.2 atm.}
    \label{fig:cs_cnf}
\end{figure}

Spectral analysis in the previous section was based on spectra simulations at ambient conditions since high-temperature spectra of the targeted species is not available in literature or any database. Therefore, high-temperature absorption cross-sections of methane, ethane, ethylene, propylene, and propyne were measured experimentally in this work. Shocks were performed over T = 1200 – 1460 K and P = 3.5 – 4.2 atm with known mole fractions for all species over 3038 – 3039.9 $cm^{-1}$. Mole fraction values were varied over 2 – 10\% in 
argon to achieve high signal-to-noise ratio in these experiments. A representative shock for determining the cross-section of methane at T = 1300 K, P = 3.78 atm, and CH4/Ar is shown in Fig. \ref{fig:rep_shock}. Incident intensity $I_0$ was determined with the shock tube being at vacuum, while transmitted intensity $I_t$ was collected behind the reflected shock wave. The inset shows a zoom-in view of one laser scan behind the reflected shock; methane absorption transition near 3038.5 $cm^{-1}$ is clearly depicted. The red shaded region indicates pre-shock conditions, and the green region indicates post-shock conditions. Time zero is marked by the step rise in pressure when the 
reflected shock wave passes the measurement location (2 cm from the end-wall). To measure the cross-sections of species, such as ethane, which are prone to decomposition at high temperatures, we focused solely on the initial scan taken at time zero (100 µs). This approach was adopted to prevent any potential interference from evolving species that occur during the decomposition process. Conversely, for species that do not undergo decomposition (evolving species), the cross-section is determined by averaging measurements over a duration of 1.5 ms, effectively reducing noise. Absorption cross-section measurements of all target species, which have been interpolated from the measured temperatures to nominal temperatures, are shown in Fig. \ref{fig:cs_cnf} over the investigated conditions. The results demonstrate that the absorption cross-sections are highly dependent on temperature and relatively insensitive to pressure variation within the range of pressures examined in this study. Uncertainty of the measured cross-sections is given in the supplementary material S5.

\section{Numerical Methods}
\subsection{Multidimensional Linear Regression}
Linear regression is a statistical method used to examine the relationship between two or more variables. When applied to multi-dimensional datasets with multiple dependent and independent variables, it allows for a comprehensive analysis of complex relationships. In the context of multi-species detection, multi-dimensional linear regression can be employed to analyze the relationship between the absorbance of a mixture containing multiple species and the concentrations of those species. This is represented as:

\begin{equation}
    y_i = a_0 + \sum_{k=1}^{n} a_k x_{ki}
\end{equation}

Here, $y_i$ represents the $i^{th}$ data point of the measured absorption, $a_k$ signifies the contribution of each absorbing component to the spectrum, where $k$ ranges from 1 to $n$ and refers to the targeted species, $x_{ki}$ is the intensity of the $k^{th}$ reference spectrum at the frequency of the $i^{th}$ data point, and $a_0$ is an offset that is set to zero.

\begin{figure}[H]
    \centering
    \includegraphics[width=\linewidth]{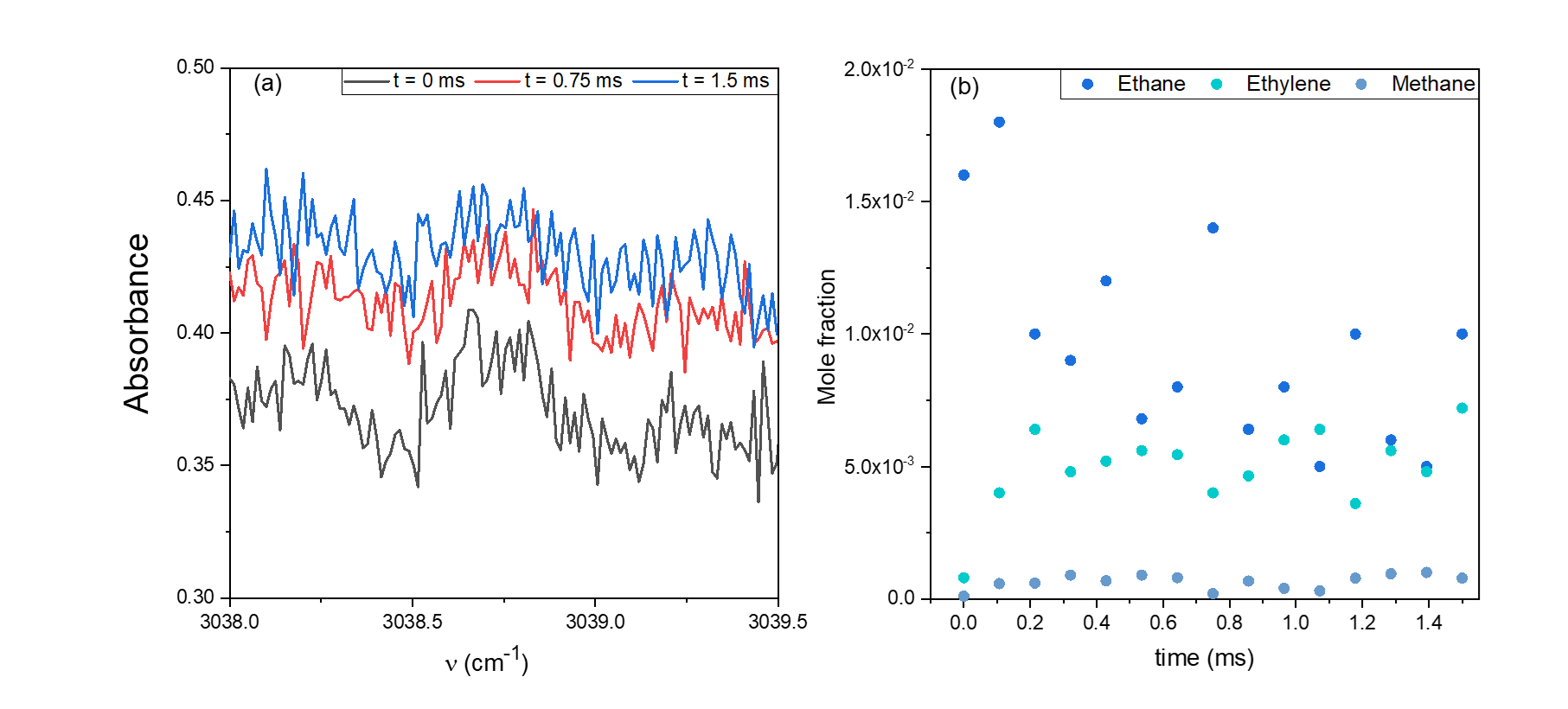}
    \caption{(a): Measured absorbance signals during the pyrolysis of 2\% ethane/Ar at T = 1312 K and P = 3.88 atm. (b): Predicted mole fractions, using the MLR model, from the composite spectra shown in Fig. \ref{MLR_cnF}(a).}
    \label{MLR_cnF}
\end{figure}

The MLR method works well when absorbing species have resolved spectra, as has been shown in previous works \cite{mhanna2022benzene_arxiv}. We applied MLR here to evaluate mole fractions of evolving species from the measured composite spectra during ethane pyrolysis; Fig.\ref{MLR_cnF} shows that the MLR model could not perform well, particularly due to the accumulated noise in the absorbance signal. There are several sources of noise here including low temporal resolution, low sampling rate of the data acquisition system, beam steering noise, and background detector noise. 

In the presence of noise, the measured data may exhibit significant random variability, which hinders discerning the underlying relationships among the variables. Under such circumstances, the performance of the MLR model may be compromised, potentially leading to biased estimates or predictions \cite{breiman1991ii}. In order to overcome the challenges posed by noisy data and to obtain reliable results, it is necessary to employ supplementary techniques such as data cleaning or regularization. In this study, we employed deep denoising autoencoders to preprocess the noisy signals obtained from shock tube experiments prior to inputting them into the MLR model.

\subsection{Deep Denoising Auto-encoders}
\begin{figure}[H]
    \centering
    \includegraphics[width=0.75\linewidth]{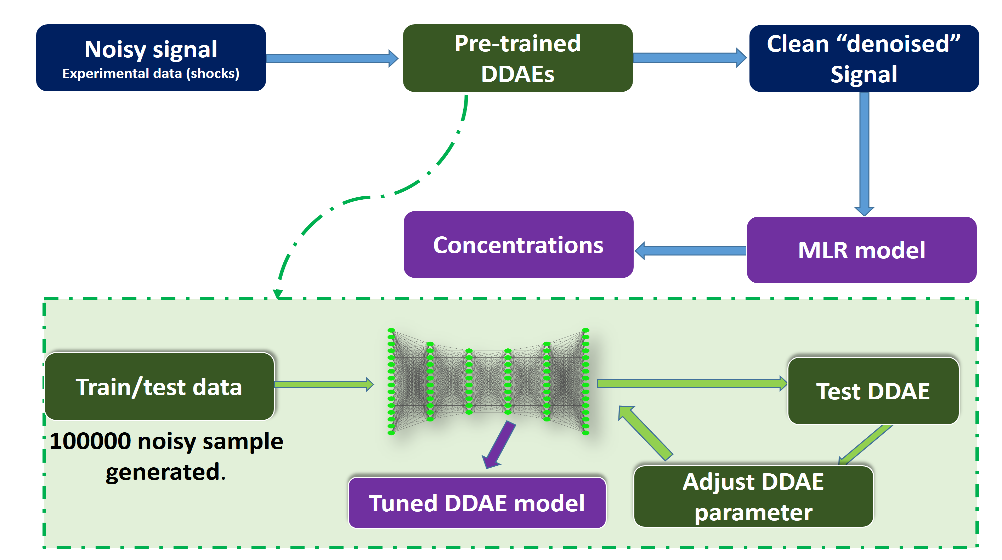}
    \caption{Flow chart of the overall process followed in this work including the training of deep denoising auto-encoders. }
    \label{fig:flow_cnf}
\end{figure}

Deep denoising autoencoders (DDAEs) are neural networks that are taught to denoise signals by learning to recreate the original, clean signal from a noisy/distorted version of it. This is done by training DDAEs on a dataset of clean signals and their corresponding noisy/distorted versions \cite{vincent2008dae}. A key advantage of using a DDAE for signal denoising over standard filters is that it can learn to denoise signals in a way that is tailored to the specific characteristics of the signals in the training dataset \cite{araki2015speech}. This can make it more effective at denoising signals compared to standard filters such as butterworth \cite{mello2007emg} or savitsky-golay \cite{savitzky1964sgolay} which use a fixed set of parameters for denoising. Additionally, DDAEs can be trained to denoise a wide variety of different types of signals, whereas standard filters are typically only effective for a specific type of signal \cite{kounovsky2017cnn}. 
In this work, autoencoders were used to create a model based on DDAEs that maps noisy absorbance signals to denoised “clean” signals. The measured high-temperature absorbance profiles of the target species were combined at various ratios, carefully selected to cover all possibilities that might be encountered in experiments, yielding 100,000 composite absorbance spectra. To scale the features within the same range and facilitate faster learning for the DDAE model, min-max normalization was applied to the composite absorbance spectra. Gaussian white noise was chosen as the noise model, as it is well-suited to represent noise from electronic components such as lasers, detectors, and data acquisition systems, often referred to as "electronic noise " \cite{boyat2015noise}. The key properties of Gaussian white noise include: (i) Gaussian amplitude distribution, where noise values follow a normal distribution with a specified mean and standard deviation, (ii) uncorrelated noise across time, meaning the noise values at different time points are independent of each other, and (iii) frequency-independent noise, where the noise's power is evenly distributed across all frequencies. Other noise models were explored and their performance is reported in the supplementary material (Fig. S4).
Gaussian white noise of varying levels was added to the spectra to obtain SNRs of 40 – 70, which matches with our experimentally observed signals. For each noise level, a DAE model was created to apply static noise-aware training based on a random 80/20 split. Each model contained six hidden layers with 256, 128, 32, 32, 128, and 256 fully connected nodes, respectively. DDEAs were tuned with these hyper-parameters and hidden layers used ReLU (Rectified Linear Unit) as the activation function \cite{agarap2018relu}. The number and size of hidden layers were selected by trial and error. During training, mean squared error (MSE) was used as the loss function and Adamax algorithm was implemented as the optimizer. Hyper-parameter tuning was achieved using Random Search CV \cite{bergstra2012random}; various folds of the training data were fed to the DDAE model using varying sets of hyper-parameters. The best model which gave the lowest root-mean-squared error (RMSE) among the reconstructed “denoised” signals was chosen. To prevent overfitting, an early stopping criterion was used with a patience of 30 epochs. Keras software package was used with Tensorflow as the backend to develop and train our DDEA models \cite{gulli2017keras}. Measured absorbance data were fed to the DDAE models, resulting in denoised output vectors, which in turn were fed to an MLR (multi-linear regression) model to split the composite absorbance into the contributions/ratios from the absorbing species \cite{mhanna2022cleos}.  A flow chart of the overall process is shown in Fig. \ref{fig:flow_cnf}. 

\begin{figure}[H]
    \centering
    \includegraphics[width=0.75\linewidth]{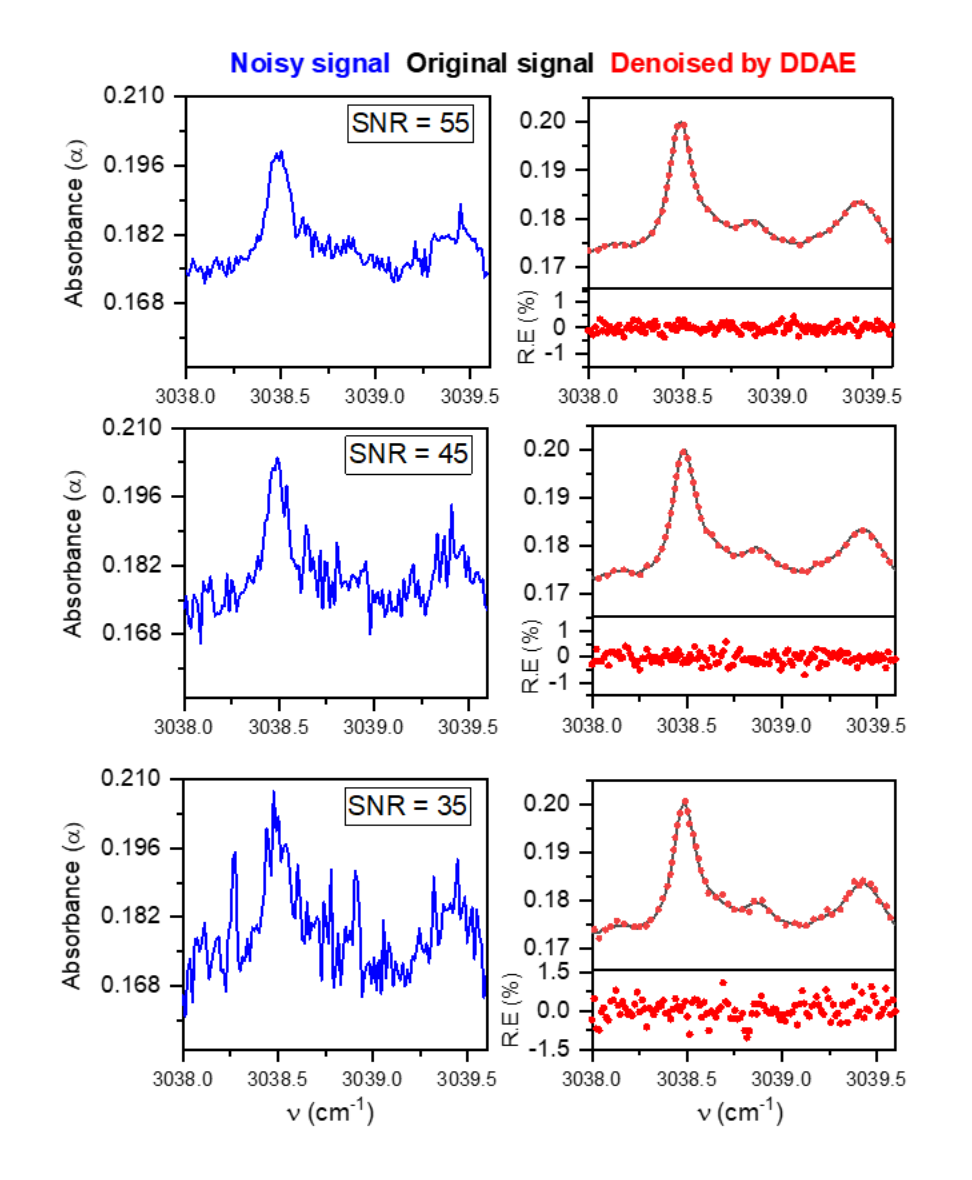}
    \caption{Results of signal cleaning with DDAEs. Blue lines show simulated noisy signals at different noise levels, while black solid lines and red dots display the original and denoised signals, respectively. The relative errors (R.E.) are also shown to evaluate the performance of DDAEs. }
    \label{fig:denoised_cnf}
\end{figure}

The DDAE model demonstrated strong performance on simulated noisy samples. The results shown in Fig.~\ref{fig:denoised_cnf} demonstrate the ability of DDAEs to accurately reconstruct the clean signal, despite varying levels of noise. A relative error of approximately 1.5\% between the original signal and the signal denoised by DDAE was achieved. This suggests that the DDAE model could be useful in real-world applications to tackle noise, such as in shock tube experiments, signal processing, or image restoration. Relative error is defined as:

\begin{equation}
    R.E.(\%) = 100 \times \frac{A_{\text{original}} - A_{\text{denoised}}}{A_{\text{original}}}
\end{equation}

where $A_{\text{original}}$ is the original or \textquotedblleft clean\textquotedblright\ signal and $A_{\text{denoised}}$ is the signal denoised by DDAEs.

\section{Results and Discussion}
\subsection{DDAE results}
\begin{figure}[h]
    \centering
    \begin{minipage}{0.49\textwidth}
           \centering
            \includegraphics[width=\linewidth]{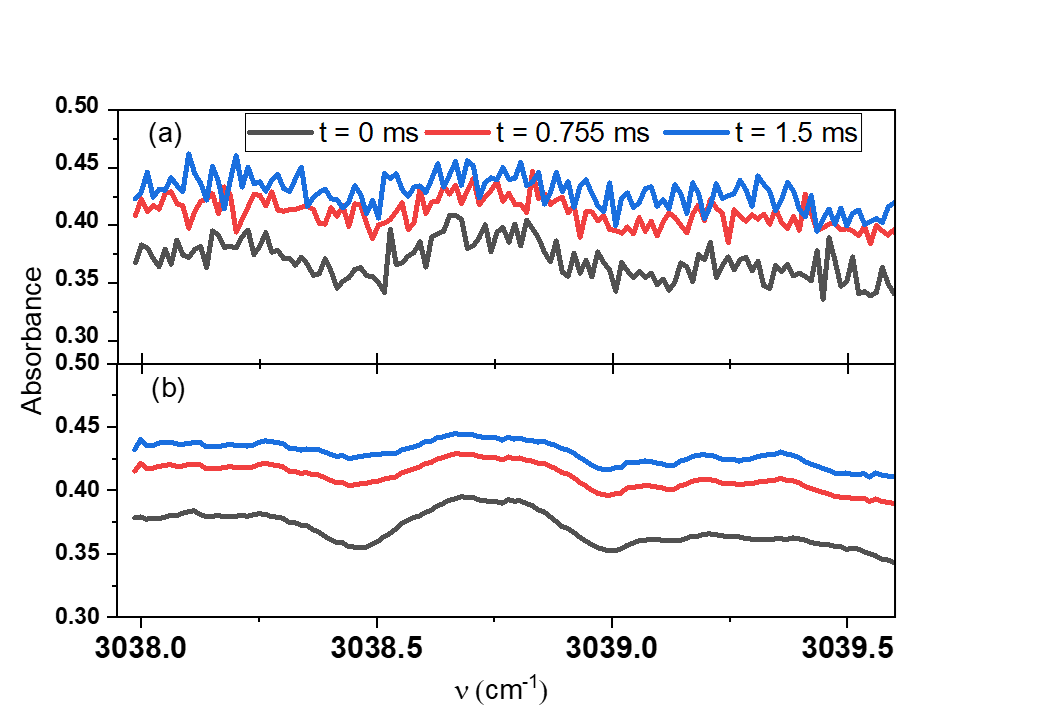}
    \end{minipage}\hfill
    \begin{minipage}{0.49\textwidth}
        \centering
            \includegraphics[width=\linewidth]{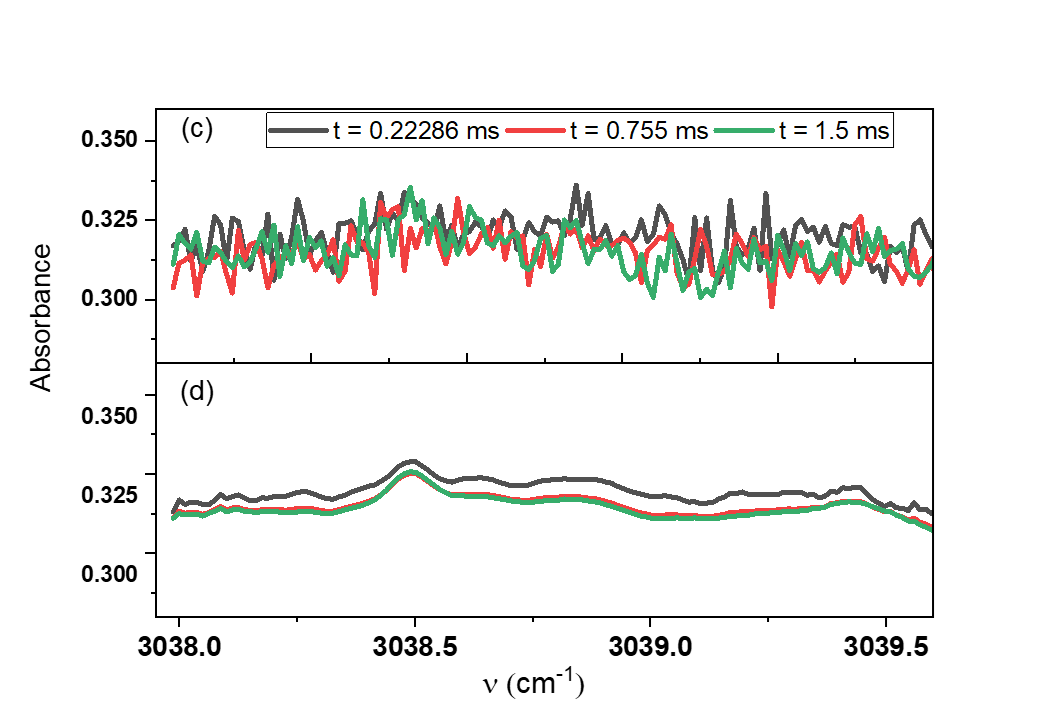}
    \end{minipage}
    \caption{Noisy composite absorbance spectra during the pyrolysis of 2\% ethane/Ar at T = 1300 K, and P = 4 atm. (b) Corresponding denoised signals. (c) Noisy composite absorbance spectra during the pyrolysis of 2\% propane/Ar at T = 1300 K, and P = 4 atm. (d) Corresponding denoised signals.}
    \label{fig:denoised_exp_cnf}
\end{figure}

DDAEs exhibited great success on simulated spectral data, reconstructing the clean “denoised” signals almost fully. DDAEs were next applied on experimental data collected from shocks to predict denoised “clean” spectra. Fig. 8(a) shows noisy composite absorbance spectra measured during the pyrolysis of 2\% ethane/Ar at T = 1312 K and P = 3.88 atm, at t = 0, 0.75 and 1.5 ms, and Fig. \ref{fig:denoised_exp_cnf}(b) shows the corresponding denoised signals. Fig. \ref{fig:denoised_exp_cnf}(c) shows noisy composite absorbance spectra during the pyrolysis of 2\% propane/Ar at T = 1300 K and P = 3.77 atm, at t = 0.22, 0.75, and 1.5 ms, and Fig. \ref{fig:denoised_exp_cnf} (d) shows the corresponding denoised signals. Significant enhancement of the SNR was achieved, which enabled detection of the hidden spectral features in the noisy absorbance spectra. Such denoising and feature extraction is essential for decomposing the composite spectra to the contributions of the absorbing species.


\subsection{Pyrolysis Results}

\begin{figure}[H]
    \centering
    \includegraphics[width=\linewidth]{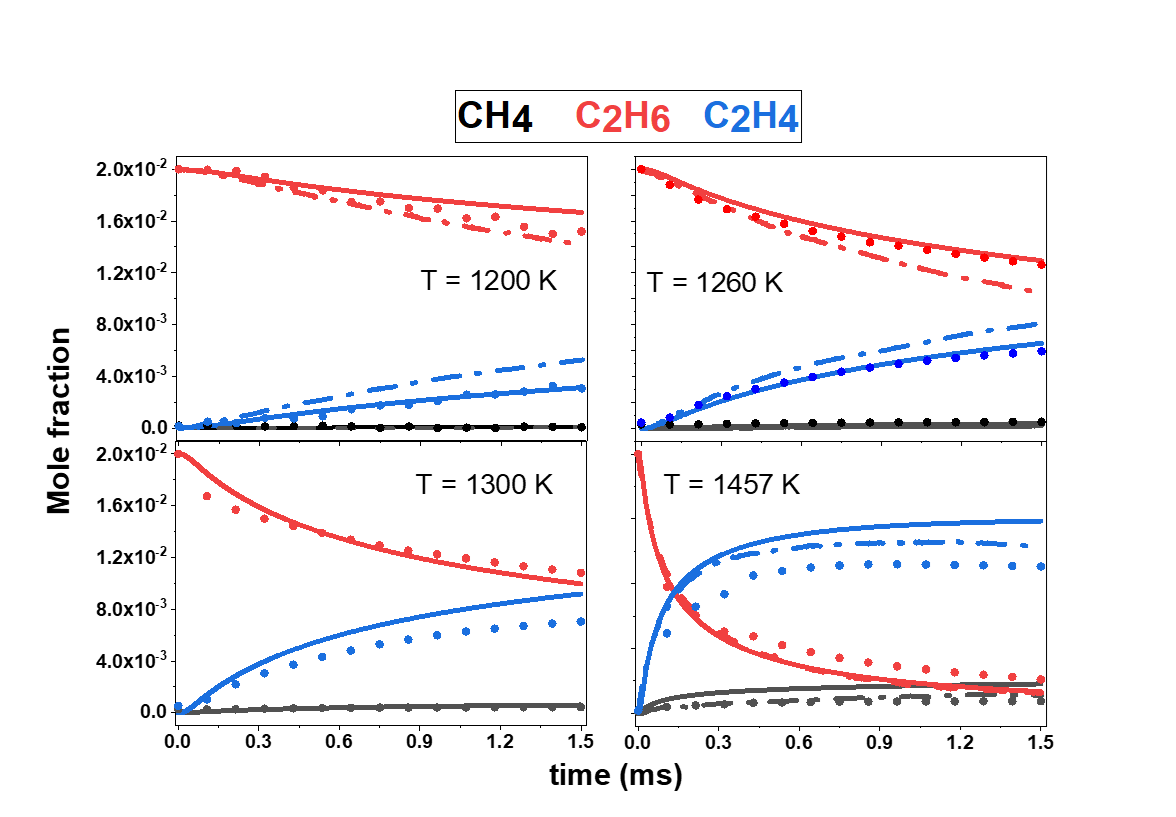}
    \caption{Mole fraction time-histories during the pyrolysis of 2\% ethane/Ar at (a) 1200 K, (b) 1260 K, (c) 1300 K, and (d) 1457 K. Comparisons to AramcoMech 3.0 \cite{zhou2018butadiene} (solid line) and Cassady et al. \cite{cassady2020ethane} (dashed line) are also shown. }
    \label{fig:res_cnf_eth}
\end{figure}

Ethane and propane combustion chemistry has been extensively studied and well-validated by various chemical kinetic models, such as AramcoMech 3.0 [38]. Our methodology was applied to measure time-histories of the evolving species during the pyrolysis of ethane and propane.  Mixtures of 2\% ethane/Ar and 2\% propane/Ar were shock-heated to reach the desired temperatures (1200 – 1460 K) and pressures (3.5 – 4.2 atm). Raw absorbance spectra were fed to our DDAE and MLR models to recover the absorbance contribution of each target species. Mole fractions of the target species were determined using MLR and our measured high-temperature absorption cross-sections. 
The key absorbing hydrocarbons (methane, ethane, and ethylene) in the selected wavelength region were detected during ethane pyrolysis. Fig. \ref{fig:res_cnf_eth} shows a comparison of the time-histories of the target species (symbols) obtained using our method that utilizes a single laser to measure multiple species, with AramcoMech 3.0 predictions (solid lines) \cite{zhou2018butadiene} and the measurements of Cassady et al. \cite{cassady2020ethane} (dashed lines) which employed a multi-color technique combining multiple single wavelength lasers, each targeting a specific species. Due to temperature decrease during pyrolysis, we simulated temperature profiles during pyrolysis using AramcoMech 3.0 model, and took the cross sections corresponding to the average temperature in each 100 s. The results of this approach are shown as triangle symbols. For the rest of the test time 0.2 – 1.5 ms, the temperature drop is minimal, which does not significantly affect the cross sections.

\begin{figure}[H]
    \centering
    \includegraphics[width=\linewidth]{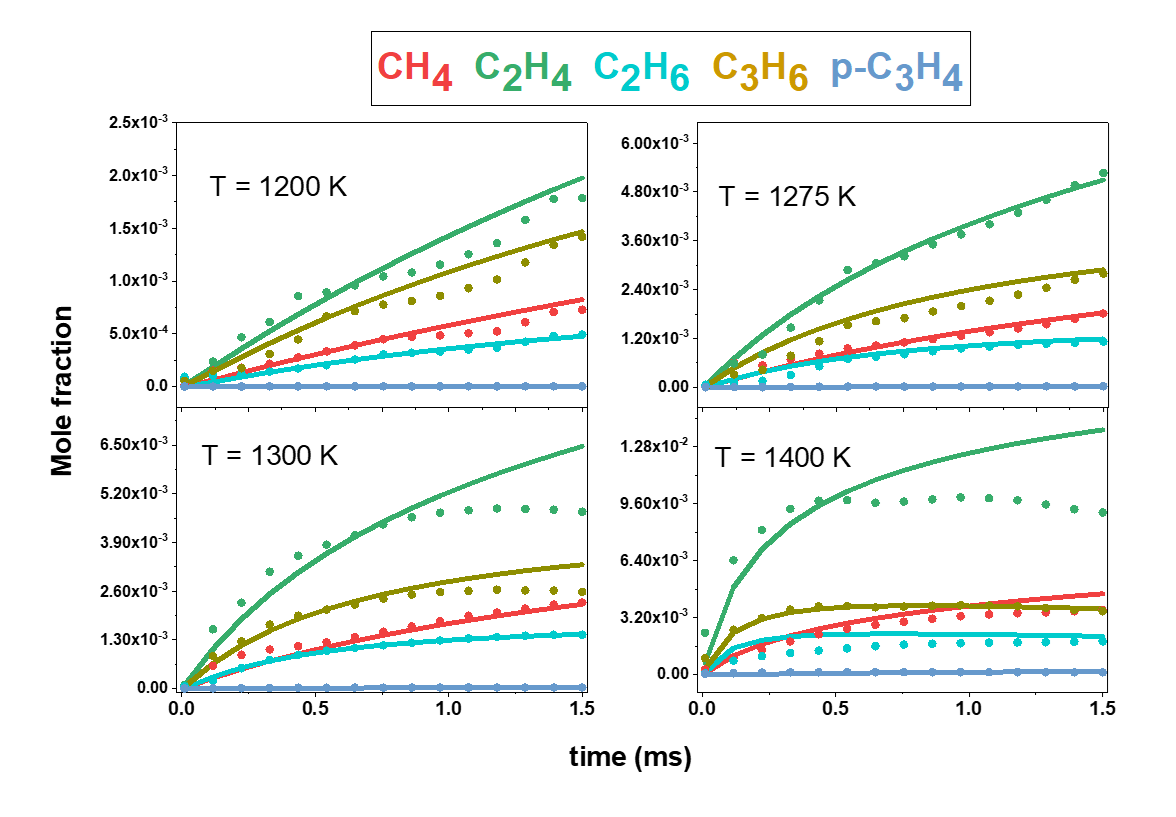}
    \caption{Mole fraction time-histories during the pyrolysis of 2\% propane in argon at (a) 1200 K, (b) 1275 K, (c) 1300 K, and (d) 1400 K. Comparisons to AramcoMech 3.0 \cite{zhou2018butadiene} (solid lines) are shown.}
    \label{fig:res_cnf_prop}
\end{figure}

For propane pyrolysis, major absorbing hydrocarbons (methane, ethane, ethylene, propylene, and propyne) in the selected wavelength region were detected using our methodology. Fig. \ref{fig:res_cnf_prop} compares time-histories of the target species (symbols) with AramcoMech 3.0 predictions (solid lines)\cite{zhou2018butadiene}. There is generally a good agreement between the results obtained by our methodology and the simulated/experimental results.

The cumulative uncertainty errors, accounted for from absorbance, temperature, pressure, mixture preparation, and denoising models, is 9.53\%.  There is a good agreement between the results obtained by our methodology and the simulated/experimental results.


    \chapter{Spectral Interference Mitigation: Technique to Tackle Noise and Unknown Interference }\label{chapter4}
This chapter is adapted from the accepted manuscript version of a paper pending publication in the \textit{Optica Sensing Congress}.

\section{Introduction\label{sec:ch4:introduction}} \addvspace{10pt}
The transition toward studying larger hydrocarbons in combustion research introduces greater complexity in chemical reaction pathways and byproduct formation~\cite{goldenstein2017infrared, farooq2022laser}. In high-temperature environments such as shock tubes and rapid compression machines (RCMs), species evolve rapidly, necessitating high-resolution diagnostics capable of capturing transient behavior on millisecond timescales~\cite{hanson2014recent, elkhazraji2024mid, de2007mass, wuilloud2004gas}. While broadband techniques such as FTIR, GC-MS, and frequency combs offer wide spectral coverage, they remain limited in time-resolved applications due to slow acquisition rates and long sampling paths.

To meet the demands of fast-reacting environments, interband cascade lasers (ICLs) and quantum cascade lasers (QCLs) have gained traction as compact, high-speed sources for tunable laser absorption spectroscopy (TLAS)~\cite{jeevaretanam2023transient, elkhazraji2024selective, mhanna2023laser}. However, their narrow tuning ranges pose challenges when target species overlap with unknown or broadband interferents. In such cases, resolving overlapping spectral features within a confined spectral window requires robust interference mitigation strategies that preserve temporal fidelity and quantification accuracy.

Various methods have been developed to address spectral interference, including multi-wavelength diagnostics~\cite{pinkowski2019multi}, convex optimization-based speciation~\cite{sy2025multi}, and modified free-induction decay (m-FID) filtering techniques~\cite{mhanna2023selective}. While effective, these techniques often rely on multiple optical channels, prior knowledge of reference spectra, or tailored spectral features. Recently, machine learning methods have shown promise for interference mitigation without explicit reference libraries~\cite{sy2024reference, sy2023multi, sy2025multi}. Nonetheless, supervised models trained on labeled data often fail to generalize when interference arises from previously unseen species. Data augmentation has been proposed as a solution to improve robustness~\cite{al2023augmentations}, yet the challenge of unknown spectral interference remains unresolved.

In this work, we introduce HT-SIMNet, an unsupervised neural network framework for interference suppression in composite absorbance spectra. HT-SIMNet builds on the Noise2Noise learning paradigm~\cite{lehtinen2018noise2noise}, originally developed for image denoising, by leveraging pairs of synthetically corrupted spectra that share the same underlying signal. Spectral augmentations—including mirroring, flipping, and dilation—are applied to simulate realistic interference and train the network to recover clean species spectra. The model is evaluated on mixtures of \(C_1\)–\(C_3\) hydrocarbons with unknown interferents in static gas cell experiments, and further validated during n-butane pyrolysis. This chapter demonstrates that HT-SIMNet effectively isolates target species and enables accurate quantification without requiring prior information about interferents.

\section{Experimental Methodology}

\subsection{Optical Setup}  

\begin{figure}[H]
\centering
\includegraphics[width=0.75\textwidth]{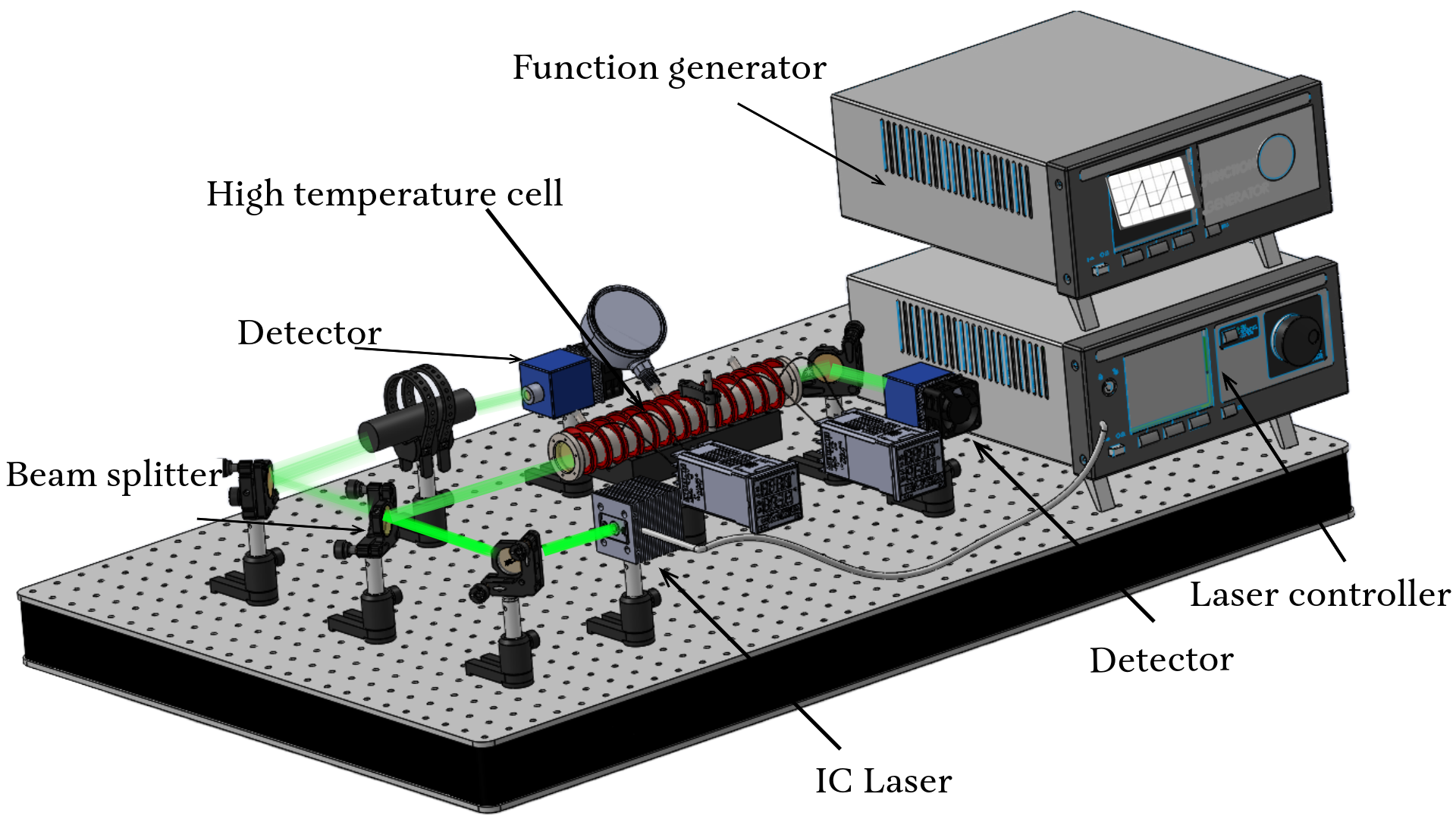}
\caption{Schematic of the experimental setup, including the ICL laser source, heated absorption cell, calibration etalon, and detector.}
\label{opticalsetup}
\end{figure}

The experimental setup (Fig. \ref{opticalsetup}) was designed for high-temperature measurements to evaluate our approach. A distributed feedback (DFB) inter-band cascade laser (ICL) from Nanoplus GmbH, operating at 3.34 µm with 10 mW output power, was used. The laser beam passed through a 21 cm heated stainless-steel cell with ZnSe windows, sealed with copper gaskets to withstand high temperatures \cite{farouki2024measurement}. Heating tapes controlled the temperature, while five K-type thermocouples monitored the thermal profile, ensuring consistency with previous setups \cite{farouki2024measurement}. The laser was modulated at 100 Hz with a sawtooth waveform, and a 7.62 cm germanium Fabry-Perot etalon provided wavenumber calibration. A Thorlabs FB-3300–300 band-pass filter minimized thermal radiation interference, and the transmitted signal was detected by a four-stage, thermoelectrically cooled photovoltaic detector (Vigo Systems) with a 1.5 MHz bandwidth.

\subsection{Wavelength Selection}
\begin{figure}[h!]
\centering
\includegraphics[width=192pt]{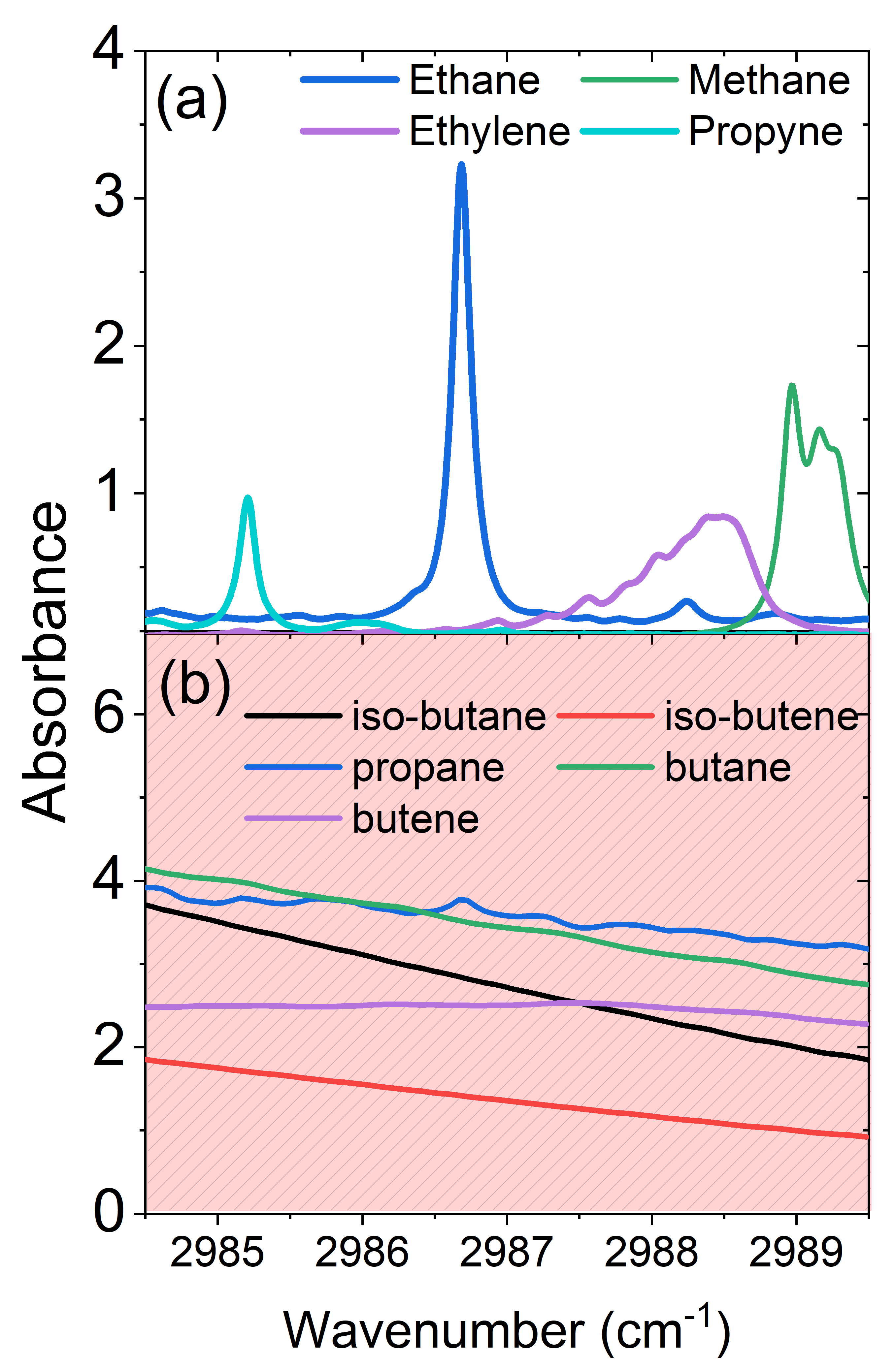}
\caption{Absorbance spectra of the target species under conditions (T = 295 K, P = 1 atm, L = 21 cm, and $\chi$ = [0.4 - 1]\%). Red shaded area shows the absorbance spectra of the unknown interferents under the same conditions.}
\label{ch4:spectra}
\end{figure}  
To rigorously evaluate the effectiveness of HT-SIMNet in mitigating spectral interference, we selected a wavelength region that encompasses absorption features from both target species and interferents. The spectral range of 2984.5–2989.5 cm\(^{-1}\) was chosen as it captures the distinct C–H stretch absorption signatures of C\(_1\)-C\(_3\) hydrocarbons, specifically methane, ethane, ethylene, and propyne~\cite{sy2025multi}. Those species are particularly relevant due to their fundamental role as intermediates in hydrocarbon decomposition and combustion chemistry.  
The chosen spectral range contains both target species and interfering byproducts, making it well-suited for evaluating HT-SIMNet’s performance under both reactive and non-reactive experimental conditions.
Figure~\ref{ch4:spectra}(a) illustrates the absorbance spectra of the target species measured under well-defined conditions (T = 295 K, P = 1 atm, L = 21 cm, and $\chi$ = [0.4 - 1]\%). Figure~\ref{ch4:spectra}(b) highlights the absorbance spectra of the unknown interferents under identical conditions. These baseline spectra and experimental conditions form the foundation for generating training data, as described next.


\section{Data and modeling}

\subsection{Training Data Generation for HT-SIMNet}
The training dataset for HT-SIMNet is designed to enhance robustness against unknown spectral interference through a structured augmentation approach. Because in practical scenarios the interfering species are often unknown or lack available spectra, we employ data augmentation to generate synthetic spectral variations that expand the training set. These transformations introduce fictitious interference, enabling the model to generalize beyond the spectral characteristics of pure target species and effectively suppress unknown interferences.

The dataset is constructed using high-temperature C\(_1\)-C\(_3\) hydrocarbon spectra as a baseline, with non-reactive measurements acquired at 673 K. It consists of 200,000 randomly generated mixture compositions, where target species concentrations are randomly varied between 0\% and 1\%. The absorbance spectrum of each mixture is computed via Eqn. \ref{eqn:beer-lambert} in accordance to Beer-Lambert law which governs Laser absorption spectroscopy (LAS). The law relates laser attenuation to absorber concentration, pressure, temperature, path length, and absorption cross-section. More details on this principle can be found in \cite{swinehart1962beer}.

\begin{equation}
\label{eqn:beer-lambert}
A(\nu) = \sum_{i=1}^{4} c_i \cdot A_i(\nu)
\end{equation}

where \( c_i \) represents the concentration of species \( i \), and \( A_i(\nu) \) is its corresponding absorbance spectrum.

To introduce spectral distortions that mimic real-world "unseen" interference, three augmentation techniques are applied\cite{al2023augmentations}:

\begin{itemize}
    \item \textbf{Flipping:} The spectrum is inverted along the wavelength axis, altering its structure while preserving key spectral features. The transformed absorbance is given by:

    \begin{equation}
    A_{\text{flip}}(\nu_n) = A(\nu_{N-n})
    \end{equation}

    where \( N \) is the total number of spectral points.

    \item \textbf{Mirroring:} A randomly selected segment of the flipped spectrum is concatenated with the original spectrum, creating an artificial interference pattern. The modified spectrum is defined as:

    \begin{equation}
    A_{\text{mirrored}} = [A_{\text{flip}}, A](\nu_x : \nu_{N+x})
    \end{equation}

    where \( x \) is a randomly chosen index.

    \item \textbf{Dilation:} Selected spectral segments are linearly up-sampled to introduce localized distortions, simulating variations observed in real interfering species. The transformation is expressed as:

   \begin{equation}
\begin{split}
    A_{\text{dilated}}(\nu_n) &= A(\nu_{n-1/2}) + \frac{1}{2} \big(A(\nu_{n+1/2}) \\
    &\quad - A(\nu_{n-1/2})\big)
\end{split}
\end{equation}

\end{itemize}

\begin{figure}[h!]
    \begin{minipage}{\linewidth}
        \centering
        \includegraphics[width=0.75\textwidth]{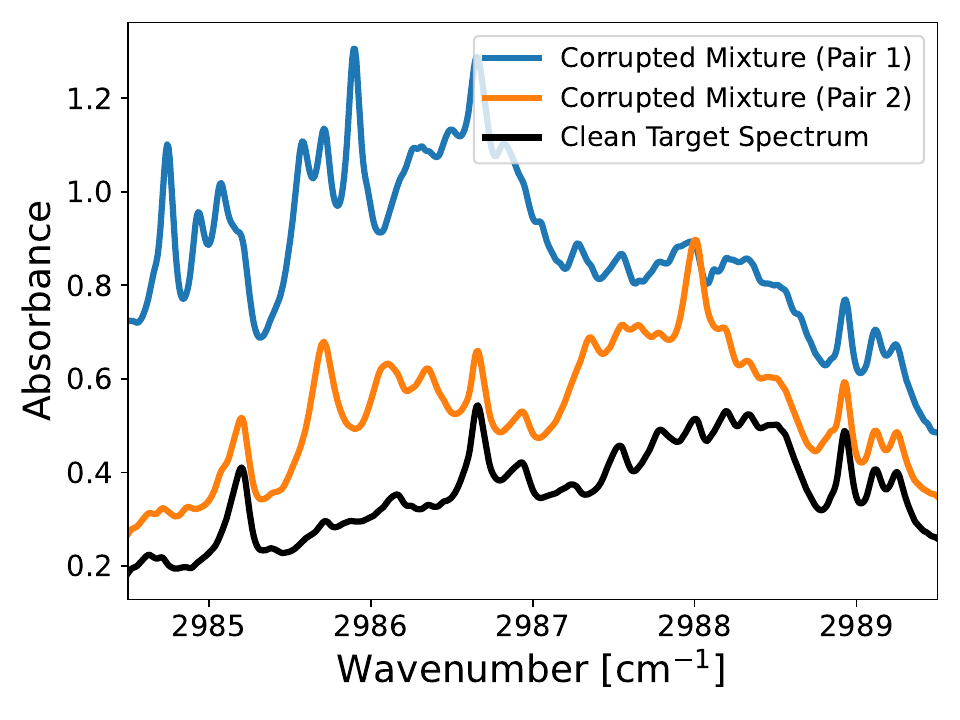}
        \caption{Absorbance spectra of a clean mixture (black) alongside a pair of corrupted spectra generated using the flip-mirror-dilate augmentation process. Both corrupted pairs share the same underlying clean signal (black), demonstrating the structured augmentation approach used for training HT-SIMNet.}
        \label{fig:pairs}
    \end{minipage}
\end{figure}

While these augmented spectra do not perfectly replicate real-world interference, they effectively enable HT-SIMNet to learn interference-resilient features, reducing its dependence on pure target spectral signatures. Figure \ref{fig:pairs} illustrates the absorbance spectra of a clean mixture (in black) alongside a pair of corrupted spectra generated by applying the random flip-mirror-dilate augmentation process to this same underlying clean spectrum. The two augmented spectra constitute an input-output pair used to train HT-SIMNet, which is tasked to reconstruct the clean underlying mixture signal.

\subsection{Modeling}

\begin{figure}[h!]
\centering

\includegraphics[width=290pt]{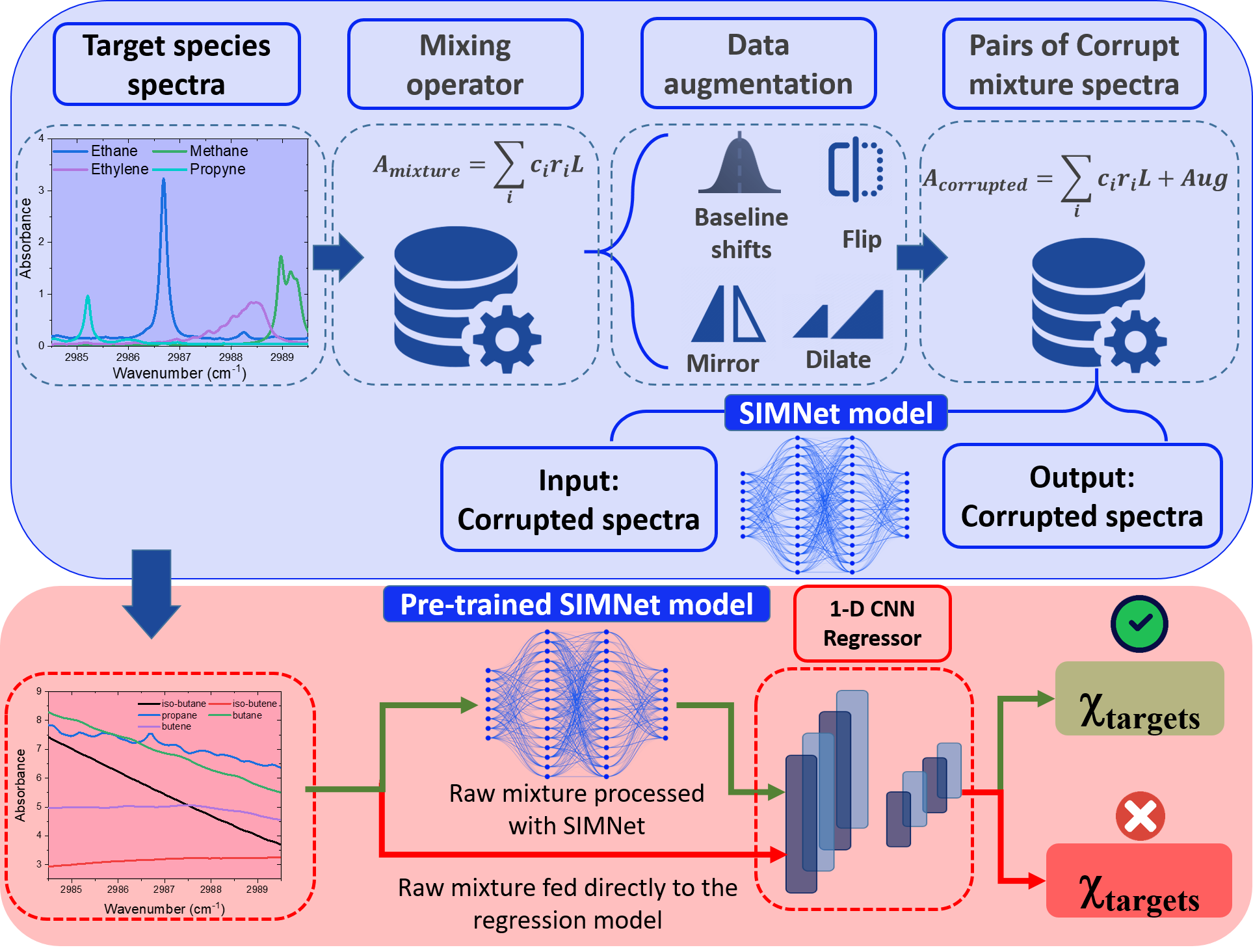}
\vspace{10 pt}
\caption{Schematic of the end-to-end training and testing workflow for HT-SIMNet and the 1D CNN regressor model. The diagram outlines key stages from data preprocessing and spectral augmentation to model optimization, highlighting the integration of HT-SIMNet for interference mitigation}
\label{fig:ch4:flowchart}
\end{figure}
For spectral interference mitigation, a U-Net model \cite{ronneberger2015u} based on the Noise2Noise framework \cite{lehtinen2018noise2noise} was used. In this approach, a pair of noisy ("corrupted") signals is provided: one used as an input and another as an output. The model learns to reconstruct the clean signal by leveraging statistical correlations between the noisy pair. The U-Net model utilizes a 1-D CNN \cite{lecun1998gradient, krizhevsky2012imagenet} with batch normalization \cite{ioffe2015batch} and dropout \cite{srivastava2014dropout} to enhance stability and prevent overfitting. It consists of three convolutional layers (32 → 64 → 128 filters) to extract hierarchical spectral features, with ReLU activations \cite{nair2010rectified} and max-pooling layers for dimensionality reduction. Adaptive pooling and skip connections, inspired by U-Net, can be incorporated to preserve fine details, while dilated convolutions \cite{yu2015multi} expand the receptive field.

A fully connected layer processes the latent representation, with a dropout layer (0.2 probability) to mitigate overfitting. HT-SIMNet learns to suppress noise, producing a denoised spectrum free from unknown interference. This output is then fed into a 1D CNN regressor, similar to \cite{sy2025multi}, to predict species concentrations. The 1-D CNN regressor begins with a convolutional layer (32 filters, kernel size = 3), followed by ReLU activation \cite{agarap2018deep} and max-pooling. This sequence repeats with a second convolutional layer (64 filters), another ReLU activation, and max-pooling. The output is flattened into a 1D vector, passed through a fully connected layer (128 neurons, ReLU), and finally through an output layer (4 neurons) for regression targets.
Figure \ref{fig:ch4:flowchart} provides a detailed schematic of the end-to-end training and testing workflow for HT-SIMNet and the regressor model, illustrating key stages from data preprocessing and spectral augmentation to model optimization.

\section{Performance Evaluation of HT-SIMNet}
\subsection{Spectral Interference Mitigation}

\begin{figure}[h!]
    \begin{minipage}{\linewidth}
        \centering
        \includegraphics[width=0.75\textwidth]{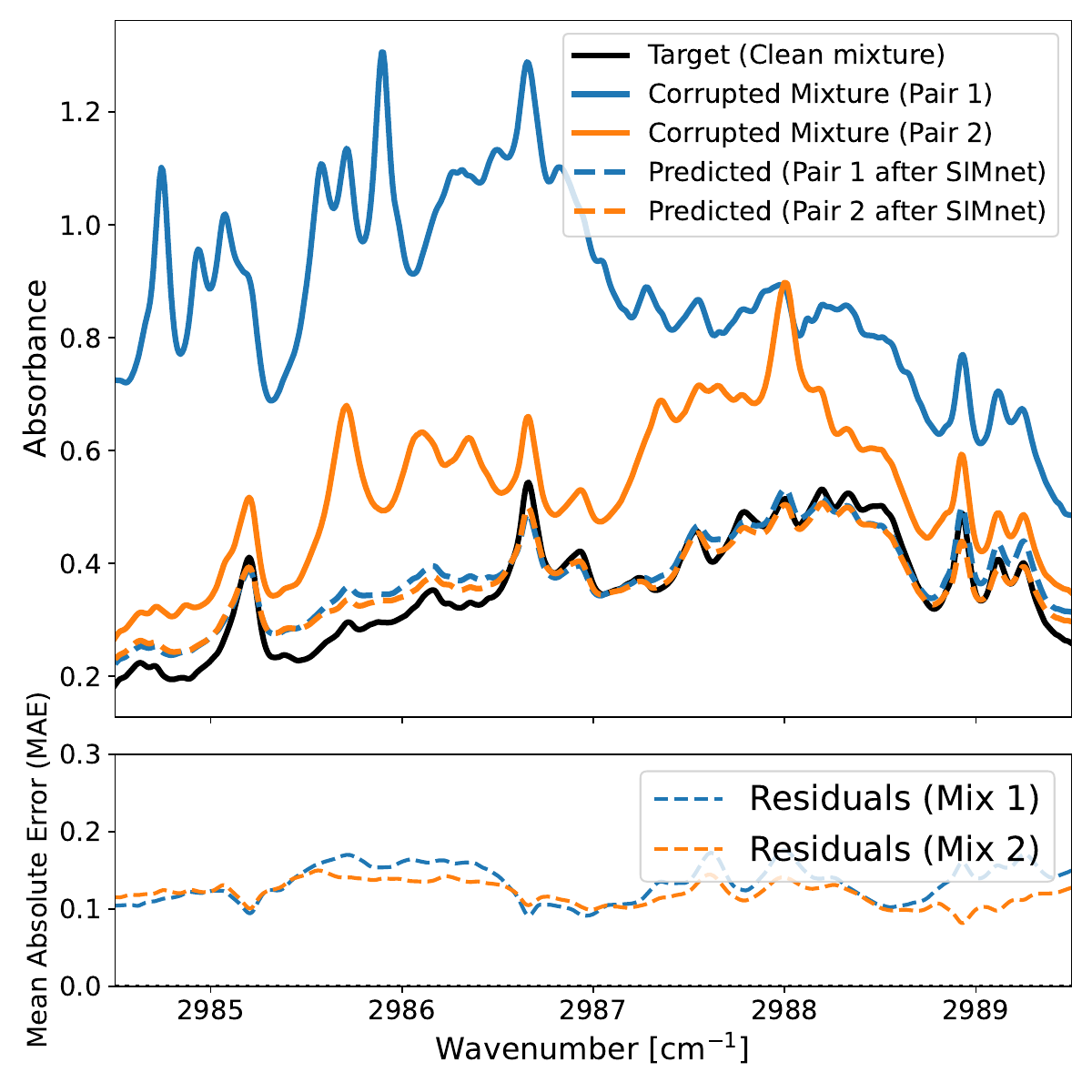}
        \caption{Comparison of corrupted spectra vs predicted spectra by HT-SIMNet (dashed lines), and the underlying clean mixture (solid black line). }
        \label{fig:predicted_spectra}
    \end{minipage}
\end{figure}
To evaluate HT-SIMNet performance, high temperature (670 K) mixtures of C$_1$–C$_3$ hydrocarbons were tested. C$_1$–C$_3$ mixtures with known mole fractions were introduced into the cell, followed by the addition of arbitrary amounts of interfering species. This produced corrupted absorbance spectra containing both target species and unknown interferents.

The corrupted spectra were processed through the pre-trained HT-SIMNet model to remove interference contributions, resulting in cleaner spectra that closely resemble the absorbance of the target species alone. Figure \ref{fig:predicted_spectra} shows the predicted spectra for two sets of corrupted mixtures (Pair 1 and Pair 2), with the same underlying clean spectra but different amounts of added interferents. The predicted spectra generated by HT-SIMNet for both pairs align closely with the clean spectra. The model achieves a mean relative error of approximately 0.1–0.3.

\subsection{Species Concentration Estimation}

\begin{figure}[h]
\centering
\includegraphics[width=\textwidth]{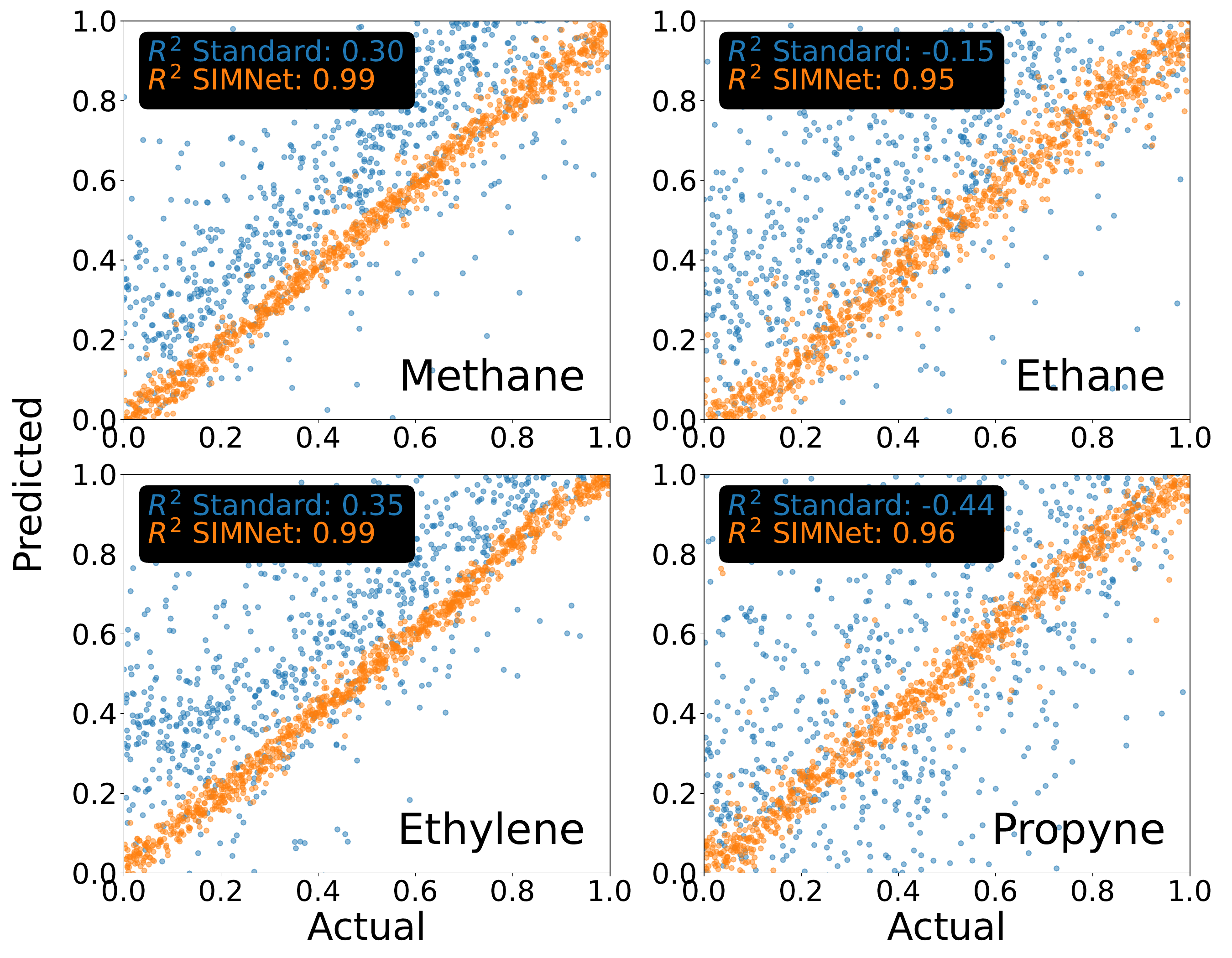}
\caption{Comparison of predicted mole fractions for methane, ethane, ethylene, and propyne. Predictions are shown for spectra fed directly into the regressor (blue scattered points) and for spectra processed with HT-SIMNet before regression (orange scattered points).}
\label{concentrations_predicted}
\end{figure}

After processing by HT-SIMNet, the mixtures were fed into a 1-D CNN regressor model to predict species concentrations. For comparison, a standard approach was also tested, where raw mixtures containing interference were directly input into the CNN regressor without prior interference mitigation.

The predicted concentrations from the standard approach were consistently higher than expected, leading to poor R² values, as the model failed to account for the unrecognized interfering species.

Figure \ref{concentrations_predicted} illustrates that direct regression on raw spectra resulted in significantly lower R² values for all species, with methane (R² = 0.30), ethane (R² = 0.15), ethylene (R² = 0.35), and propyne (R² = 0.44), due to the presence of unaccounted spectral interference. Conversely, when the mixtures spectra were first processed by HT-SIMNet to mitigate unknown interference, the predicted concentrations showed strong agreement with actual values, yielding R² values of 0.99 for Methane, 0.95 for Ethane, 0.99 for Ethylene, and 0.96 for Propyne. 
\subsection{Impact of Augmentation Strategies}
\begin{figure}[h!]
    \begin{minipage}{\linewidth}
        \centering
        \includegraphics[width=0.75\textwidth]{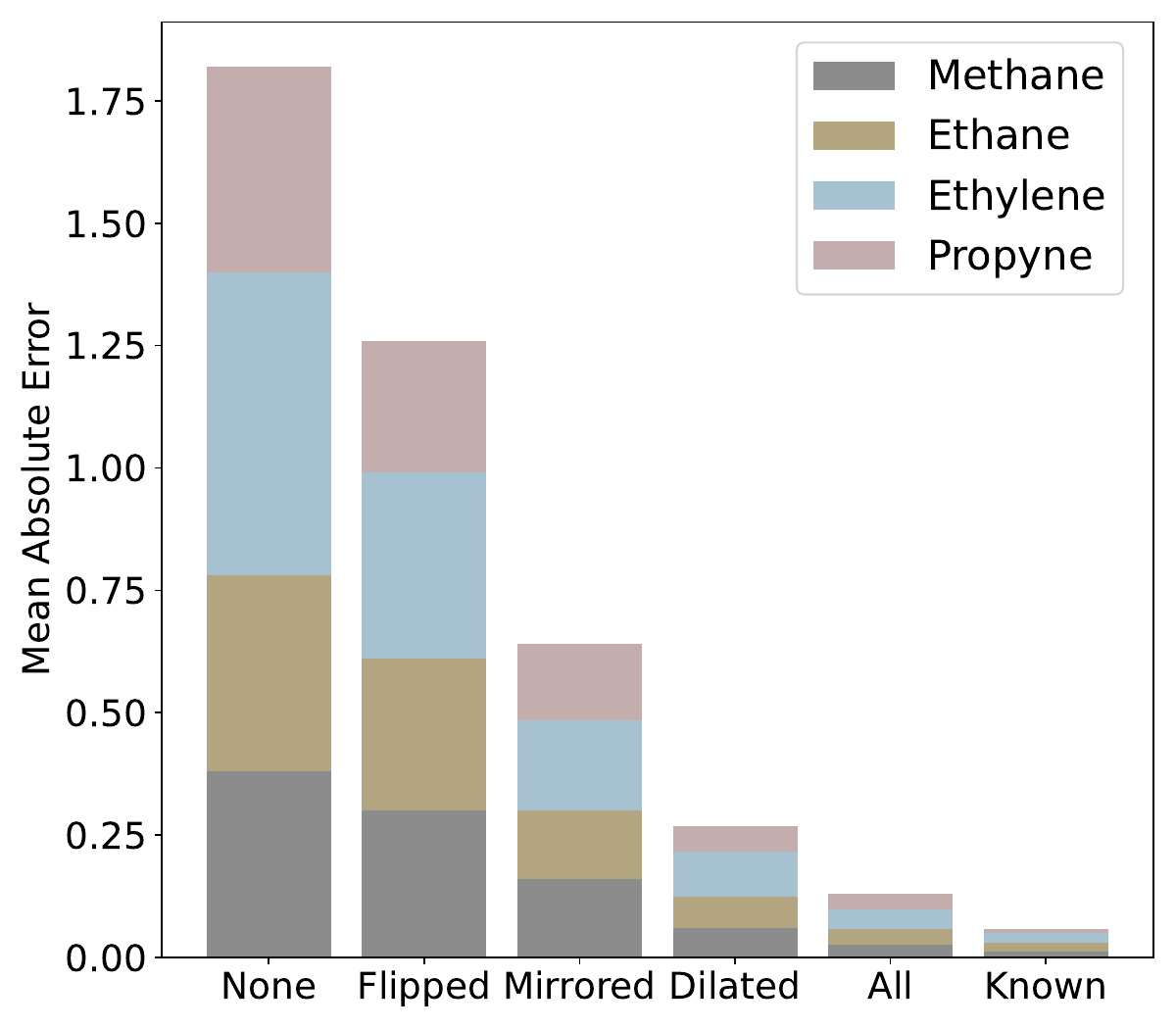}
        \caption{Stacked bar plot comparing the mean absolute error of different augmentation strategies (flipped, mirrored, dilated) and their combined approach.}
        \label{fig:effect_aug}
    \end{minipage}
\end{figure}
Selecting effective augmentation strategies is essential to enhance HT-SIMNet’s ability to handle spectral interference. Training on unaltered mixtures provides a baseline but often fails to generalize. Three augmentation strategies are considered: flipped, mirrored, and dilated spectra. Flipping the spectra along the wavenumber axis introduces some variation, though its impact is limited by predictable feature locations. Mirroring about the y-axis offers more diverse interference patterns, improving generalization, while dilation—randomly stretching or compressing spectral features—generates a broader range of interference and increases model robustness.

Figure \ref{fig:effect_aug} presents a stacked bar plot comparing the baseline model (without augmentation) to models trained with each individual strategy and a model using the combined approach. The combined strategy produces the most diverse training set and significantly enhances HT-SIMNet’s performance in resolving spectral mixtures under varying conditions. These results highlight the important role of structured augmentation in improving robustness against unseen spectral interference.

\section{Application to Reactive Systems}
The diagnostic technique was applied to investigate the pyrolysis of n-butane. A 1\% n-butane mixture in nitrogen was heated to 923 K under controlled conditions (T, P). To ensure stable thermal profiles, the cell was stabilized at 923 K for one hour before initiating 30-minute pyrolysis experiments. During the experiments, methane, ethane, ethylene, propane, and propyne were identified as pyrolysis products. Among these, methane and ethylene dominated, exhibiting significantly higher absorbance levels, while the contributions from other species remained below 1\%.

To assess the proposed spectral interference mitigation method, the analysis focused on detecting the reactants and the two major byproducts. Absorbance spectra were collected every 10 ms, resulting in 180,000 absorbance vectors over the 30-minute observation period at 923 K. These spectra contained not only the target species but also interferents such as the reactant (n-butane) and other byproducts. HT-SIMNet was employed to remove the unknown interferents, yielding spectra that reflected only the target species.

\begin{figure}[h!]
    \begin{minipage}{\linewidth}
        \centering
        \includegraphics[width=0.75\textwidth]{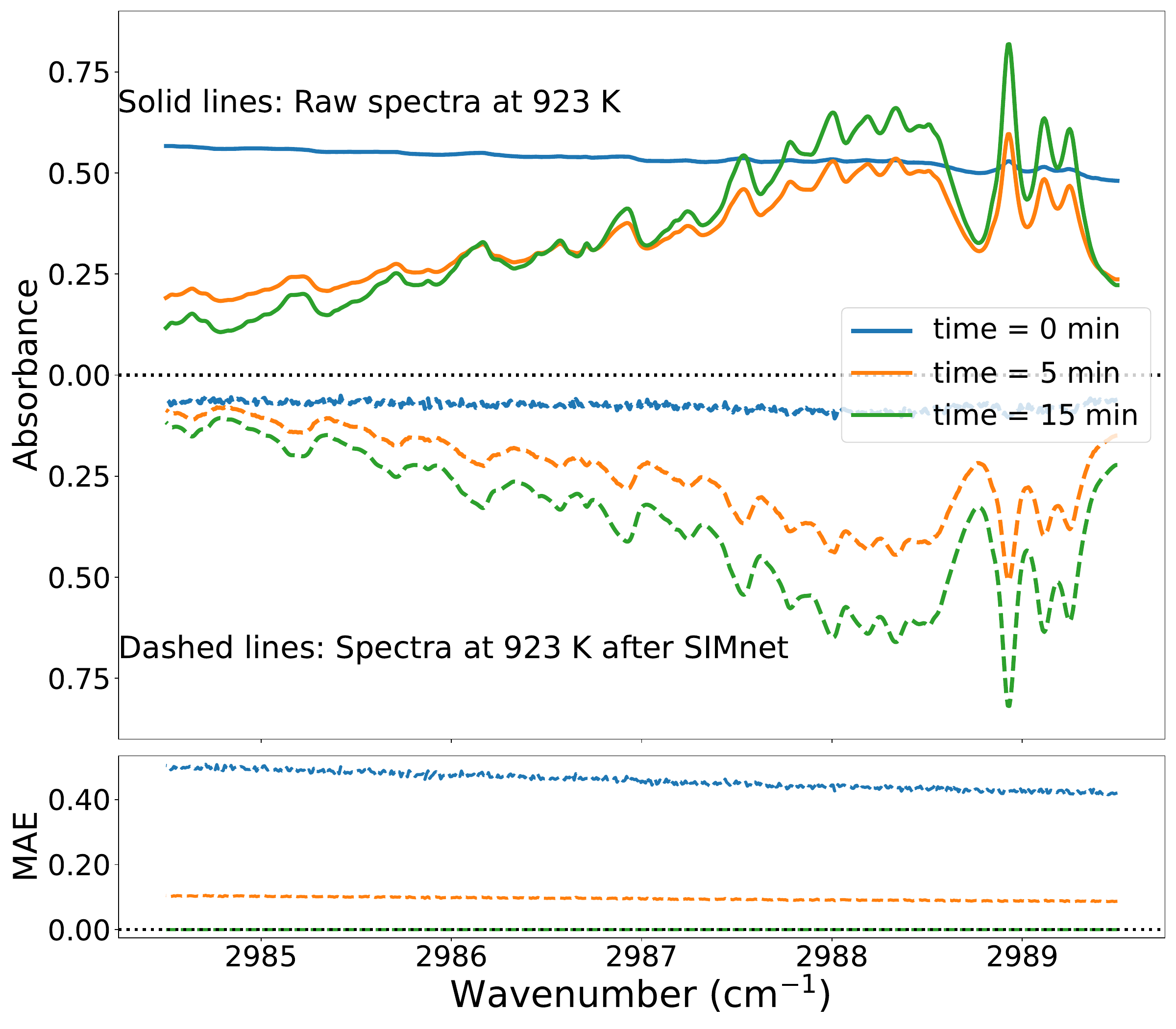}
        \caption{Absorbance spectra collected during n-butane pyrolysis at different times, alongside the corresponding spectra processed by HT-SIMNet.}
        \label{fig:pyrolysis}
    \end{minipage}
\end{figure}

Figure \ref{fig:pyrolysis} illustrates the composite spectra collected during pyrolysis at different times, alongside the spectra processed by HT-SIMNet. At early stages, n-butane is clearly dominant, but HT-SIMNet corrects the spectra to near zero, as the contributions of target species are minimal. As n-butane undergoes consumption over time, the collected and HT-SIMNet-processed spectra converge. This is also evident from the mean absolute errors between the collected composite spectra and those processed by HT-SIMNet.

\begin{figure}[h!]
    \begin{minipage}{\linewidth}
        \centering
        \includegraphics[width=0.75\textwidth]{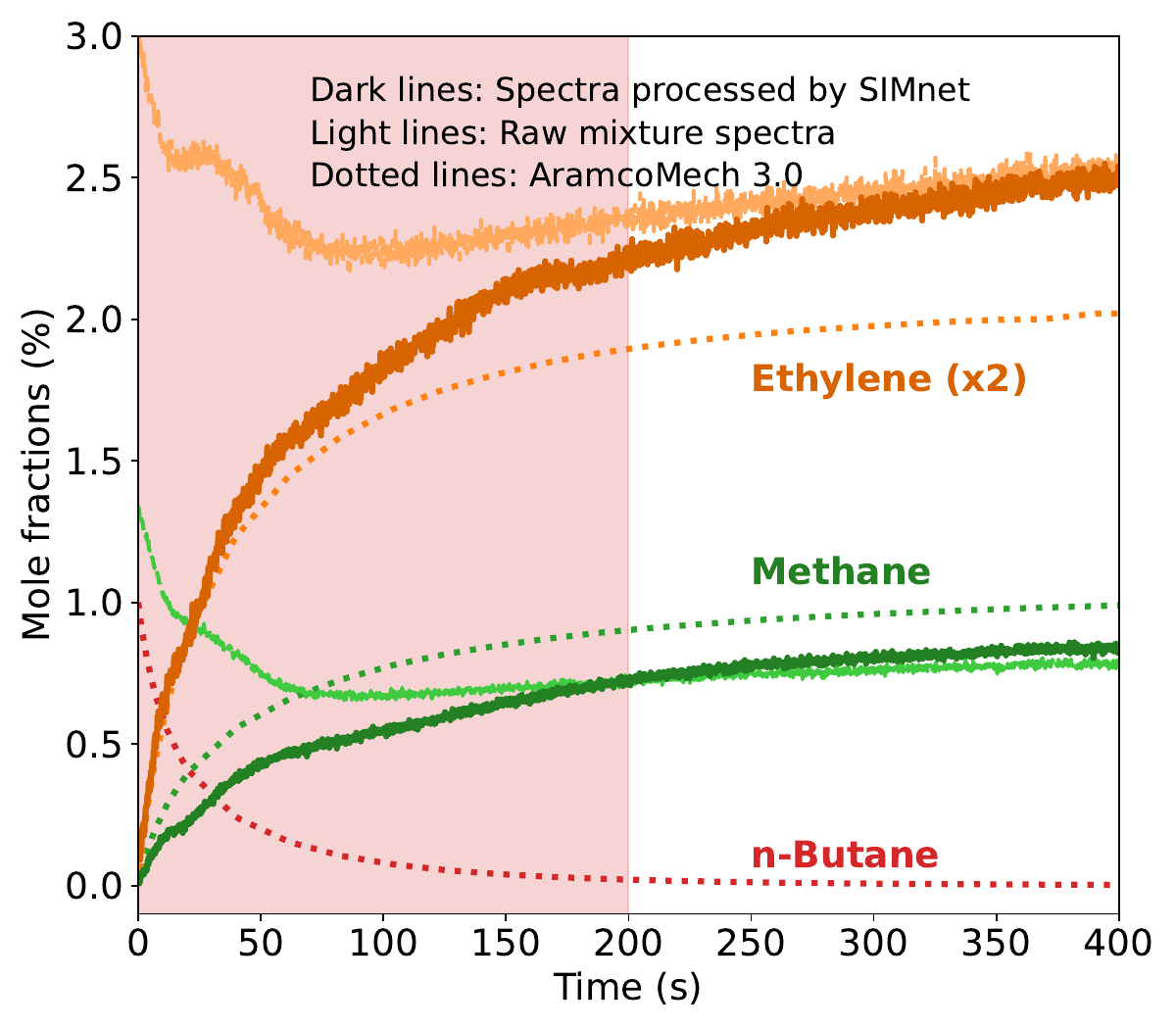}
        \caption{Predicted concentrations of methane and ethylene during n-butane pyrolysis. Comparisons are shown between predictions using raw composite spectra (light-colored lines) and HT-SIMNet-processed spectra (dark-colored lines) to AramcoMech 3.0 simulations (dotted lines). The red-shaded area indicates early reaction times when n-butane is still present in the system.}
        \label{fig:con_pyro}
    \end{minipage}
\end{figure}

The processed spectra from HT-SIMNet were then fed into the 1-D CNN regressor to predict the concentrations of the target species (methane, ethylene). Figure \ref{fig:con_pyro} presents the results from over 15 minutes of n-butane pyrolysis. At early reaction times (shaded red area), when n-butane is actively decomposing, feeding the composite spectra directly into the regressor leads to erroneous predictions, consistently overestimating the concentrations. This is due to the presence of n-butane, which acts as an unaccounted interferent that the regressor is not trained to handle. 

As the reaction progresses and n-butane undergoes complete decomposition, the predictions improve, since only methane and ethylene remain in the system. In contrast, when the composite absorbance spectra are processed through HT-SIMNet, the model effectively removes interference from n-butane and other byproducts before feeding the cleaned spectra into the regressor. This allows the regressor to predict the concentrations of methane and ethylene even at early reaction times. At later stages, the performance of both approaches converges, as n-butane is fully consumed and the contribution of other unknown interferents becomes negligible.

\chapter{Unsupervised Spectral Source Separation: A Reference-free Technique for Multi-species Detection}\label{chapter5}

This chapter is adapted from the accepted manuscript versions of articles by Sy et al., accepted for publication in IEEE Sensors and presented at CLEO, with one manuscript currently pending publication in Combustion and Flame~\cite{sy2024reference, sy2024unsupervised}.
\section{Introduction}
\label{sec:ch5:intro}
Infrared laser absorption spectroscopy (IR-LAS) enables high-fidelity, time-resolved measurements of gas-phase species in high-temperature environments such as shock tubes and rapid compression machines (RCMs)~\cite{goldenstein2017infrared, farooq2022laser, hanson2014recent, elkhazraji2024mid, hanson2016spectroscopy}. Its sensitivity and selectivity stem from the unique vibrational signatures of molecules, making IR-LAS particularly well-suited for studying pyrolysis and oxidation of hydrocarbons under extreme conditions.

Quantification with IR-LAS typically requires high-temperature absorption cross-section data for each species~\cite{pinkowski2019multi, elkhazraji2024selective, sy2023multi, jeevaretanam2023transient, mhanna2023laser}. However, existing databases such as HITRAN~\cite{gordon2022hitran2020}, HITEMP~\cite{rothman2010hitemp}, and PNNL~\cite{sharpe2004gas} offer limited species coverage at elevated temperatures and pressures. Generating new cross-section data under controlled combustion conditions is labor-intensive and resource-demanding~\cite{cassady2020thermal, sy2024laser, sy2025multi}, restricting the scalability of traditional quantification models that depend on pre-characterized reference spectra.

Blind source separation (BSS) methods provide an alternative route for decomposing mixture spectra into individual species without requiring reference spectra~\cite{belouchrani1997blind}. Classical BSS techniques such as independent component analysis (ICA)~\cite{hyvarinen2000independent} and non-negative matrix factorization (NMF)~\cite{lee1999learning} have been applied to linear mixtures but suffer from scaling ambiguities~\cite{wang2012nonnegative} and are poorly suited for high-temperature diagnostics where non-linear interference, baseline drift, and noise are prevalent.

Recent developments in machine learning have enhanced the flexibility of BSS. Variational autoencoders (VAEs)~\cite{neri2021unsupervised} and other deep unsupervised methods have demonstrated success in decomposing complex signals without ground truth references. Building on these advances, our previous work introduced an autoencoder-based BSS (AE-BSS) model capable of learning species components from room-temperature mixtures~\cite{sy2024reference, sy2024unsupervised}.

In this chapter, we present UnblindMix, an unsupervised blind source separation framework designed for high-temperature combustion diagnostics. UnblindMix learns to infer species concentrations and reconstruct high-temperature reference spectra directly from composite mixture spectra, eliminating the dependence on external reference databases. The model is trained exclusively on experimental mixture data and leverages synthetic augmentations to account for baseline drift, spectral noise, and unknown interference.

We demonstrate the performance of UnblindMix in both non-reactive and reactive scenarios. For non-reactive conditions, the model accurately recovers species concentrations in static mixtures of \(C_1\)–\(C_3\) hydrocarbons, despite strong spectral overlap. Under pyrolysis conditions, the model captures the temporal evolution of species during shock tube experiments involving \(n\)-butane and \textit{iso}-butane at 923~K. These results establish UnblindMix as a scalable and reference-free framework for multi-species detection in chemically and spectrally complex environments.


\section{Experimental Setup}\addvspace{0pt}
\subsection{Optical Setup}
\begin{figure}[h!]
\centering
\includegraphics[width=0.55\textwidth]{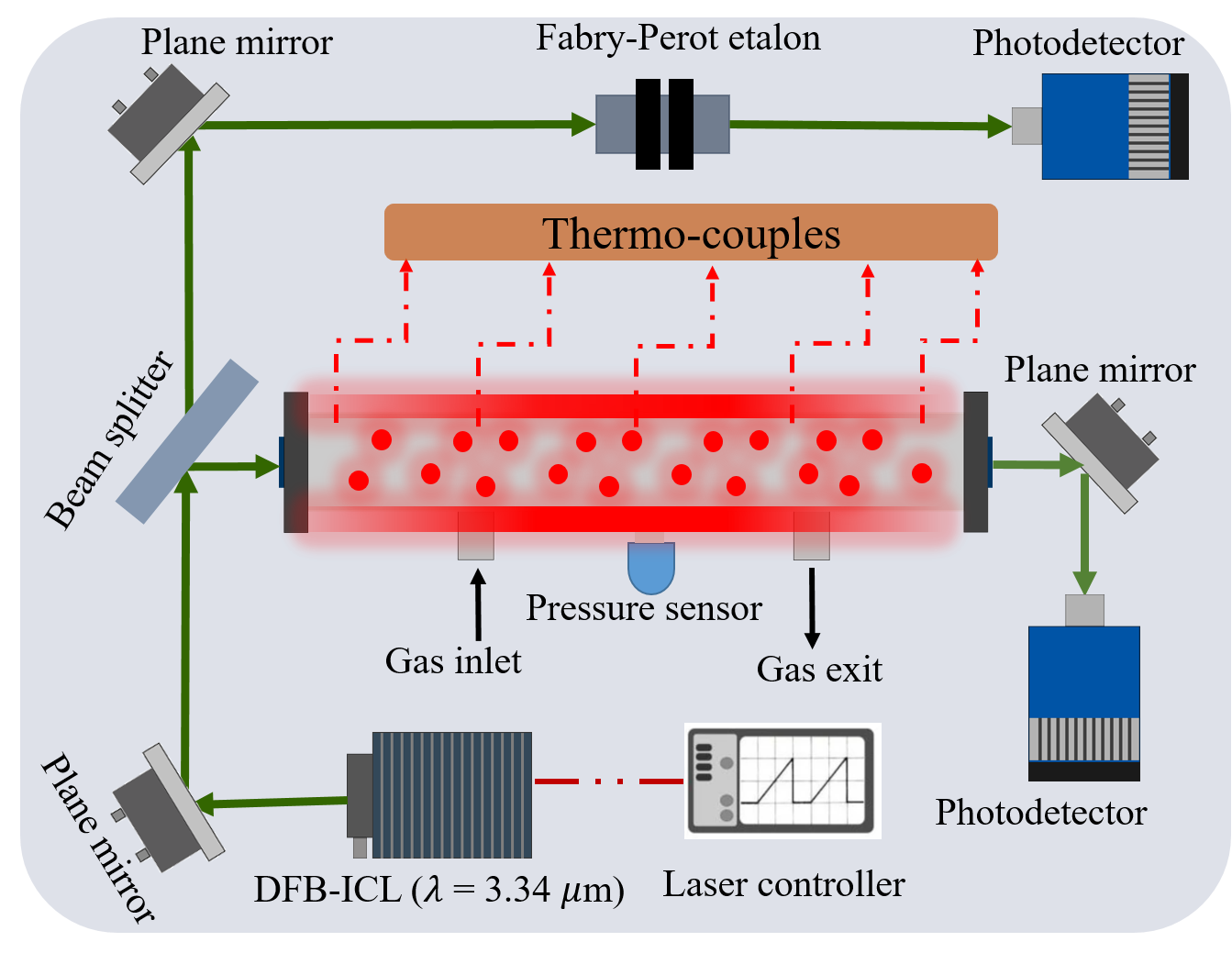}
\caption{\footnotesize Experimental setup.}
\label{fig:opticalsetup}
\end{figure}

The experimental setup, illustrated in Fig. \ref{fig:opticalsetup}, was employed for high-temperature measurements to test the effectiveness of our proposed approach. A distributed feedback (DFB) inter-band cascade laser (ICL) from Nanoplus GmbH, emitting 10 mW at 3.34 µm, was used to target the 2984.5–2989.5 cm\(^{-1}\) range. The laser beam was guided through a 21 cm heated stainless-steel cell equipped with ZnSe windows and sealed with copper gaskets to maintain integrity under high-temperature operation\cite{farouki2024measurement}. Heating tapes were used to maintain the cell at the desired temperature, while five K-type thermocouples were strategically positioned along its length to monitor temperature distribution. Similar thermal gradient profiles to those reported in \cite{farouki2024measurement} were observed.  The laser was modulated using a sawtooth signal from a function generator at a repetition rate of 100 Hz. Time was mapped to wavenumber using a 7.62 cm germanium Fabry-Perot etalon. To minimize interference from thermal emissions, a Thorlabs FB-3300–300 band-pass filter was positioned in the optical path, while a four-stage, thermoelectrically-cooled photovoltaic detector with a 1.5 MHz bandwidth (Vigo Systems) recorded the laser intensity signal.

\subsection{Wavelength Selection}
The wavelength region between 2984.5 and 2989.5 cm\(^{-1}\) was chosen to capture  C–H stretch absorption features of C\(_1\)-C\(_3\) hydrocarbons, including methane, ethane, ethylene, and propyne\cite{sy2025multi}. These hydrocarbons are key indicators in the decomposition pathways of larger molecules and are commonly observed as intermediates in pyrolysis and combustion reactions, making them ideal candidates for developing and validating a multispecies detection methodology. 

\begin{figure}[h!]
\centering
\includegraphics[width=0.65\textwidth]{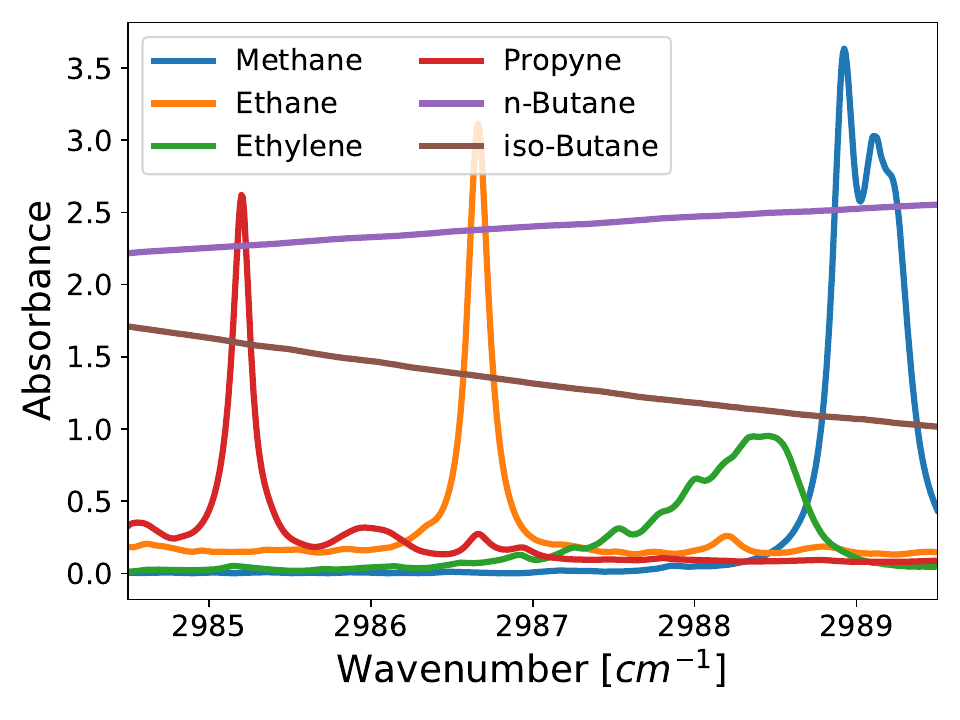}
\caption{\footnotesize Measured absorption spectra of the target species at T = 295~K, P = 1~atm, L = 21~cm, and mole fractions of 1.1\% (methane), 0.5\% (ethylene), 0.28\% (ethane), 1\% (propyne), 0.8\% (\textit{n}-butane), and 0.8\% (\textit{iso}-butane).}
\label{ch5:spectra}
\end{figure}

For the non-reactive static cell experiments, mixtures of C\(_1\)-C\(_3\) hydrocarbons were prepared at varying concentrations to evaluate model performance. In the reactive experiments, we studied n-butane and iso-butane pyrolysis for two key reasons: they absorb within the chosen spectral range, and their decomposition produces  C\(_1\)-C\(_3\) species observed in the non-reactive tests. This makes them ideal for demonstrating the effectiveness of the proposed unsupervised diagnostic approach under reactive conditions. Figure \ref{ch5:spectra} displays the measured absorbance spectra of the target species at conditions: T = 295 K, P = 1 atm, L = 21 cm, and $\chi$ = [0.02 - 1]\%.


\newpage
\section{Model Framework and Dataset Preparation} \addvspace{6pt}
\subsection{Model Framework}
\vspace{2mm}

The UnblindMix model is based on a one-dimensional convolutional autoencoder for blind source separation of gas-phase mixture spectra. The architecture follows a standard encoder–decoder structure. The encoder transforms the high-dimensional composite spectrum into a compact, normalized latent space \( Z_i \), which encodes the concentration estimates of each species. This latent space has a shape of (batch size, \( n \)), where \( n \) corresponds to the number of species, and it encodes their concentrations.

The encoder comprises six convolutional layers. The first layers use 128 filters to extract detailed spectral features, followed by max-pooling layers that reduce dimensionality while preserving relevant structure. The number of filters is reduced in subsequent layers (64, 32, and 16) to compress the spectral representation. The input mixture spectrum \( x \) has a shape of (batch size, 1, 600), where 600 is the number of wavenumber points. The final compressed features are passed through a fully connected layer to produce the latent vector \( Z \). A dropout layer is applied after the compression layer to improve generalization and prevent overfitting by limiting co-adaptation of feature detectors.

The decoder reconstructs the input spectrum from the latent representation using a linear combination of learned spectral bases. It applies the function \( F(Z) = \sum Z_i A_i \), where \( A_i \) represents the predicted spectrum of species \( i \). The decoder directly learns a matrix\( A \in \mathbb{R}^{n \times m} \), where \( n \) is the number of species and \( m = 600 \) is the number of wavenumber points. Each row \( A_i \) corresponds to the reference spectrum of species \( i \). The reconstructed spectrum \( F(Z) \) has shape (batch size, \( m \)) and is compared against the input spectrum to evaluate performance.

During training, the decoder and encoder are jointly optimized by minimizing the mean squared error (MSE) between the reconstructed and original spectra. This loss ensures that the predicted species concentrations and learned spectral features accurately reproduce the measured absorbance. Equation~\ref{eq:mse_loss} defines the objective function used for training.

\begin{equation}  
\mathcal{L}_{\text{real}} = \mathbb{E}_{x \sim p_{\text{meas}}} \left[ \| x - d_\phi(e_\theta(x)) \|^2 \right]  
\label{eq:mse_loss}  
\end{equation}  

\noindent
where:
\begin{itemize}
    \item \( x \) is the input measured mixture spectrum sampled from the empirical distribution \( p_{\text{meas}} \),
    \item \( e_\theta(\cdot) \) is the encoder parameterized by \( \theta \), mapping input spectra to mole fractions \( Z \),
    \item \( d_\phi(\cdot) \) is the decoder parameterized by \( \phi \), reconstructing spectra from \( Z \),
    \item \( \mathcal{L}_{\text{real}} \) quantifies the loss of spectral reconstruction.
\end{itemize}

To improve the robustness of training, an additional synthetic loss term was introduced. This auxiliary loss enables the model to reinforce internal consistency by reusing its own decoder outputs during training. Specifically, the decoder’s predicted spectra are treated as pseudo ground truth and reintroduced as input to the encoder. The resulting latent representation is then passed through the decoder a second time, and the reconstructed spectra are compared to the originals. This mechanism encourages consistency in the learned spectral and concentration representations over repeated forward passes.
The synthetic loss is implemented as a mean squared error (MSE) between the original decoded spectrum and its reconstruction after re-encoding. It serves as a regularization strategy to mitigate overfitting and reduce sensitivity to noise and initialization artifacts in the training data. The loss is formally defined as:

\begin{equation}
\mathcal{L}_{\text{synth}} = \mathbb{E}_{x_{\text{synth}} \sim p_{\text{synth}}} \left[ \| x_{\text{synth}} - d_\phi(e_\theta(x_{\text{synth}})) \|^2 \right]
\label{eq:synth_loss}
\end{equation}

\noindent
where:
\begin{itemize}
    \item \( x_{\text{synth}} \in \mathbb{R}^{1 \times m} \) is the synthetic mixture spectrum sampled from the learned decoder distribution \( p_{\text{synth}} \),
    \item \( e_\theta(\cdot) \) is the encoder parameterized by weights \( \theta \), which maps \( x_{\text{synth}} \) to a latent concentration vector \( Z \in \mathbb{R}^n \),
    \item \( d_\phi(\cdot) \) is the decoder parameterized by \( \phi \), which reconstructs a spectrum from \( Z \),
    \item \( \mathcal{L}_{\text{synth}} \) is the synthetic loss, computed as the mean squared error (MSE) between the input synthetic spectrum and its reconstruction,
    \item \( \mathbb{E}_{x_{\text{synth}} \sim p_{\text{synth}}}[\cdot] \) denotes the expectation over synthetic mixture spectra generated by the decoder.
\end{itemize}

\begin{figure}[h!]
\centering

\includegraphics[width=0.8\textwidth]{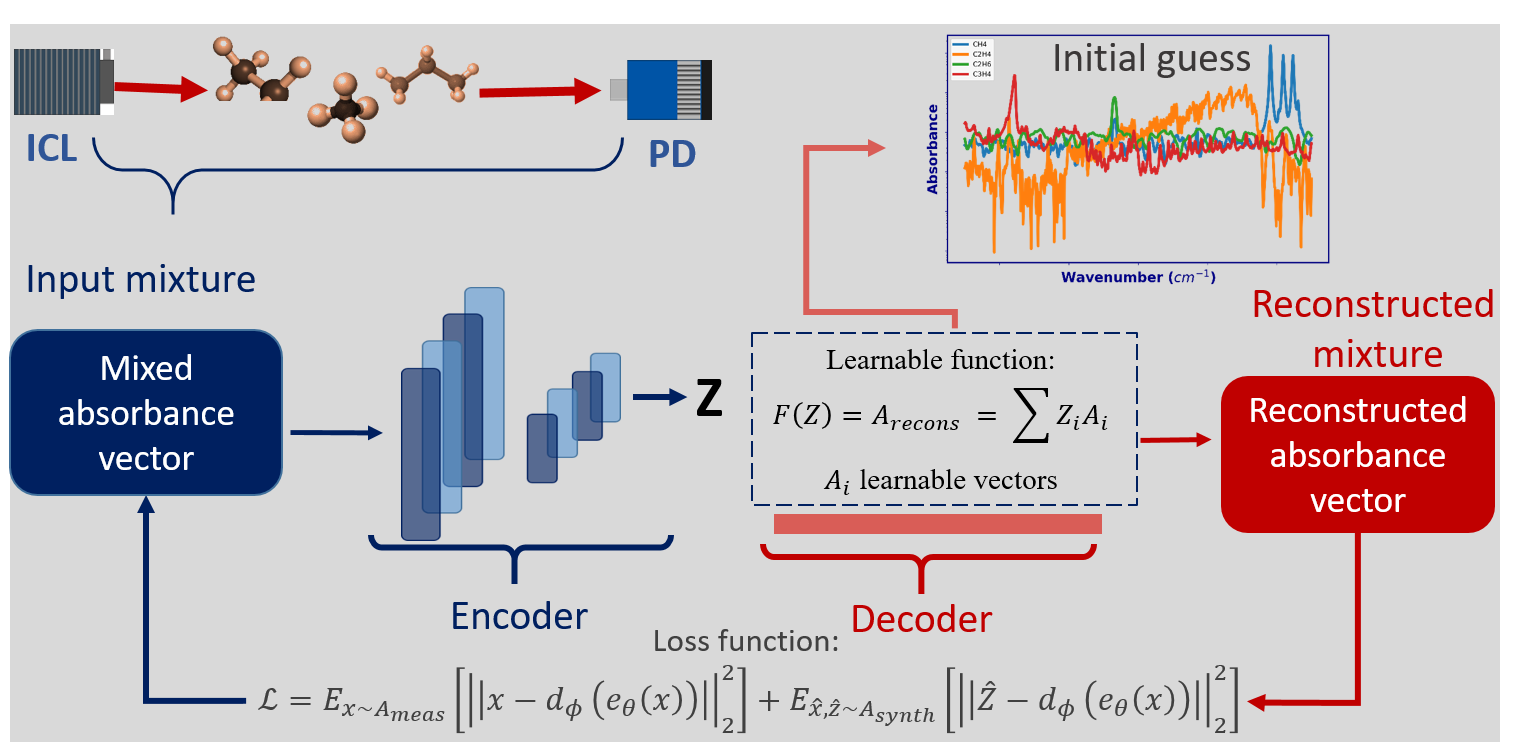}
\vspace{10 pt}
\caption{\footnotesize Flowchart of the training process for the UnblindMix model.}
\label{UnblindMIx}
\end{figure}

This loss term is incorporated into the total loss function, which combines contributions from both real and synthetic losses, achieving balanced learning. The training uses Adam optimizer with an initial learning rate of $10 \times e^{-3}$ and batch size of 32, typically converging within 200 epochs. Further details on the autoencoder architecture and training methodology can be found in \cite{sy2024unsupervised,sy2024reference}. Figure \ref{UnblindMIx} illustrates the flowchart of the training process of the UnblindMix model.

\subsection{Dataset Preparation}

\subsubsection{Training Data}

\begin{figure}[H]
    \centering
    \includegraphics[width=0.75\textwidth]{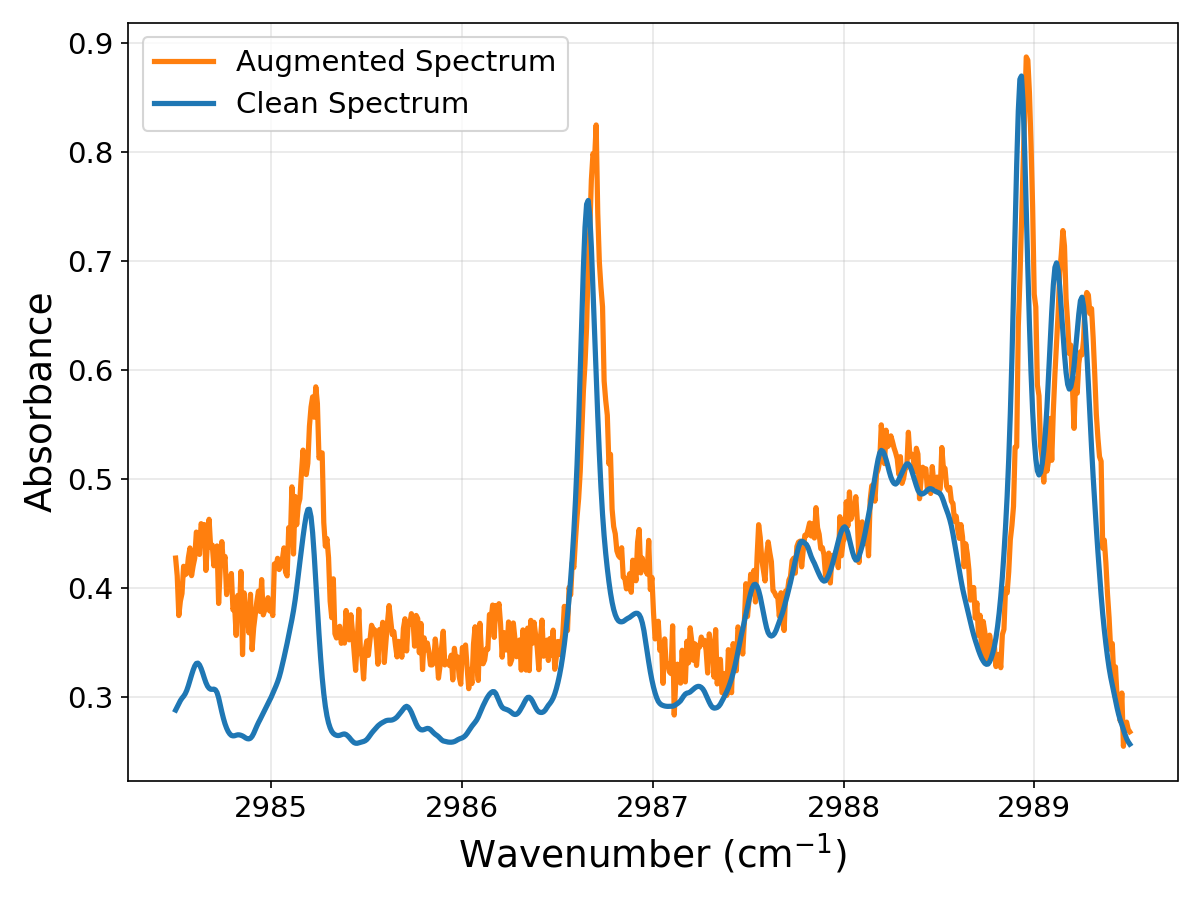}
    \caption{\footnotesize Comparison between a clean mixture spectrum and an augmented spectrum after applying synthetic training data transformations}
    \label{fig:augmented_spectrum}
\end{figure}

In this work, simulated datasets closely mimicking experimental conditions were generated. For the non-reactive case, training data consisted of simulated spectra for mixtures of $C_1–C_3$ species at various concentrations, following the Beer–Lambert law. Noise and baseline shifts were added to represent realistic experimental effects. Figure \ref{fig:augmented_spectrum} compares a clean mixture spectrum to its augmented counterpart, which includes horizontal shifts, non-linear baseline drift, and additive Gaussian noise. These augmentations are designed to mimic experimentally observed distortions.

A total of 500 synthetic mixtures were generated for model training. The non-reactive case was tested using spectral data acquired at 673 K. This temperature was chosen because the targeted species remain stable without pyrolysis during the experimental test duration. Concentrations for simulated mixtures ranged from 0–1\% for methane, 0–0.5\% for ethane, 0–3\% for ethylene, and 0–2\% for propyne. Algorithm \ref{algo} outlines the training data generation procedure.

\begin{algorithm}[h!]
\small
\caption{Training Data for UnblindMix}
\begin{algorithmic}[1]
\State \textbf{Input:} High-Temperature \(C_1-C_3\) spectra
\State \textbf{Output:} Training data
\State Set temperature $T = 673$ K for non-reactive measurements

\For{$i = 1$ to $500$}
    \State Generate concentration for \(C_1-C_3\)
    \State Linear combination
    \vspace{-2mm}
    \[
    \text{Mixture}_i = \sum_{j=1}^{3} \text{Concentration}_{j} \times \text{Spectrum}_{j}
    \]
    \vspace{-2mm}
    \State Add non-linear effects:
    \vspace{-2mm}
    \begin{itemize}
        \item Add random noise to $\text{Mixture}_i$
        \vspace{-2mm}
        \item Add non-linear baseline shifts
    \end{itemize}
    \vspace{-2mm}
\EndFor
\State \textbf{Return:} Generated training data

\end{algorithmic}
\label{algo}
\end{algorithm}

\newpage
\subsubsection{Temperature-dependent absorption cross-sections}

\begin{figure}[h!]
\centering
\includegraphics[width=0.7\textwidth]{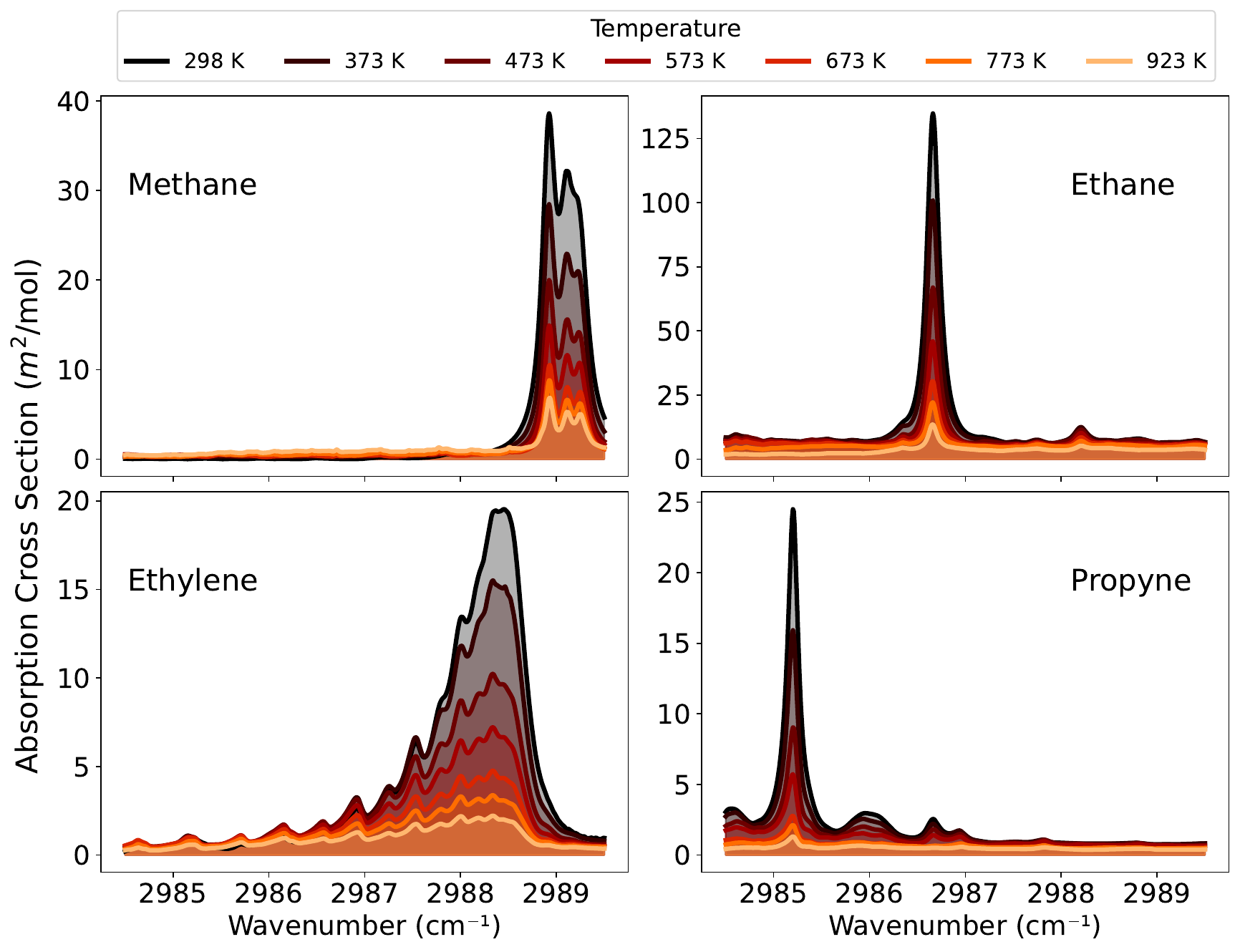}

\caption{\footnotesize Measured temperature-dependent absorption cross-sections of methane, ethane, ethylene and propyne over T = 298 – 923 K and P = 1 atm.}
\label{CSdata}
\end{figure}

High-temperature absorption cross-sections for \(C_1–C_3\) hydrocarbons, including methane, ethane, ethylene, and propyne, were experimentally measured over a temperature range of 298–923 K at 1 atm using the high-temperature static cell. The results are presented in Fig. \ref{CSdata}. The cell was pre-heated for one hour to ensure thermal stability, and measurements were completed in 10 ms to extent of fuel decomposition. Mixtures of the target species were prepared using nitrogen as the bath gas, with mole fractions optimized to achieve high signal-to-noise ratios. The absorption cross-sections were obtained using the methodology described in\cite{farouki2024measurement}, which accounts for thermal gradients over the length of the optical cell and other sources of uncertainty. This high-temperature spectral data provides a basis for validating the UnblindMix model. 


\section{Results and Discussion}
\subsection{Comparison Between UnblindMix and Linear-BSS Models}
\label{sec:comparison_clean}

\begin{figure}[H]
    \centering
    \includegraphics[width=0.7\linewidth]{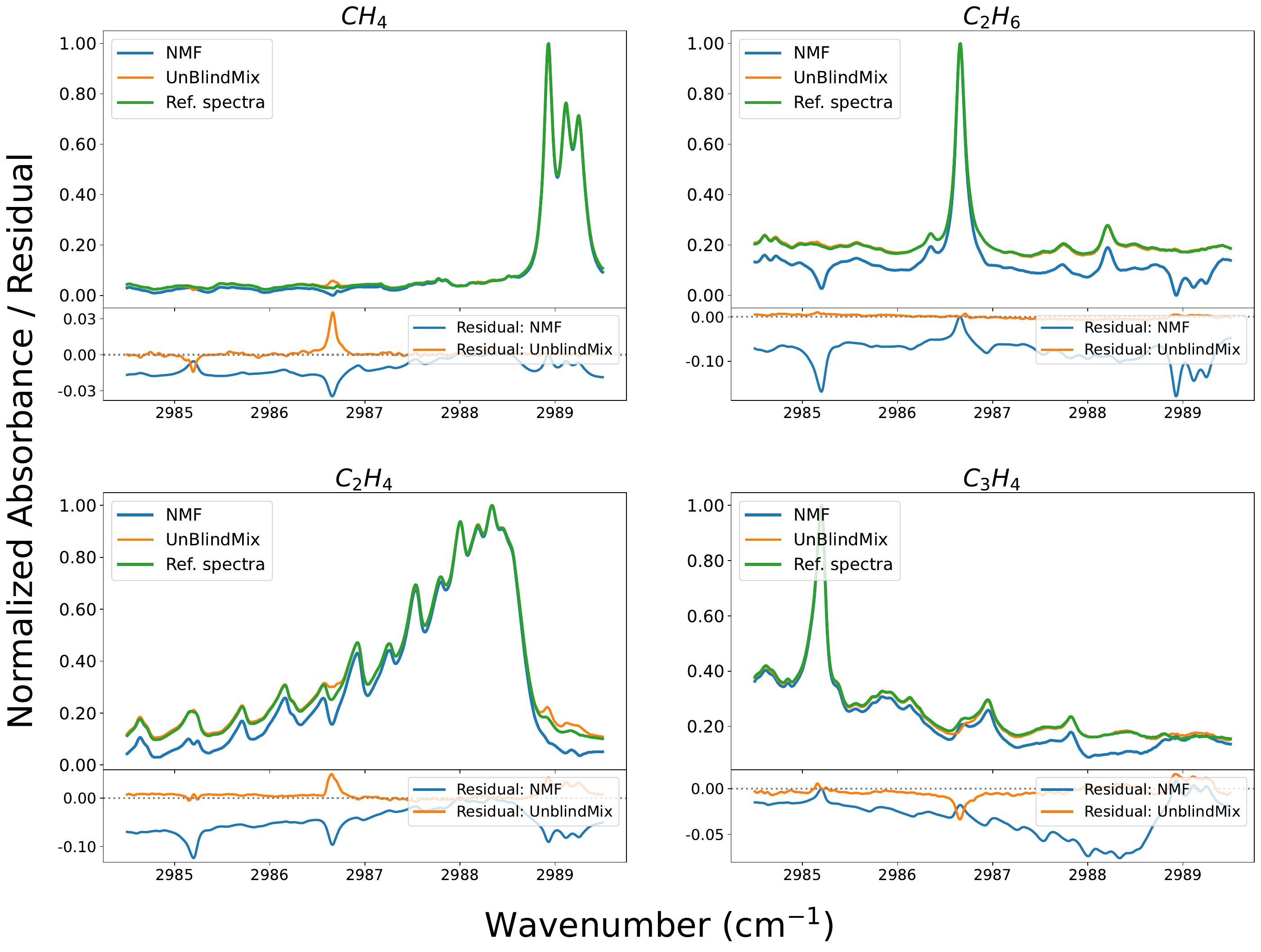}
    \caption{Comparison of reconstructed spectra and residuals for NMF and UnblindMix trained on clean, noise-free mixture spectra.}
    \label{fig:nmf_unblindmix_clean}
\end{figure}

To establish a performance baseline, UnblindMix and non-negative matrix factorization (NMF) were both trained and tested using clean mixture spectra, free from noise and baseline distortions. Figure~\ref{fig:nmf_unblindmix_clean} compares the reconstructed reference spectra for methane, ethylene, ethane, and propyne obtained from both methods. Residuals, computed as the absolute difference between the reconstructed and measured spectra, highlight the relative performance of the two approaches. The residual at each wavenumber point \( i \) is given by:

\begin{equation}
r_i = \left| A_i^{\text{meas}} - A_i^{\text{recon}} \right|
\label{eq:residual}
\end{equation}

\noindent
Here, \( A_i^{\text{meas}} \) denotes the measured absorbance at index \( i \), and \( A_i^{\text{recon}} \) is the corresponding predicted value.

Under ideal conditions, NMF residual amplitudes are approximately $2 \times 10^{-1}$, while UnblindMix residuals remain below $3 \times 10^{-2}$. Both models perform nearly the same, with minor differences in reconstruction errors across species.

\begin{figure}[H]
    \centering
    \includegraphics[width=0.7\linewidth]{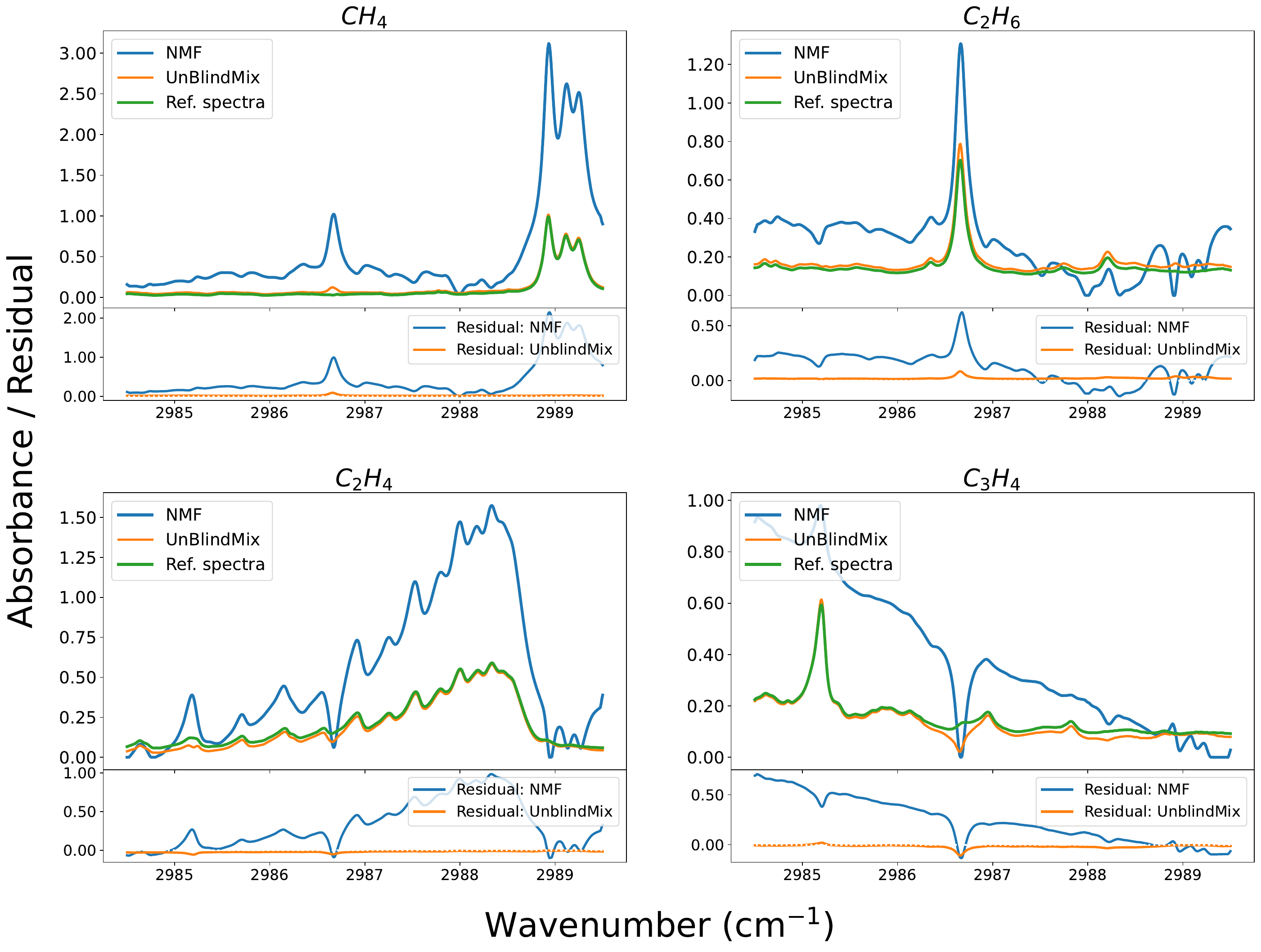}
    \caption{Comparison of reconstructed spectra and residuals for NMF and UnblindMix trained on noisy, baseline-shifted, and augmented mixture spectra.}
    \label{fig:nmf_unblindmix_noisy}
\end{figure}

Under realistic conditions, both models were retrained using augmented spectra with randomized baseline shifts, additive noise, and spectral distortions. Figure~\ref{fig:nmf_unblindmix_noisy} shows the results.

UnblindMix gives residual amplitudes below $5 \times 10^{-1}$ for all species. NMF residuals are higher, up to $2.0$ for methane and around $1.0$ for ethylene. This shows a mismatch between reconstructed and true spectra. UnblindMix was trained on augmented data. This helps it handle noise and distortions. NMF uses a simpler model and does not account for these effects.

\subsection{Experimental Validation} \addvspace{4pt}
To evaluate the performance of the UnblindMix model for the non-reactive case, 500 simulated spectra were generated at various concentrations and species using Algorithm \ref{algo} and used to train the model. For testing, 24 experimental mixtures of methane, ethane, ethylene, and propyne were carefully prepared, with mole fractions ranging from 0 to 1\%. These mixtures were introduced into a heated cell at 673 K, and their absorbance spectra were collected for evaluation.

\begin{figure}[h!]
    \begin{minipage}{\linewidth}
        \centering
        \includegraphics[width=0.6\textwidth]{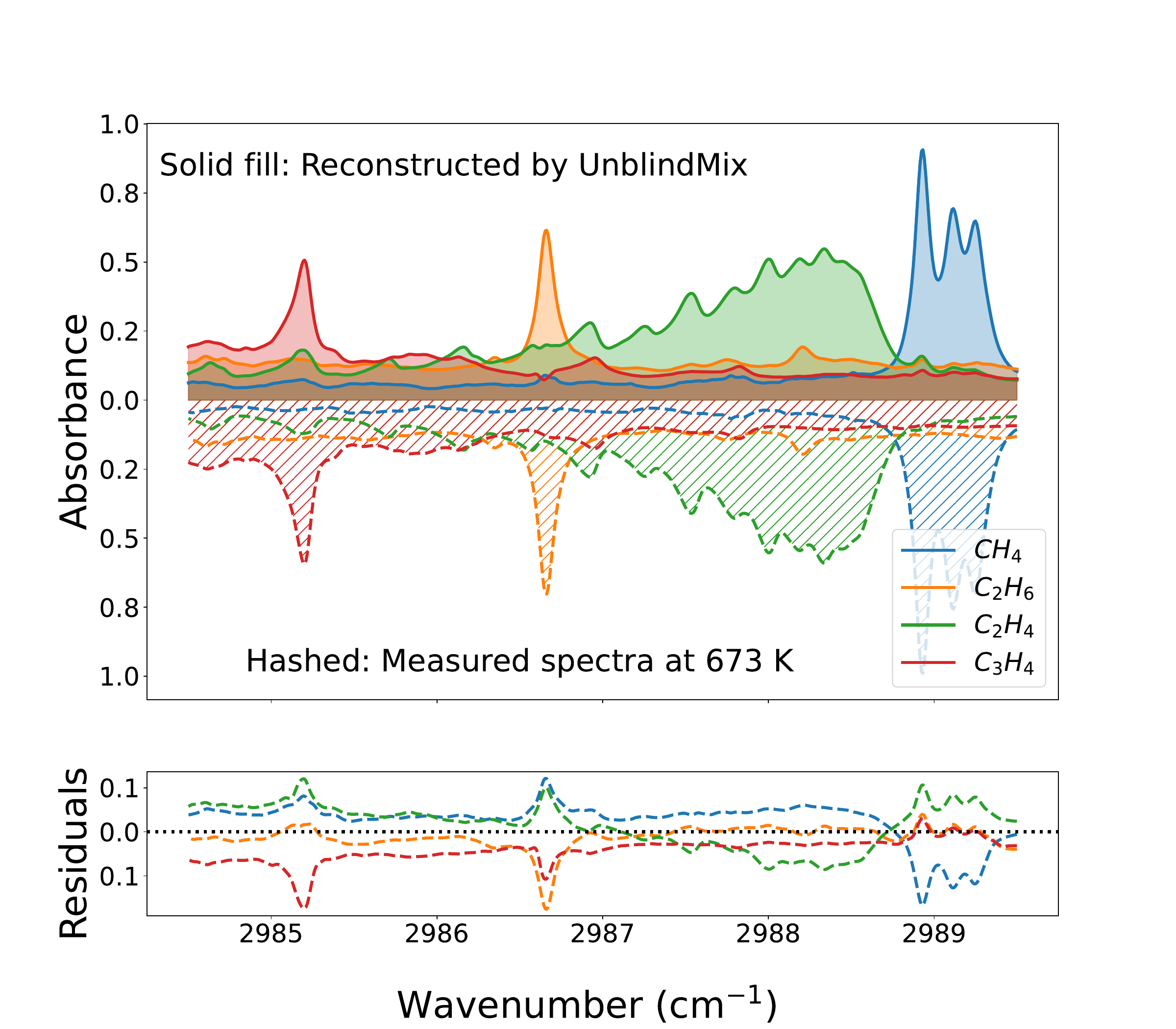}
        \caption{Comparison of measured spectra and those predicted by UnblindMix. Measured spectra refer to experimentally recorded high-temperature absorbance spectra of individual species.}
        \label{predicted_non_reactive}
    \end{minipage}
\end{figure}

\begin{figure}[h!]
    \begin{minipage}{\linewidth}
        \centering
        \includegraphics[width=0.6\textwidth]{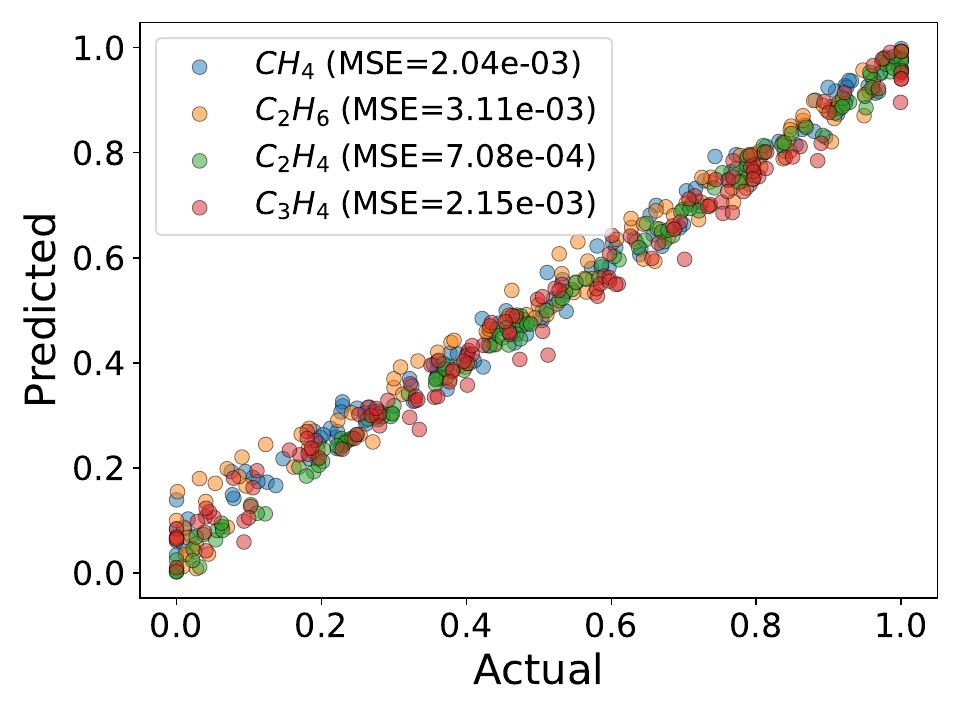}
        
    \end{minipage}
    \caption{Comparison of measured mole fractions and those predicted by AE-BSS (UnblindMix) model.}
    \label{predicted_con_non_reactive}
\end{figure}

UnblindMix demonstrated reliable performance in predicting species concentrations and reconstructing high-temperature reference spectra based solely on the composite mixture spectra as input. Figure 
 \ref{predicted_non_reactive} compares the reconstructed spectra (solid fill) with the experimentally measured high-temperature reference spectra (hashed fill) at 673~K, within the 2985–2989~cm\(^{-1}\) range. The reconstructed spectra exhibit strong agreement with the reference data, with absolute residual errors consistently below 0.15. 
Figure \ref{predicted_con_non_reactive} presents the predicted versus actual mole fractions for methane, ethane, ethylene, and propyne. The mean squared error (MSE) values for these predictions are \(2.4 \times 10^{-3}\), \(3.11 \times 10^{-3}\), \(7.08 \times 10^{-3}\), and \(2.15 \times 10^{-4}\), respectively.

\subsection{Effect of Initial Guess}\addvspace{3pt}

The selection of the initial guess in an unsupervised source separation framework is as important as the model architecture itself. While Gaussian noise is a common and straightforward choice, it often leads to slower convergence and necessitates a larger training dataset to achieve satisfactory performance. In the UnblindMix framework, room-temperature spectral data is employed as the initial guess for the decoder, given its widespread availability for most combustion-relevant species. These initial spectra are selected at a fixed mole fraction of 1\% and  at 1~atm.

\begin{figure}[H]
    \begin{minipage}{\linewidth}
        \centering
        \includegraphics[width=0.65\textwidth]{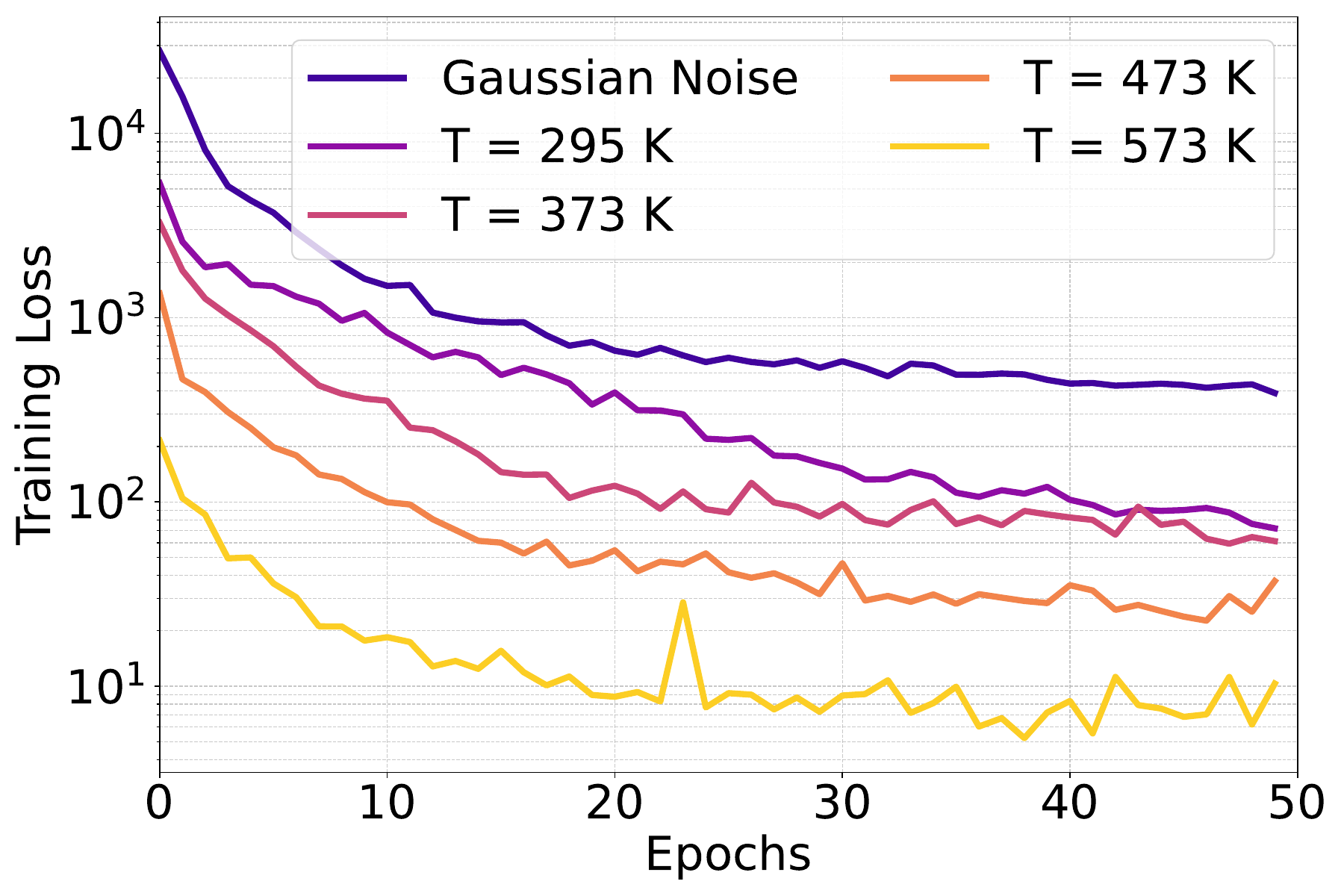}
        \caption{ Effect of the initial guess on model convergence.}
        \label{fig:initialguess}
    \end{minipage}
\end{figure}

To evaluate the influence of initial guesses on the performance of UnblindMix, we conducted comparisons using Gaussian noise and spectral data at temperatures increasingly closer to the target temperature. The results demonstrated that initial guesses closer to the target temperature significantly enhance the convergence rate of the model. Figure~\ref{fig:initialguess} illustrates the effect of different initial guesses on convergence rate, comparing Gaussian noise and spectral data at various temperatures.
Results show that using initial guesses closer to the target temperature (673~K) improves convergence and reduces training loss. For example, an initial guess at 573~K, which is 100~K below the target, leads to a final training loss of lower than $10$ within 50 epochs.When the initial guess is at Gaussian noise, the final loss increases to about $10^3$, three orders of magnitude higher. These results indicate that the proximity of the initial guess to the target temperature affects the efficiency of training in high-temperature conditions.

\subsection{Effect of Training Size}\addvspace{4pt}
Acquiring high-temperature spectral data is both challenging and resource-intensive. To optimize the training process, we investigated the performance of UnblindMix across varying training dataset sizes to determine the minimum number of observations required for reliable predictions. 

\begin{figure}[H]
    \begin{minipage}{\linewidth}
        \centering
        \includegraphics[width=0.65\textwidth]{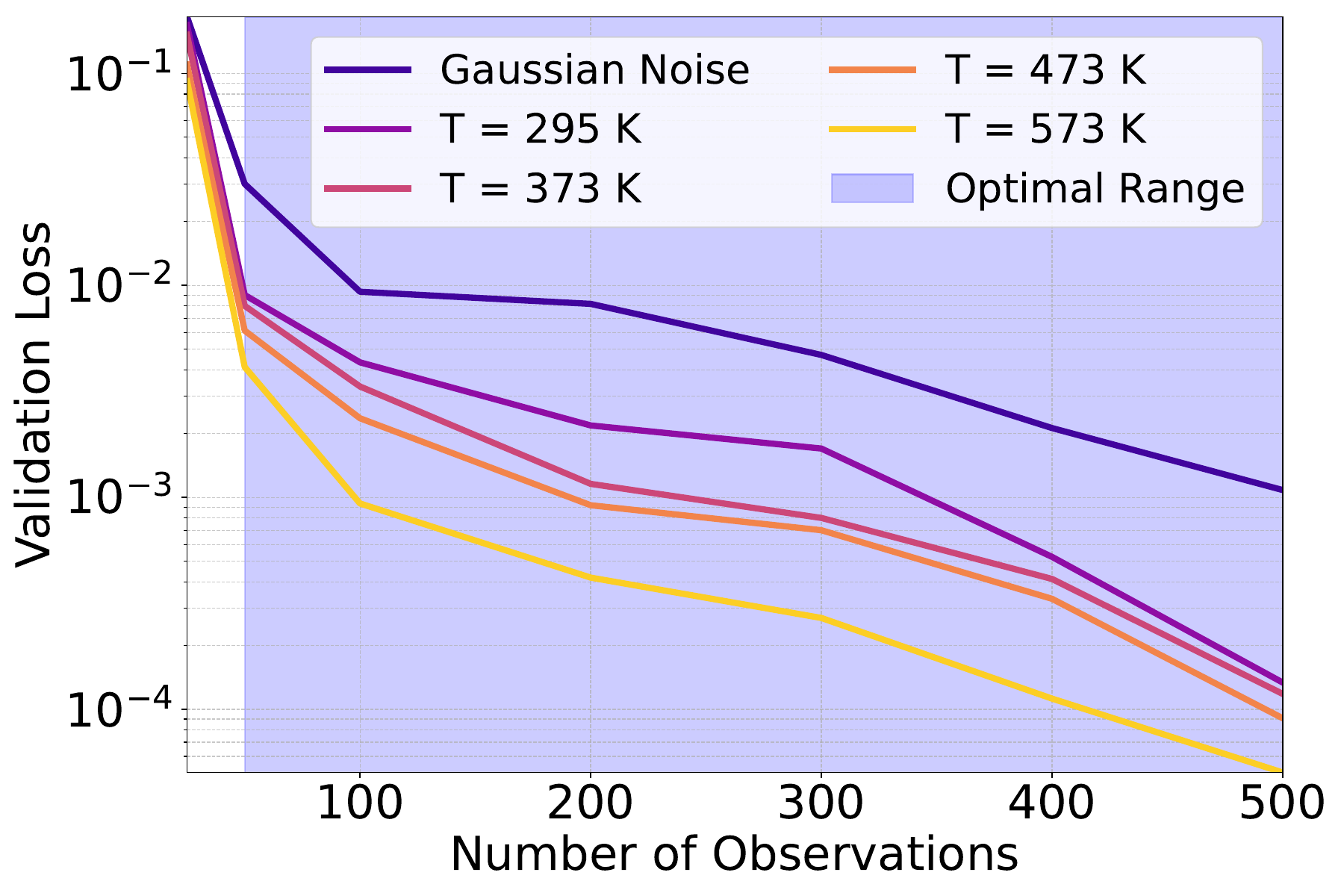}
        \caption{Effect of training data size on model convergence. Validation loss is plotted for various cases of initial guess.}
        \label{fig:trainingsize}
    \end{minipage}
\end{figure}

The results presented in Fig. \ref{fig:trainingsize} demonstrate a clear relationship between training data size and model accuracy. When the training dataset was limited to only 20 mixtures, the MSE validation error increased significantly, indicating that the model struggled to learn effectively with such a small sample size. However, when the training set included more than 50 mixtures, UnblindMix achieved an error of around \(10^{-2}\), particularly when using spectral data from other conditions as an initial guess. Figure~A.1 in the \textit{Appendix A} shows how validation losses vary when the model is trained using lower-dimensional input spectra, demonstrating the effect of input resolution on model performance.

\subsection{Uncertainty Quantification}\addvspace{3pt}
\label{sec:uncertainty}

A similar high-temperature static cell was used in absorption cross-section measurements conducted by Farouki et al.\cite{farouki2024measurement}, where multiple sources of uncertainty were identified. Key factors included fluctuations in laser intensity, variations in temperature and pressure, path-length inconsistencies within the cell, and mole fraction deviations in the prepared mixtures. In our experiments, laser intensity was found to contribute approximately 0.5\% to the absorbance uncertainty, which was calculated using the following formula:

\[
\frac{\partial A}{A} = \left[ \left( \frac{\partial I_0}{I_0} \right)^2 + \left( \frac{\partial I_t}{I_t} \right)^2 \right]^{\frac{1}{2}}
\]

When combined with other uncertainties—temperature variations (8\%), pressure deviations (1\%), path-length variability (1.4\%), and mole fraction differences (0.2\%)—these factors resulted in a cumulative uncertainty in the absorption cross-sections of approximately 11.1\%, calculated as:

\[
\frac{\partial \sigma}{\sigma} = \left[ \left( \frac{\partial A}{A} \right)^2 + \left( \frac{\partial P}{P} \right)^2 + \left( \frac{\partial \chi}{\chi} \right)^2 + \left( \frac{\partial T}{T} \right)^2 + \left( \frac{\partial L}{L} \right)^2 \right]^{\frac{1}{2}}
\]

To reflect the uncertainty introduced by the UnblindMix model, we also incorporated the residual error from the predicted mixture spectra \( \hat{x} \). The mean residual between measured and reconstructed spectra was found to be consistently on the order of 5-10\% in the pyrolysis experiments. This additional contribution is captured in the total propagated uncertainty expression as:

\[
\frac{\partial \sigma}{\sigma} = \left[ \left( \frac{\partial A}{A} \right)^2 + \left( \frac{\partial P}{P} \right)^2 + \left( \frac{\partial \chi}{\chi} \right)^2 + \left( \frac{\partial T}{T} \right)^2 + \left( \frac{\partial L}{L} \right)^2 + \left( \frac{\partial \hat{x}}{\hat{x}} \right)^2 \right]^{\frac{1}{2}}
\]

Here, \( \frac{\partial \hat{x}}{\hat{x}} \) quantifies the contribution from the model's reconstruction error and is based on the validation residuals obtained during training.

\subsection{Applications to Reactive Environments}\addvspace{7pt}

The diagnostic technique was applied to investigate the pyrolysis of iso-butane and n-butane. A 1\% mixture of either hydrocarbon in nitrogen was heated to 923 K under controlled conditions. To ensure consistent thermal profiles, the cell was stabilized at 923 K for one hour before initiating 30-minute pyrolysis experiments. During these experiments, methane, ethane, ethylene, propane, and propyne were identified as the products of pyrolysis. Among these, methane and ethylene were observed to dominate, exhibiting significantly higher absorbance levels, while the contributions from other species remained below 1\%. 

To evaluate the proposed diagnostic method under pyrolysis conditions, the analysis focused on detecting the primary reactants and two major byproducts (methane and ethylene). Absorbance spectra were collected every 10~ms over a 30-minute period at 923~K, resulting in 180{,}000 mixture spectra. These data were used to train a dedicated instance of the UnblindMix model for high-temperature conditions. The model is fully unsupervised and learns directly from the measured mixture spectra to infer both reference spectra and species mole fractions, without requiring any labeled data or prior knowledge of species identities.

Although the same UnblindMix architecture had previously been applied at 673~K, the training is temperature-specific. The model trained at 673~K outputs spectra and mole fractions valid only at that temperature and cannot be applied to 923~K due to the temperature-dependence absorption cross sections. For this reason, a new model was trained independently using only the experimental data collected at 923~K.

The outputs included predicted high-temperature spectra for the reactants, methane, and ethylene at 923~K, along with their time-resolved mole fraction profiles. These predictions were compared to simulated mole fraction profiles from AramcoMech 3.0~\cite{zhou2018experimental}, assuming 1\% \textit{n}-butane or \textit{iso}-butane pyrolysis under constant temperature and pressure. The predicted reference spectra were also evaluated against experimentally measured high-temperature spectra at 923~K. The comparisons to kinetic simulations are provided solely as reference baselines to contextualize the UnblindMix predictions. No attempt is made to calibrate, validate, or assess the accuracy of the kinetic mechanism itself.

\subsubsection{n-Butane Pyrolysis:}

\begin{figure}[H]
    \begin{minipage}{\linewidth}
        \centering
        \includegraphics[width=0.65\textwidth]{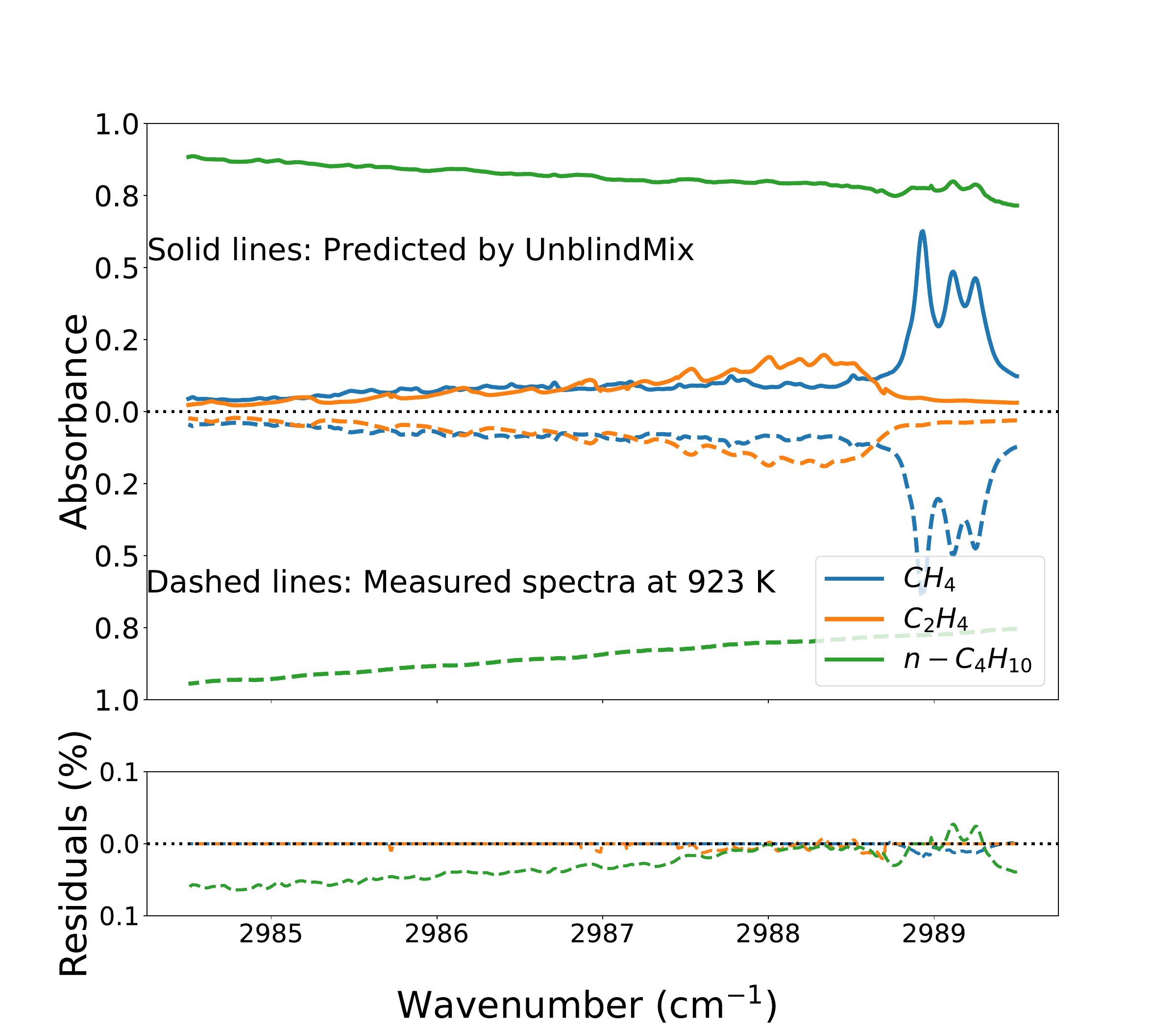}
    \end{minipage}
    \caption{Measured reference spectra vs. predicted by UnblindMix at 923 K during pyrolysis of n-butane.}
    \label{fig:n_butane}
\end{figure}

Figure~\ref{fig:n_butane} compares the predicted and experimentally measured reference spectra of \textit{n}-butane at 923~K. The spectra reconstructed by the UnblindMix decoder exhibit good agreement with experimental measurements, with deviations generally within the uncertainty bounds of the absorption cross-section data. Some discrepancies are observed, particularly in flatter regions of the spectrum. These deviations are likely due to a combination of factors, including baseline drift, nonlinear spectral distortions, and the presence of interfering species such as ethane, propyne, and 1-butene, whose contributions may not be fully captured in the training spectra.

\begin{figure}[H]
    \centering
    \begin{minipage}[t]{0.48\linewidth}
        \centering
        \caption*{(a)}
        \includegraphics[width=\linewidth]{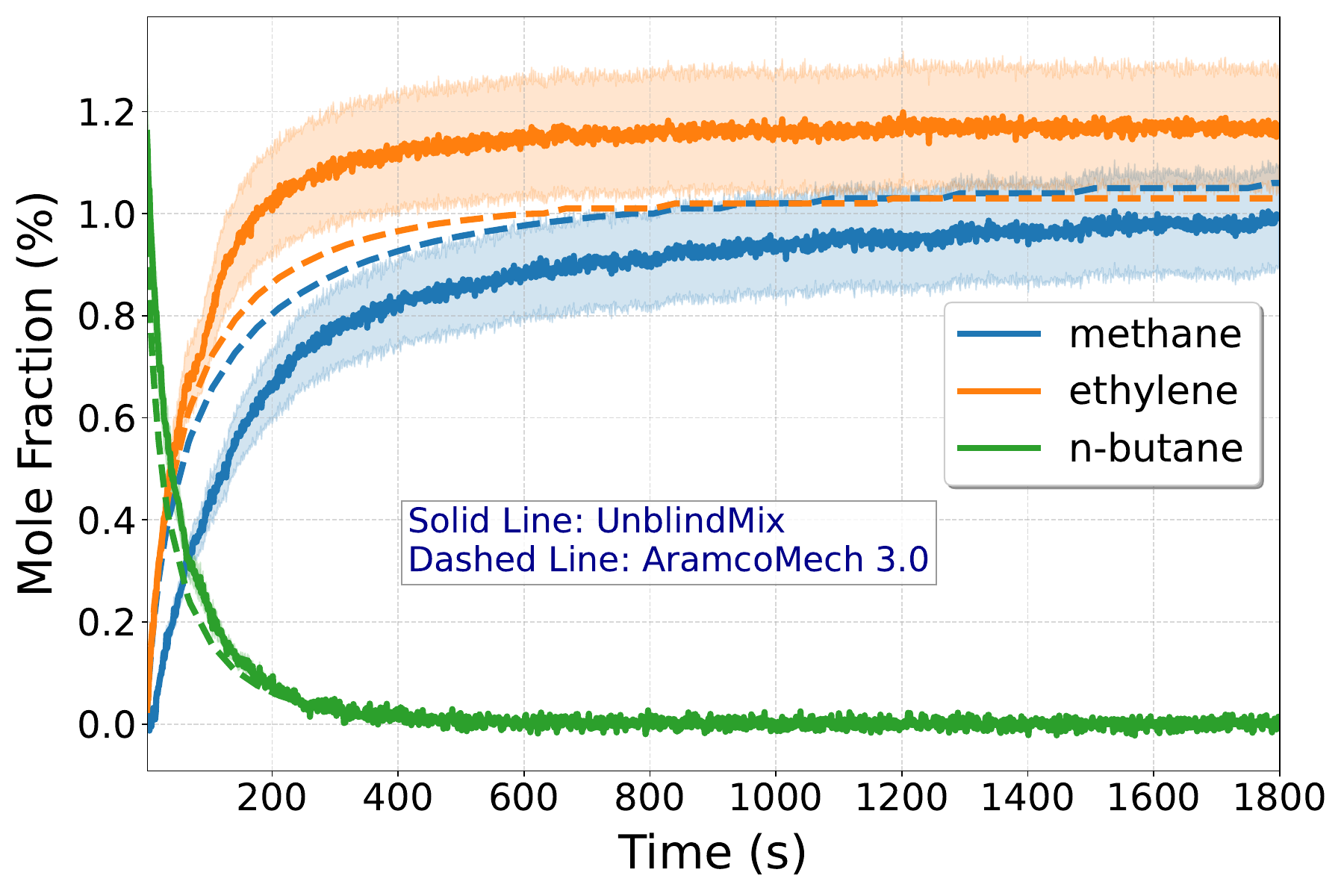}
    \end{minipage}
    \hfill
    \begin{minipage}[t]{0.48\linewidth}
        \centering
        \caption*{(b)}
        \includegraphics[width=\linewidth]{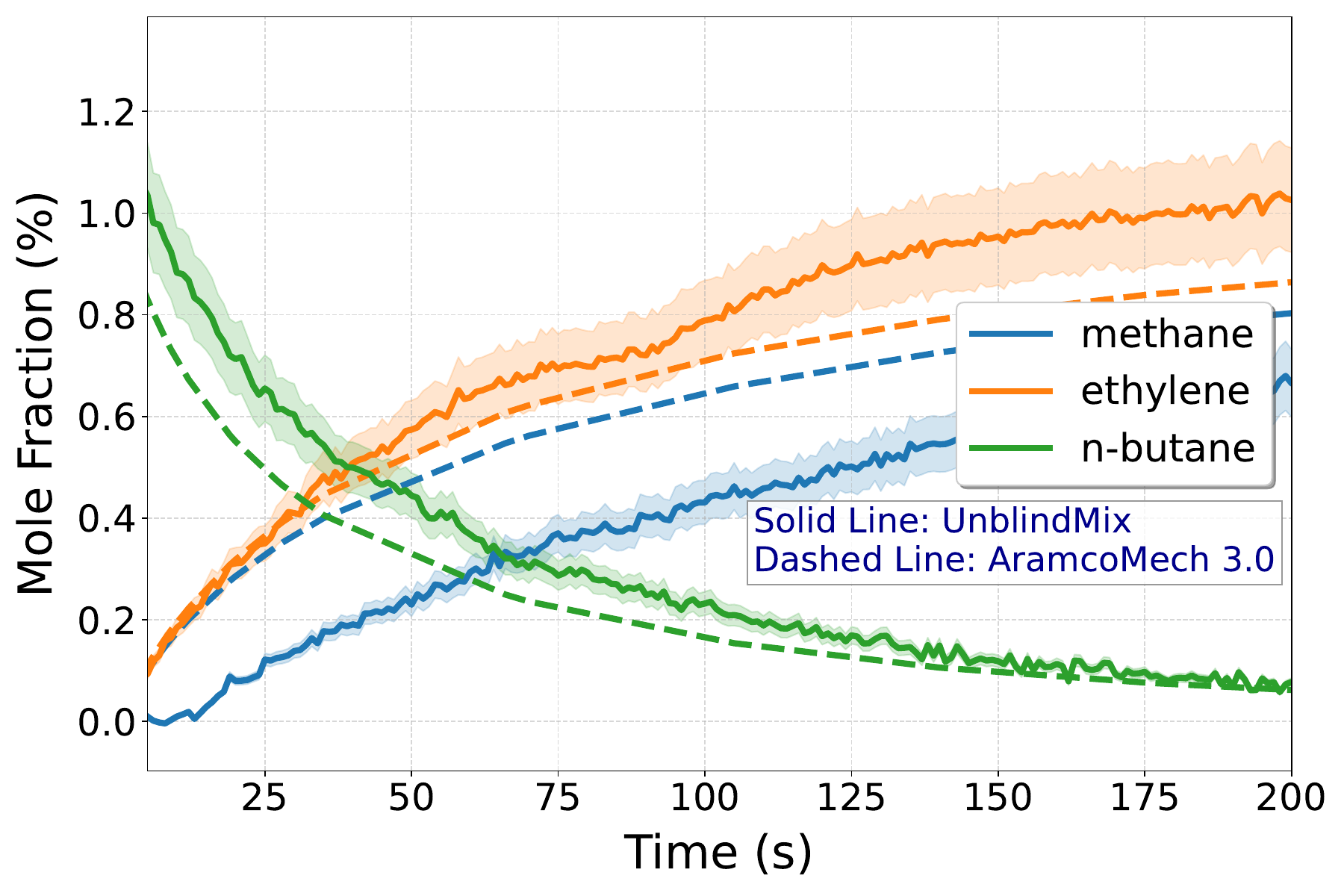}
    \end{minipage}
    \vspace{-5mm}

    \caption{(a) Actual vs. predicted mole fraction time histories by UnblindMix during n-butane pyrolysis. The shaded regions represent cumulative uncertainties from both the UnblindMix model and absorption cross-section variations. (b) Zoomed-in view of the first 200 seconds.}
    \label{fig:butane_comparison}
\end{figure}

In Fig.~\ref{fig:butane_comparison}, the time-resolved mole fraction predictions show consistent trends with those from AramcoMech 3.0. While the agreement is generally good, mismatches are present and can be attributed to several sources of uncertainty. First, the UnblindMix model, while trained to be robust to noise, still operates on measured spectra that may contain residual distortions due to temperature gradients, or optical misalignment. Second, uncertainties in the thermodynamic conditions, particularly fluctuations in temperature and pressure, can affect both spectral line intensities and chemical evolution, contributing to variability in both the measured data and kinetic model predictions. Third, the kinetic mechanism itself may introduce errors; deviations between UnblindMix and model simulations could reflect limitations in reaction rates, species pathways, or thermochemical parameters used in kinetic model.

The shaded regions in Fig.~\ref{fig:butane_comparison} represent cumulative uncertainties that combine reconstruction residuals from UnblindMix, measurement variability in absorption cross-sections, and potential deviations arising from kinetic model predictions. These results demonstrate that while UnblindMix provides a physically consistent reconstruction of the system dynamics, some differences with simulation data are expected and can be explained by well-understood sources of experimental and modeling uncertainty.

To improve the reliability of the predictions, we implemented a physically-constrained variant of UnblindMix.

\textbf{Carbon balance:} Conservation of atomic carbon was enforced through the following constraint:

\begin{equation}
\sum_i C_i \cdot x_i(t)  \leq C_{\text{fuel}}
\end{equation}
where \(C_i\) is the number of carbon atoms in species \(i\), \(x_i(t)\) is its mole fraction at time \(t\), and \(C_{\text{fuel}}\) is the total number of carbon atoms initially present in the fuel.

\textbf{Monotonicity of key products:} For major pyrolysis products such as CH\textsubscript{4} and C\textsubscript{2}H\textsubscript{4}, monotonic growth over time was enforced:
\begin{equation}
x(t) \leq x(t+1)
\end{equation}
This reflects the expected temporal evolution of stable end products under isothermal pyrolysis.

\textbf{Mole fraction bounds:} Predicted mole fractions were constrained not to exceed species-specific upper limits:
\begin{equation}
x_i(t) \leq x_{i,\text{max}}
\end{equation}
where \(x_{i,\text{max}}\) was derived from kinetic model predictions. This constraint narrows the optimization domain and prevents the model from converging toward unrealistic concentration values.

The impact of these physical constraints is illustrated in Figure~C.1 of \textit{Appendix C}.

\subsubsection{iso-Butane Pyrolysis}

Similarly, for \textit{iso}-butane pyrolysis, the model effectively captured the evolution of species; however, it overpredicted ethylene and underpredicted methane. These discrepancies are likely due to baseline shifts caused by the presence of additional species such as ethane, propyne, and 1-butene, which are known byproducts of \textit{iso}-butane. Other contributing factors may include uncertainties in temperature and pressure measurements, limitations in the absorption cross-section data at high temperatures, and potential uncertainty in the kinetic model used for comparison.

\begin{figure}[H]
    \begin{minipage}{\linewidth}
        \centering
        \includegraphics[width=0.6\textwidth]{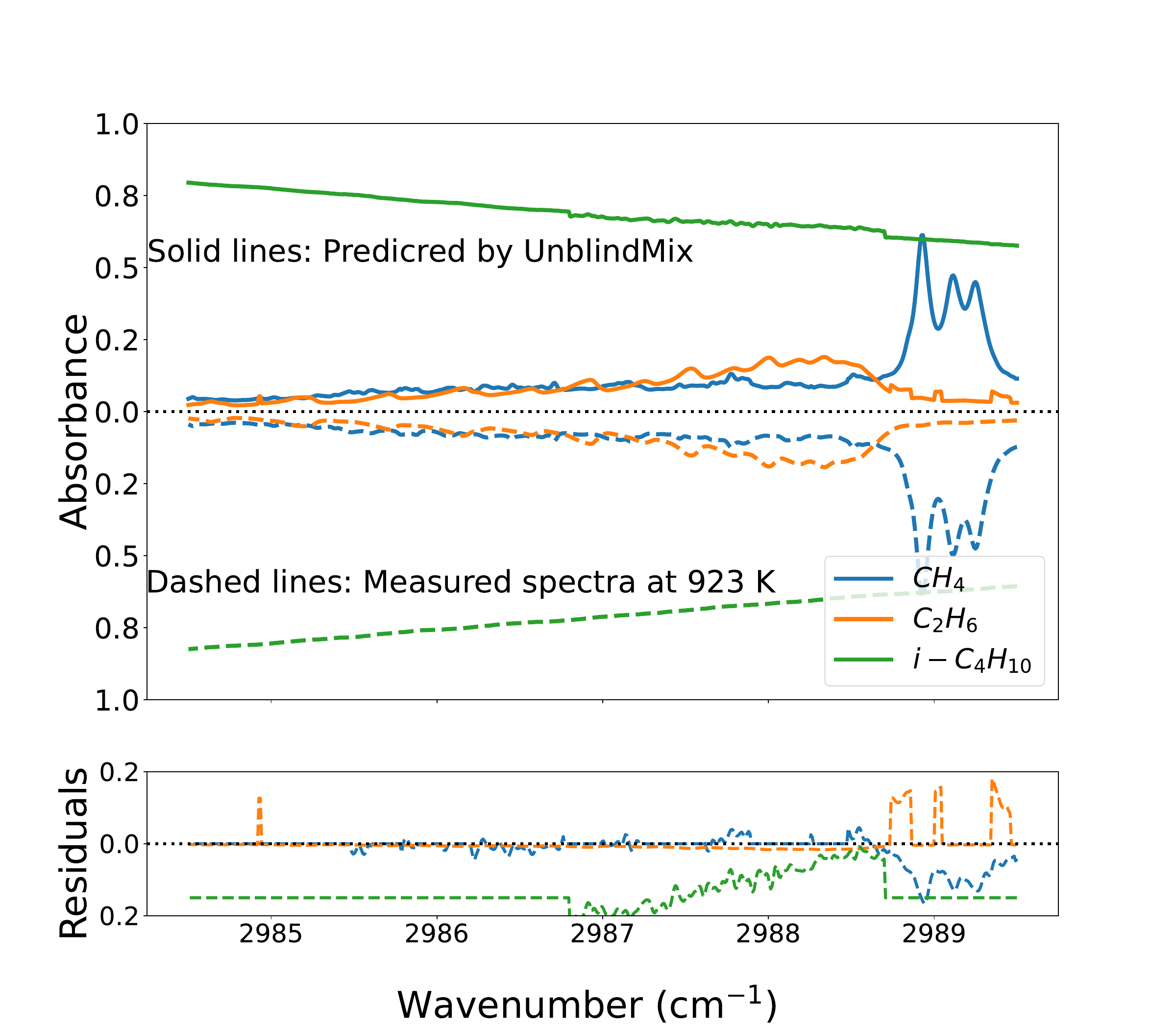}
        \caption{Measured reference spectra vs. predicted by UnblindMix at 923 K during pyrolysis of iso-butane. Measured spectra refer to experimentally recorded high-temperature absorbance spectra of individual species.}
        \label{fig:isobutane}
    \end{minipage}
\end{figure}

\begin{figure}[H]
    \centering
    \begin{minipage}[t]{0.48\linewidth}
        \centering
        \caption*{(a)}
        \includegraphics[width=\linewidth]{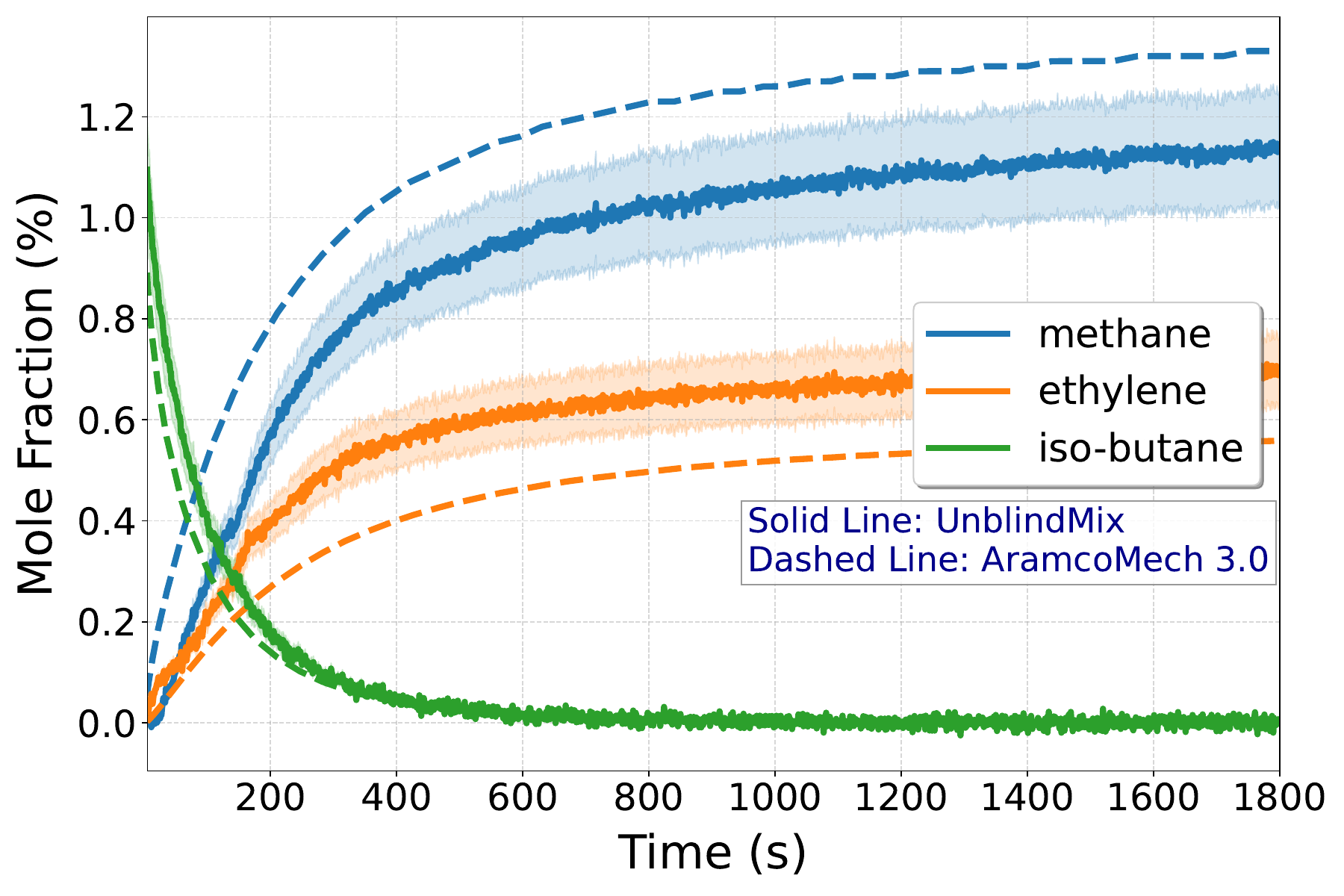}
        
    \end{minipage}
    \hfill
    \begin{minipage}[t]{0.48\linewidth}
        \centering
        \caption*{(b)}
        \includegraphics[width=\linewidth]{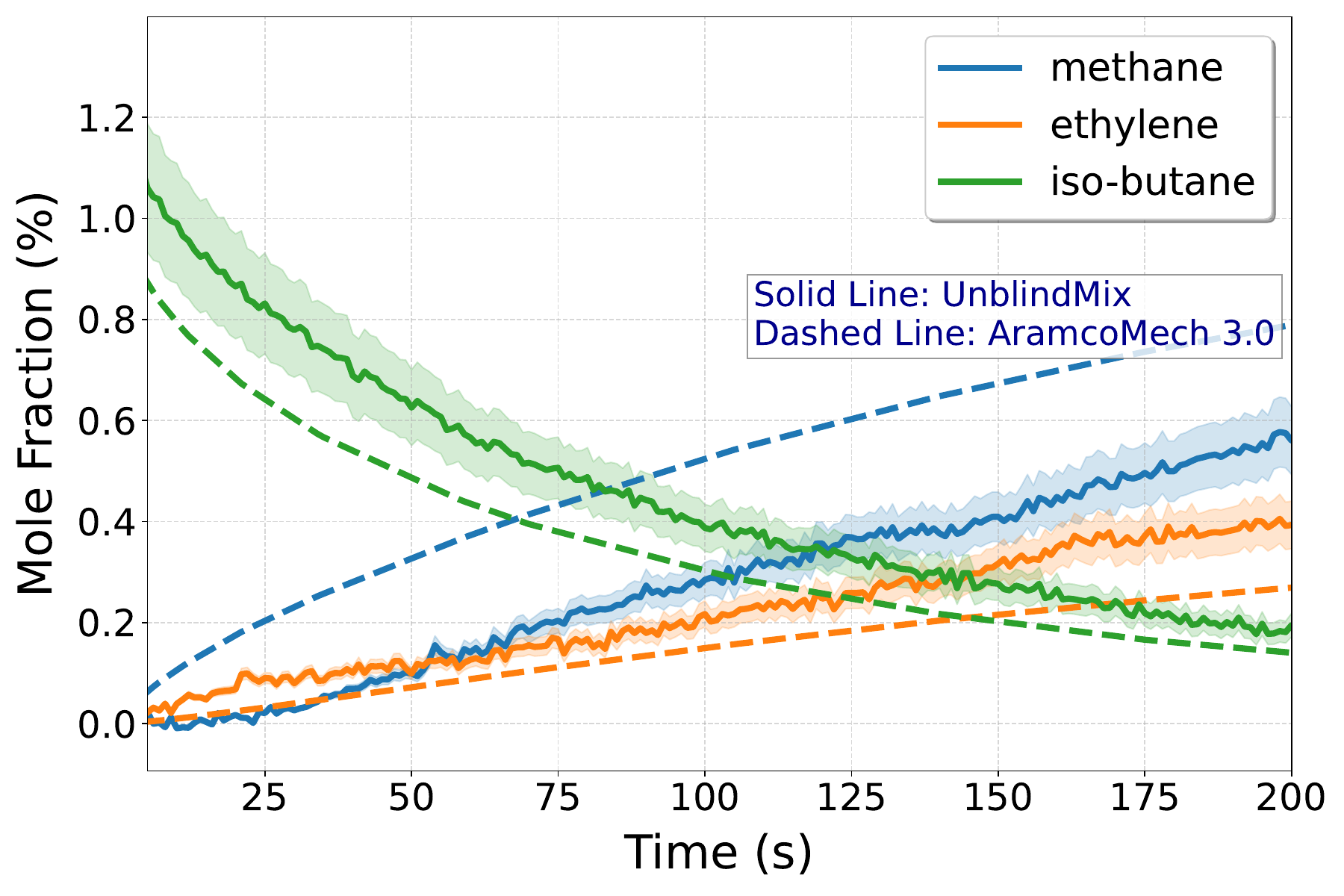}
    \end{minipage}
    \vspace{-5mm}
    \caption{(a) Actual vs. predicted mole fraction time histories by UnblindMix during iso-butane pyrolysis. The shaded regions represent cumulative uncertainties from both the UnblindMix model and absorption cross-section variations. (b) Zoomed-in view of the first 200 seconds.}
    \label{fig:ibutane_comparison}
\end{figure}

Figure~\ref{fig:isobutane} compares the predicted and measured reference spectra for \textit{iso}-butane, methane, and ethylene at 923~K. In the predicted spectra, some abrupt variations are observed. These features are a result of the non-negativity constraints applied during training. Figure~\ref{fig:ibutane_comparison} shows the comparison between the simulated and predicted mole fraction time-histories over the 30-minute \textit{iso}-butane pyrolysis experiment.

\subsection{UnBlindMix Limitations}\addvspace{6pt}

While UnblindMix provides a flexible and reference-free approach for spectral decomposition, several limitations remain:

\begin{itemize}
    \item \textbf{Dependence on measured mixture spectra:} Although the framework does not require isolated reference spectra, it must be trained on a set of measured composite spectra at the target temperature and pressure. The accuracy of the predictions is therefore linked to the quality and representativeness of these measurements.

    \item \textbf{Limited generalization outside training conditions:} The model performs reliably within the range of its training data but may not generalize well to conditions involving untrained temperatures, pressures, or unaccounted interfering species.

    \item \textbf{Assumption of spectral linearity:} UnblindMix assumes that measured spectra are linear combinations of species spectra and does not account for spectral line shape effects. In environments where pressure broadening or other nonlinear phenomena are significant, this assumption may reduce reconstruction accuracy.

    \item \textbf{Sensitivity to unknown species:} If a species is present in the mixture but not represented in the training data, its contribution may be misattributed to other species in the model’s learned basis, particularly in regions with strong spectral overlap.

    \item \textbf{Constraint-induced bias:} The physically constrained variant improves stability by incorporating chemically-informed regularization, including carbon balance and monotonic formation trends. However, these constraints rely on prior knowledge of the species and reaction behavior, which may limit applicability in poorly characterized kinectics models.
\end{itemize}
    \chapter{Spectral Feature Engineering : Feature Transformation for an Enhanced Multi-species Detection}\label{chapter6}
This chapter is adapted from the accepted manuscript version of an article by Sy et al., accepted for publication in \textit{Sensors and Actuators B: Chemical}~\cite{sy2025multi}.

\section{Introduction}
\label{sec:ch6:intro}
\label{sec:ch6:introduction}

Infrared absorption spectroscopy is widely used for gas mixture analysis due to its specificity, non-invasiveness, and suitability for real-time measurements~\cite{farooq2022laser}. While broadband instruments such as FTIR spectrometers and frequency combs provide extensive spectral coverage~\cite{waxman2017intercomparison}, their limited portability restricts use in field-deployable sensing platforms. Compact sources such as interband cascade lasers (ICLs) and quantum cascade lasers (QCLs)~\cite{mhanna2023selective2, elkhazraji2023laser} offer fast scanning and high stability, making them suitable for portable applications, albeit with restricted tuning ranges that complicate multi-species detection when absorption features are broad or overlapping.

Quantitative analysis of absorbance spectra typically relies on regression methods such as multiple linear regression (MLR), partial least squares regression (PLSR), or multivariate curve resolution (MCR)~\cite{jeevaretanam2023transient, elkhazraji2024selective, mhanna2023selective, o2016multi}. However, these approaches degrade in performance when spectra are highly correlated, spectrally featureless, or when one of the analytes has comparatively weak absorbance. Recent studies have explored deep learning models, including feed-forward and convolutional neural networks (CNNs), to estimate species concentrations directly from raw absorbance spectra~\cite{mhanna2022laser, sy2023multi, castorena2021deep, sy2023multi2}. Although promising, these models often underperform when tasked with identifying low-concentration or weakly absorbing species~\cite{wang2020improved, ren2024laser, sy2024voc}.

To address these limitations, we explore a spectral feature engineering strategy designed to enhance model robustness and sensitivity to subtle absorbance features. Feature engineering plays a critical role in enabling models to learn more effectively from structured transformations of raw input~\cite{zheng2018feature}, and has been widely used in other domains such as image analysis through techniques like SIFT and HOG~\cite{lindeberg2012scale, dalal2005histograms}.

In this work, we develop and evaluate a convolution-based feature extraction method that combines the composite absorbance spectra and their first derivatives to emphasize underlying spectral structures. This approach improves the model’s ability to detect weak absorbers and distinguish species with overlapping or flat spectral features. The methodology is demonstrated using experimental data acquired with an ICL scanned over 2984.5–2989.5~cm\(^{-1}\), applied to mixtures containing CH\(_4\), C\(_2\)H\(_4\), C\(_2\)H\(_6\), C\(_3\)H\(_8\), and propyne (C\(_3\)H\(_4\)-P).

\section{Sensor description}

\subsection{Wavelength selection}
\begin{figure}[ht!]
\centering
\includegraphics[width=0.85\linewidth]{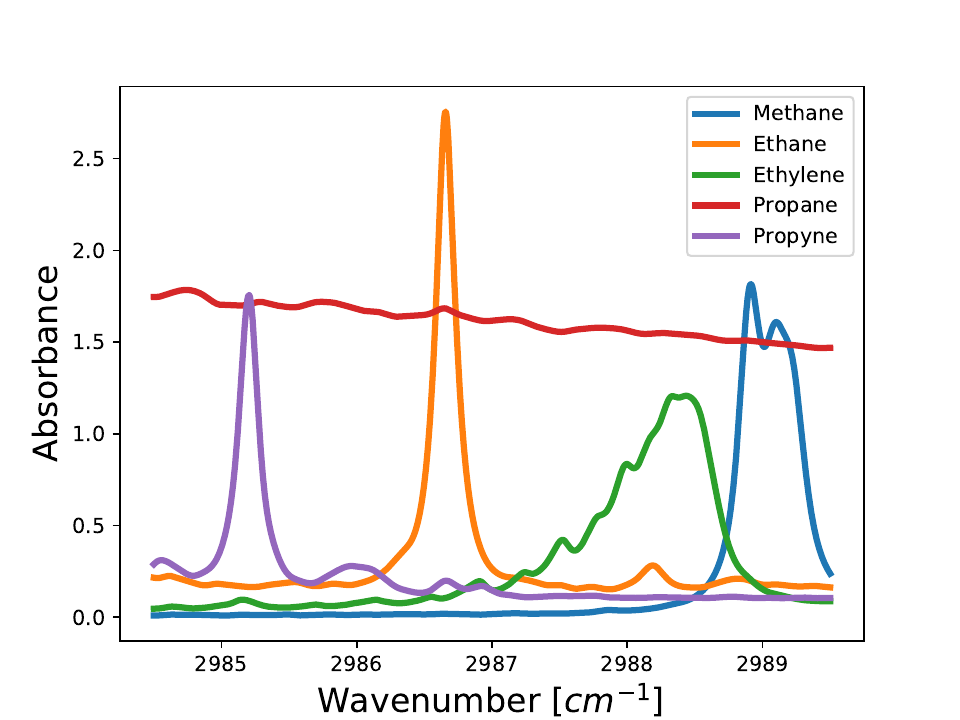}
\caption{Measured absorption spectra of the target species at T = 295 K, P = 1 atm, L = 10 cm and \(\chi\) = 1\%.}
\label{fig:spectra}
\end{figure}

The C–H stretching band, particularly near 3.3 $\mu m$, is highly effective for detecting hydrocarbons, in particular \(C_1-C_3\) hydrocabons, due to its strong and distinct absorption characteristics\cite{goldenstein2017infrared, sy2024laser}. This vibrational mode is advantageous because it exhibits significant intensity, enabling sensitive detection even at low concentrations. Furthermore, the position of the C–H stretching bands varies with the hybridization state of the carbon atom, which allows for the differentiation of various hydrocarbon types based on their molecular structure\cite{hanson2016spectroscopy}. In this study, a wavenumber range from 2984.5 to 2989.5 $cm^{-1}$ is selected, which can be accessed using a single ICL (interband cascade laser), to target \(C_1-C_3\) species. Each of these hydrocarbons shows distinct spectral features within this range, making them easily distinguishable. Figure \ref{fig:spectra} displays the measured absorbance spectra of the target species at conditions: T = 295 K, P = 1 atm, L = 10 cm, and $\chi$ = 1\%.

\subsection{Optical setup}
\begin{figure}[ht!]
\centering
\includegraphics[width=\linewidth]{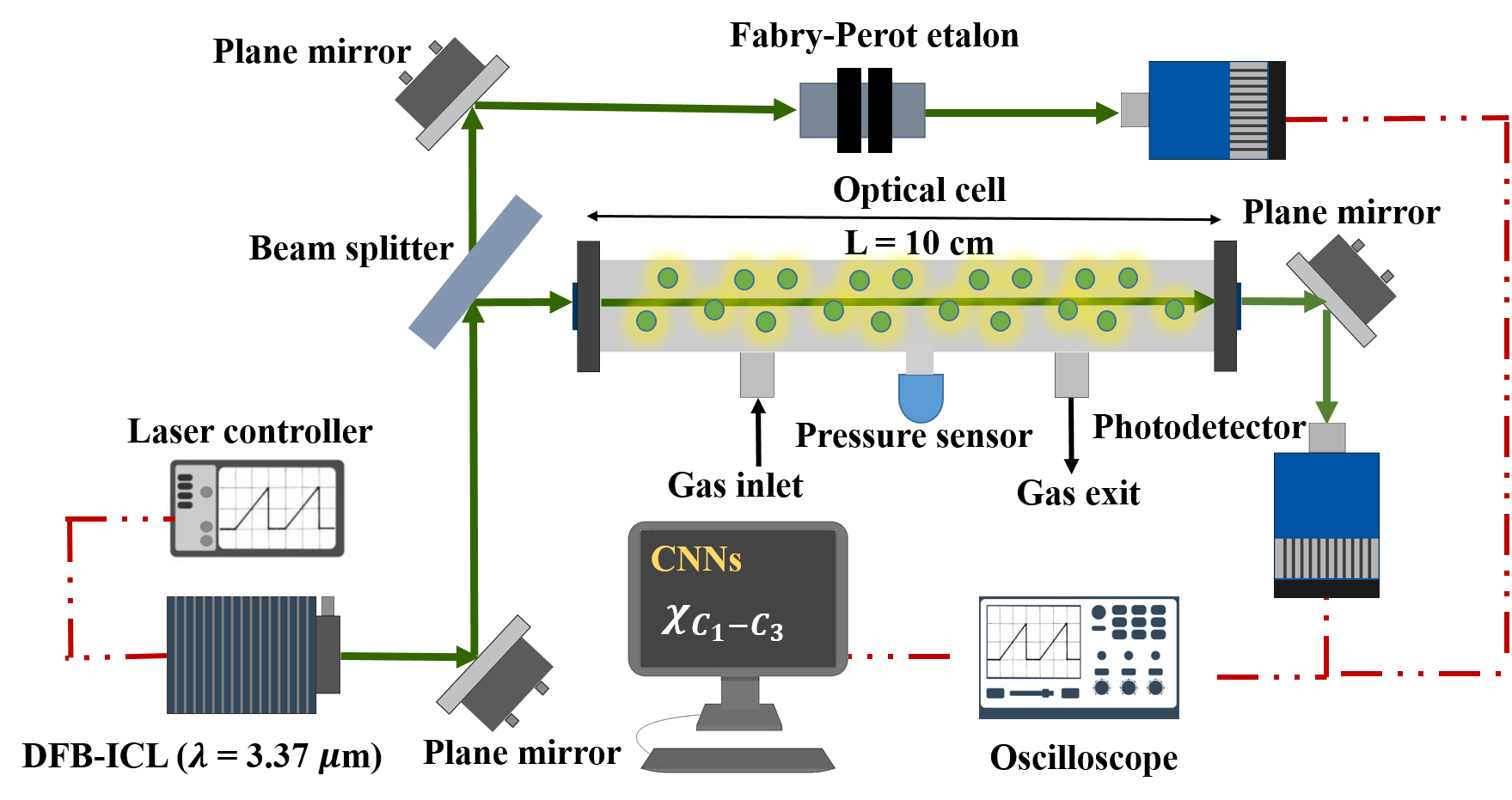}
\caption{Schematic of the optical sensor setup.}
\label{fig:setup}
\end{figure}

To access the desired wavelength region, a distributed feedback inter-band cascade laser (DFB-ICL, Nanoplus) was employed. This laser, operating near 3.34 $\mu$m with an output power of 10 mW, enabled precise tuning across the range of 2984.5 to 2989.5 $cm^{-1}$. The wavelength tuning was achieved by modulating the injection current using a sawtooth waveform at a scan rate of 100 Hz. To convert the scan time into wavelength, a 7.62-cm Germanium Fabry-Perot etalon was utilized. Measurements were performed in a compact 10 cm stainless steel cell, selected solely for demonstration purposes, while acknowledging the availability of better alternatives (e.g., multipass cell or cavity-enhanced absorption) for achieving lower detection limits. A four-stage thermo-electrically (TE) cooled photovoltaic (PV) detector from Vigo was used to record the signal. The optical setup is schematically depicted in Figure \ref{fig:setup}, illustrating the arrangement and components used for wavelength selection and measurement. All measurements were carried out at a static pressure of 1 atm.

\section{Spectral feature engineering}
\subsection{Feature transformation process}
Feature transformation is a common technique in machine learning designed to highlight new features and improve model learning efficiency\cite{zheng2018feature}. In the context of spectral feature engineering, this can be useful particularly for addressing the challenge of detecting species with low absorbance that can be overshadowed by those with high absorbance. By taking the first derivative of the absorbance spectra and convolving it with the original spectra, we can effectively isolate these low-absorbing species\cite{mhanna2024method}. This approach eliminates flat regions—where there is minimal variation—since the derivative of a constant is zero, thus reducing the y-axis (absorbance) value range and enhancing the visibility of subtle features. The first derivative of the mixture absorbance can be expressed as:
\[
\frac{d\alpha_\nu}{d\nu} = \frac{d}{d\nu} \left( \sum_{i=1}^n \sigma_i(T, P, \nu) \cdot n_i \cdot L \right)
\]
where $\alpha_\nu$ is the mixture absorbance, $\sigma_i$ is the absorption cross-section for species $i$, $n_i$ is the number density, and $L$ is the path length.
Next, the convolution of the first derivative with the original absorbance spectrum is given by:
\[
C(\alpha_\nu, \frac{d\alpha_\nu}{d\nu}) = \int \alpha_\nu(t) \cdot \frac{d\alpha_\nu}{d\nu}(t-\tau) d\tau
\]
where $C(\alpha_\nu, \frac{d\alpha_\nu}{d\nu})$ represents the convolved spectra, $\alpha_\nu(t)$ is the original absorbance signal, and $\frac{d\alpha_\nu}{d\nu}(t-\tau)$ is the first derivative of the absorbance.

\begin{figure}[ht!]
\centering
\includegraphics[width=\linewidth]{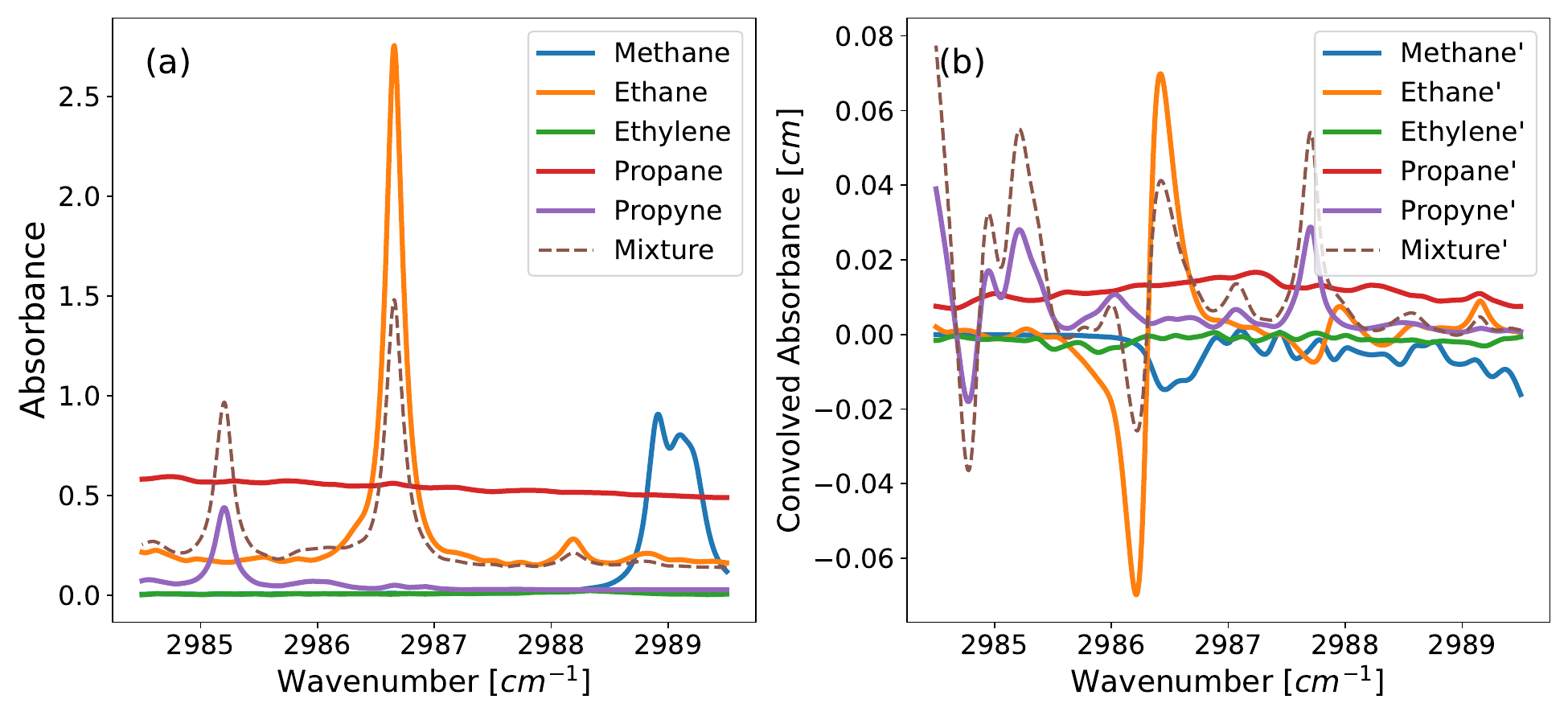}
\caption{((a) Original measured spectra of C\(_1\)-C\(_3\) hydrocarbons, where ethylene absorbance is kept very low (0.05), including a mixture with ethane and propyne at 1\% and ethylene at 200 ppm. (b) Feature engineered spectra highlighting key features after spectral transformation.}
\label{fig:convolvedvsstandard}
\end{figure}

Figure \ref{fig:convolvedvsstandard}\textcolor{blue}{(a)} illustrates a representative reference spectra (referred as 'original spectra'), including a mixture of \(C_1-C_3\) hydrocarbons with ethane and propyne at 1\% and ethylene at 200 ppm, where the ethylene absorbance is deliberately kept very low—nearly 20 times lower than the other components. Figure \ref{fig:convolvedvsstandard}\textcolor{blue}{(b)} displays the convolved spectra, obtained by taking the first derivatives of the original spectra and performing convolutions with the original spectra.  The convolved spectra reveal subtle features of ethylene. Small sinusoidal structures, observed in the convolved spectra, arise from the first derivative operation and small variations in the raw spectra. While the absorbance values of the original spectra ranged from 0 to 2.7, the range of the convolved spectra was significantly reduced, varying from -0.02 to  0.06, allowing the feature-engineered model to learn more effectively and efficiently. This highlights the importance of scaling and normalizing data, as these strategies can enhance model performance by making it easier to detect and learn from small but important features.
\newpage
\subsection{Model description}
\begin{figure}[ht]
\centering
\includegraphics[width=0.75\linewidth]{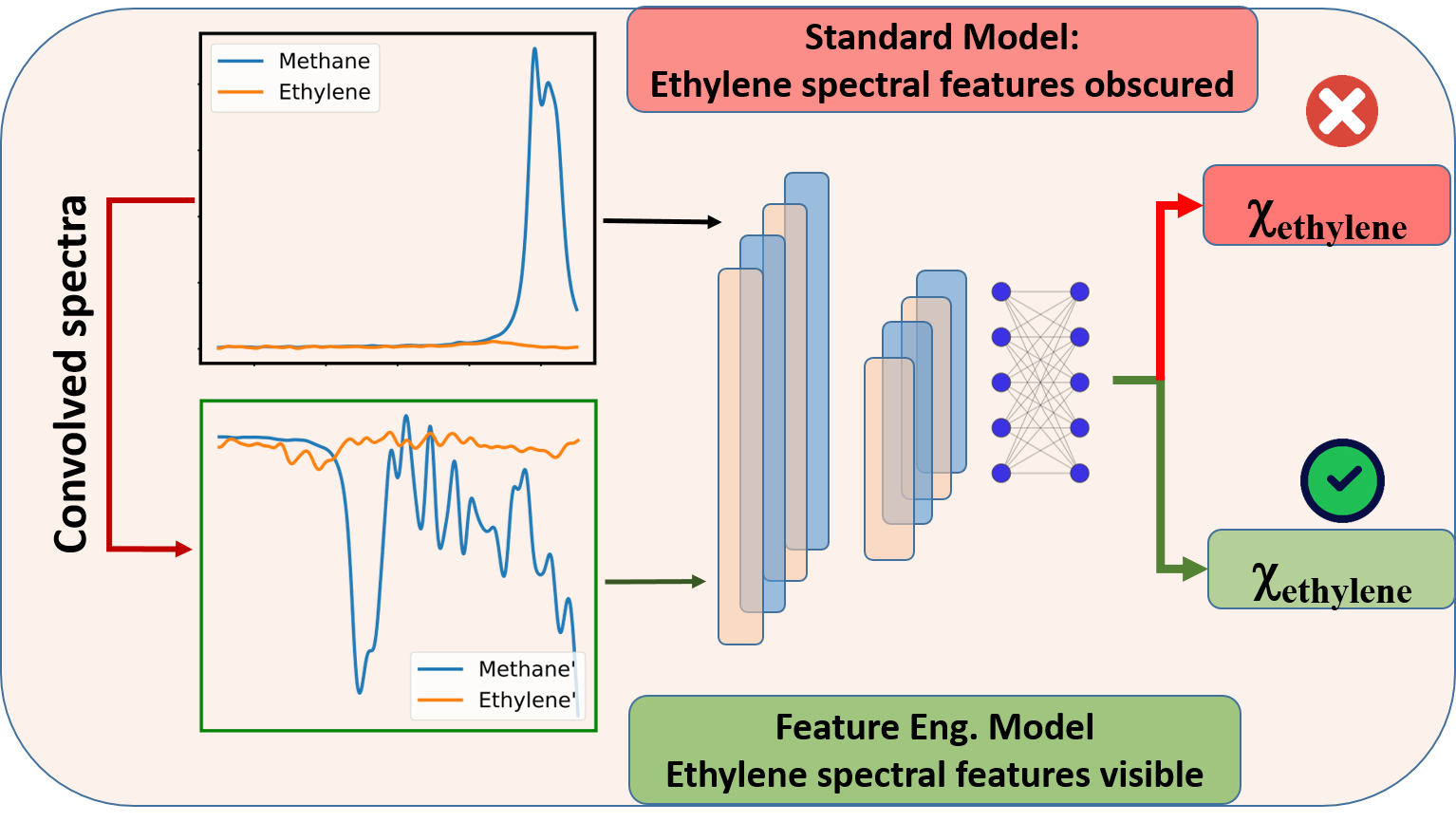}
\caption{Flowchart of the standard and feature-engineered models.}
\label{fig:ch6:flowchart}
\end{figure}

Convolutional neural networks (CNNs) have proven effective in extracting valuable information from corrupted spectra, making them a popular choice in various studies \cite{lecun1998gradient}. CNNs are known for their ability to learn task-specific features, potentially offering more generally applicable solutions compared to other architectures \cite{al2023augmentations}. In this study, we compared two models with identical CNNs architectures but differ in input data. The standard model directly processes the full 1-D mixture spectra, while the feature-engineered model uses convolutions of the first derivatives and the composite spectra. Both models are based on a 1-D convolutional neural network (CNN) architecture. This architecture begins with a convolutional layer consisting of 32 filters with a kernel size of 3, followed by a ReLU activation function\cite{agarap2018deep} to introduce non-linearity. A max-pooling layer is then applied to reduce the spatial dimensions and capture the most significant features. This sequence is repeated with a second convolutional layer containing 64 filters, another ReLU activation, and a second max-pooling layer. The resulting output is flattened into a 1D vector, which is passed through a fully connected layer with 128 neurons and another ReLU activation. The final output layer, also fully connected, consists of 5 neurons corresponding to the regression targets. Figure \ref{fig:ch6:flowchart} provides a flow chart that illustrates the process for both the standard and feature-engineered models.

\subsection{Training procedure}
The training dataset for both the standard and feature-engineered models consists of 1,000 simulated mixture spectra, generated from measured reference spectra. Ethylene mole fractions vary from 0 to 200 ppm, resulting in a maximum absorbance of 0.05, while the mole fractions of other hydrocarbons range from 0 to 1\%, yielding a maximum absorbance of 2.7. For the standard model, the training data are generated by directly combining reference spectra with various component mole fractions, resulting in composite spectra based on the Beer-Lambert law. In contrast, the feature-engineered model utilizes convolutions of these composite spectra with their first derivatives. The data are split into 80\% for training and 20\% for validation. During testing, various \(C_1-C_3\) hydrocarbon mixtures are analyzed under atmospheric conditions using the developed optical sensor.

\begin{figure}[H]
\centering
\includegraphics[width=\linewidth]{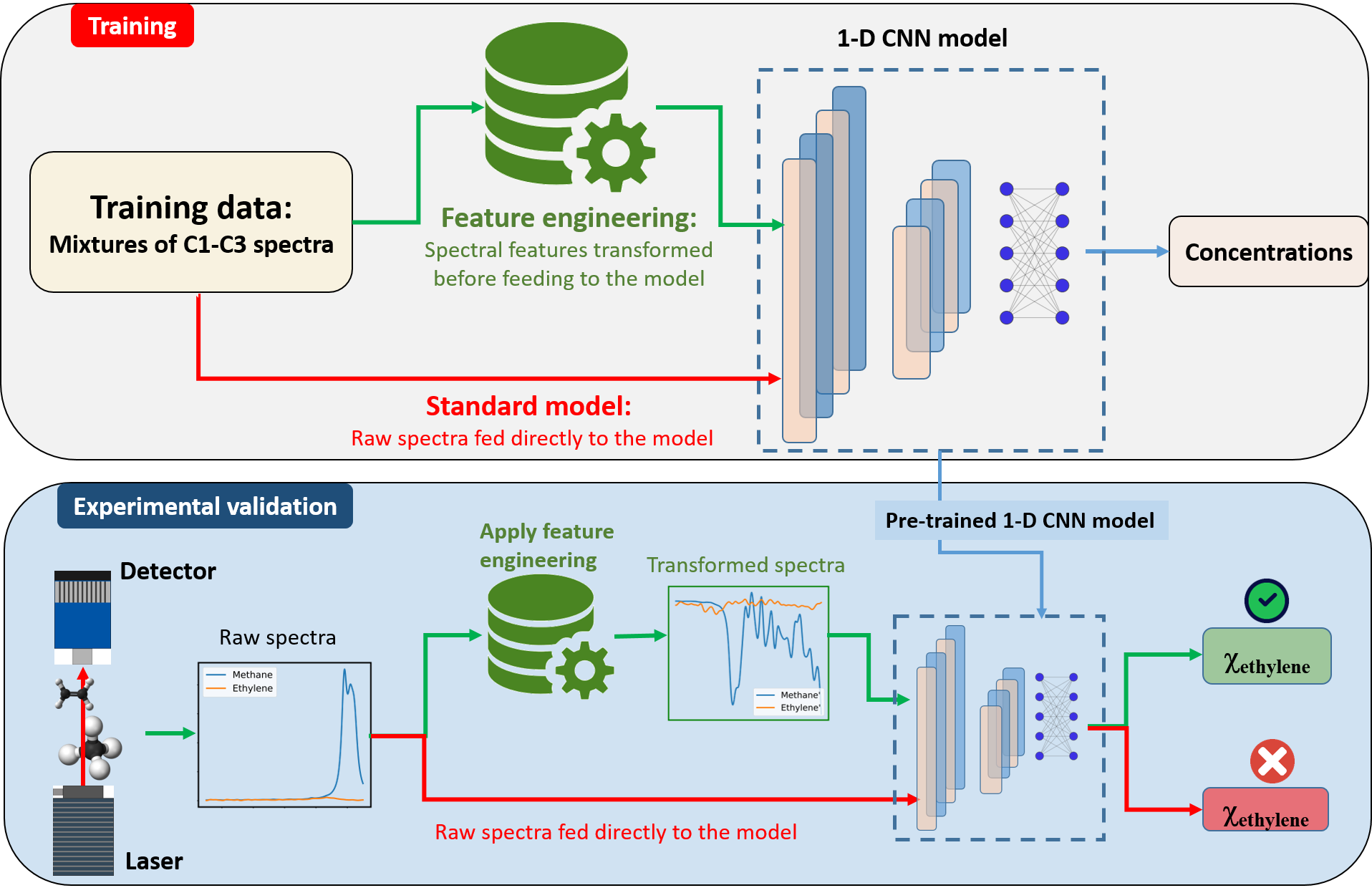}
\caption{Flowchart of the training process: the top light yellow box represents the training phase, while the bottom light blue box represents the testing phase via experiments . Red arrows indicate the standard model steps, and green arrows show the feature-engineered model steps.}
\label{fig:training_full}
\end{figure}

The CNN model is trained using mean squared error (MSE) as the loss function and the Adam optimizer\cite{zhang2018improved} with a learning rate of 0.001. Training spans 50 epochs, with batches of size 32 processed. For each batch, the model computes the loss, backpropagates the error, and updates the parameters to minimize the loss. After training, the model weights are saved and used as a pre-trained model to evaluate experimental data. For experimental testing, the same data processing protocol applied during training is used. Raw experimental \(C_1-C_3\) absorbance data are input into the pre-trained standard model, while convolved \(C_1-C_3\) absorbance data are fed into the feature-engineered model for quantifying \(C_1-C_3\) concentrations. Figure \ref{fig:training_full} illustrates the training process flow chart, where the top light yellow box represents the training phase and the bottom light blue box represents the testing phase. The red arrows correspond to the steps followed by the standard model, while the green arrows depict the processes specific to the feature-engineered model.
\subsection{Model evaluation}
\begin{figure}[H]
\centering
\includegraphics[width=0.85\linewidth]{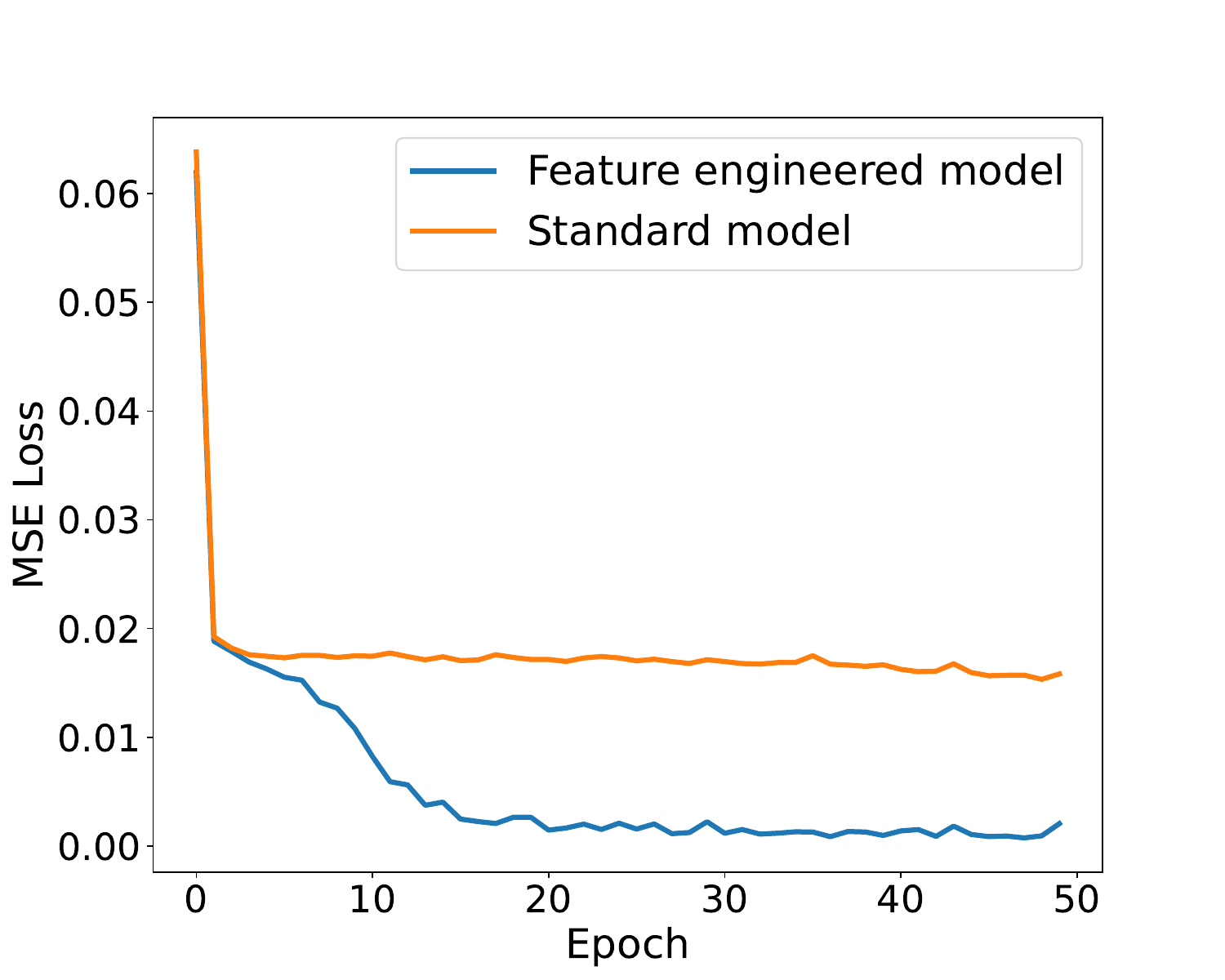}
\caption{MSE loss of the standard and feature engineered models.}
\label{fig:training}
\end{figure}

During the training process, the performance of both models was evaluated using mean squared error (MSE) as the loss function. 
Figure \ref{fig:training} illustrates the MSE loss for both the standard and feature-engineered models across the epochs. Initially, both models show a decrease in MSE, indicating effective learning. However, around epoch 10, the two models begin to diverge.
The MSE loss for the standard model stabilizes and reaches a plateau at this point. This suggests that the standard model has achieved its maximum performance with the given data. On the other hand, the feature-engineered model continues to show a decrease in MSE beyond epoch 10. The use of feature-transformed absorbance as opposed to raw absorbance allowed the feature engineering model to capture more subtle and complex features in the data, leading to further improvements in accuracy.

\section{Experimental validation}

\subsection{Allan deviation}
\begin{figure}[H]
\centering
\includegraphics[width=0.75\linewidth]{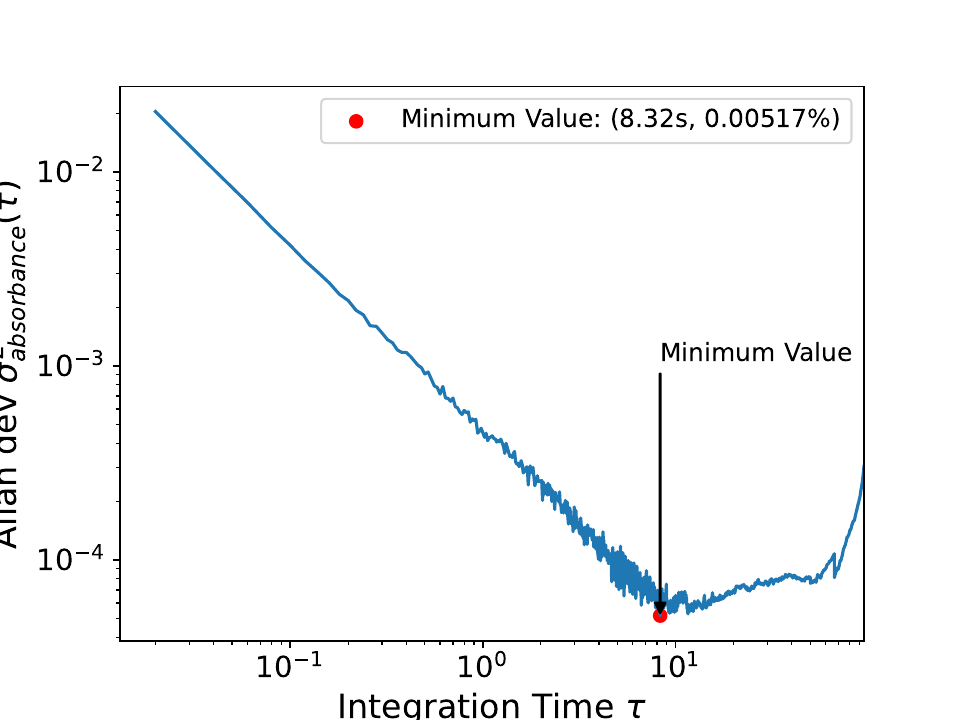}
\caption{Allan deviation plot of the sensor. Optimal integration time is 8.32 seconds with a minimum absorbance of 0.00005.}
\label{fig:allan}
\end{figure}

To evaluate the effectiveness of our feature-engineered model, we kept ethylene mole fraction lower than other target hydrocarbons in the prepared mixtures. Allan deviation analysis \cite{allan1966statistics} was performed to identify the lowest detectable absorbance and to confirm that the target ethylene absorbance remained detectable, even at relatively low levels. Figure \ref{fig:allan} illustrates that the sensor achieves a lowest detectable absorbance of 0.00005 at an integration time of 8.32 seconds. \hl{This sensitivity corresponds to detection limits of 400 ppb for methane, 150 ppb for ethane, 300 ppb for propyne, and 750 ppb for ethylene. This demonstrates that an absorbance of 0.01-0.05 (50 - 200 ppm) for ethylene would be easily detectable.}

\subsection{\(C_1-C_3\) concentration prediction}
To experimentally test the models, 48 mixtures of \(C_1-C_3\) hydrocarbons were prepared. Methane, ethane, propane and propyne mole fractions were varied between 0-1\%, while ethylene mole fraction was kept below 200 ppm. To minimize noise, 800 absorbance signals were averaged, corresponding to the sensor’s optimum integration time, where each absorbance signal takes 10 ms, corresponding to laser scan rate of 100 Hz.

\begin{figure}[H]
\centering
\includegraphics[width=1\linewidth]{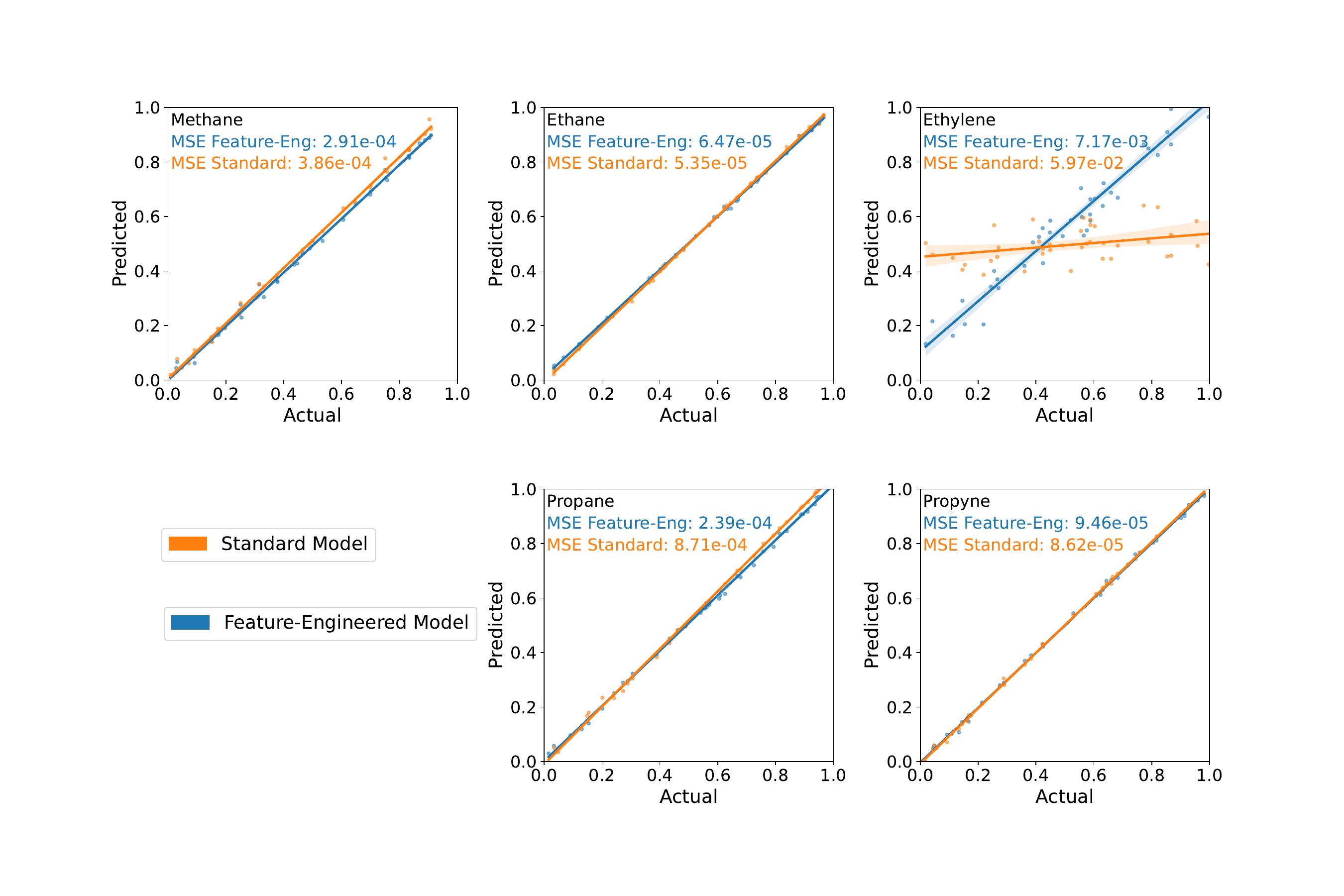}
\caption{ Comparison of predicted mole fractions with manometric (actual) values for 48 experimental mixtures with the standard and feature-engineered models. (Results are normalized to a maximum value of 1 for visualization purposes.)}
\label{fig:false-color}
\end{figure}

The results are summarized in Table \ref{tab:mse_results} and visualized in Figure \ref{fig:false-color}, which compares the known mole fractions with those predicted by the two 1-D CNN models. Both models demonstrated strong performance for methane, ethane, propane, and propyne, with average MSEs of around $10^{-4}$. Notably, the feature-engineered model achieved MSEs of $2.91 \times 10^{-4}$ for methane, $6.47 \times 10^{-5}$ for ethane, $2.39 \times 10^{-4}$ for propane, and $9.46 \times 10^{-5}$ for propyne, which were either comparable to or marginally better than the MSEs of the standard model. However, the most significant difference was observed in the detection of ethylene. The standard model consistently underperformed, giving mole fractions of about half of the known values, with an average MSE of $5.97 \times 10^{-2}$. In contrast, the feature-engineered model demonstrated a significantly improved performance with an MSE of $7.17\times 10^{-3}$, representing almost a tenfold improvement over the standard model.

\begin{table}[H]
    \centering
    \caption{Summary of MSE results for standard and feature-engineered models.}
    \label{tab:mse_results}
    \begin{tabular}{c|c|c}
    \hline
    \textbf{Species} & \textbf{MSE: Standard Model} & \textbf{MSE: Feature-Eng. Model} \\
    \hline
    Methane & $3.86 \times 10^{-4}$ & $2.91 \times 10^{-4}$ \\
    \hline
    Ethane & $5.35 \times 10^{-5}$ & $6.47 \times 10^{-5}$ \\
    \hline
    Ethylene & $5.97 \times 10^{-2}$ & $7.17 \times 10^{-3}$ \\
    \hline
    Propane & $8.71 \times 10^{-4}$ & $2.39 \times 10^{-4}$ \\
    \hline
    Propyne & $8.62 \times 10^{-5}$ & $9.46 \times 10^{-5}$ \\
    \hline
    \end{tabular}
\end{table}

The improvement in ethylene detection highlights the benefits of the feature engineering methodology. By applying convolutions to the first derivatives and composite spectra, previously obscured features within the spectral data were revealed. This process allows the model to learn more effectively and accurately, particularly when spectral features are flat, overlapping and/or relatively weak. 

\section{Effect of noise}
A sensitivity analysis was conducted to evaluate the robustness of the feature-engineered model across varying signal-to-noise ratio (SNR) levels. Noise levels were set relative to the ethylene absorbance, given its weaker absorbance, to examine the model's capacity to capture subtle spectral features in realistic, noisy environments. The SNR was progressively lowered from "clean" (no added noise) to 80, 50, 10, and 1 dB.

\begin{figure}[H]
    \centering
    \includegraphics[width=0.85\linewidth]{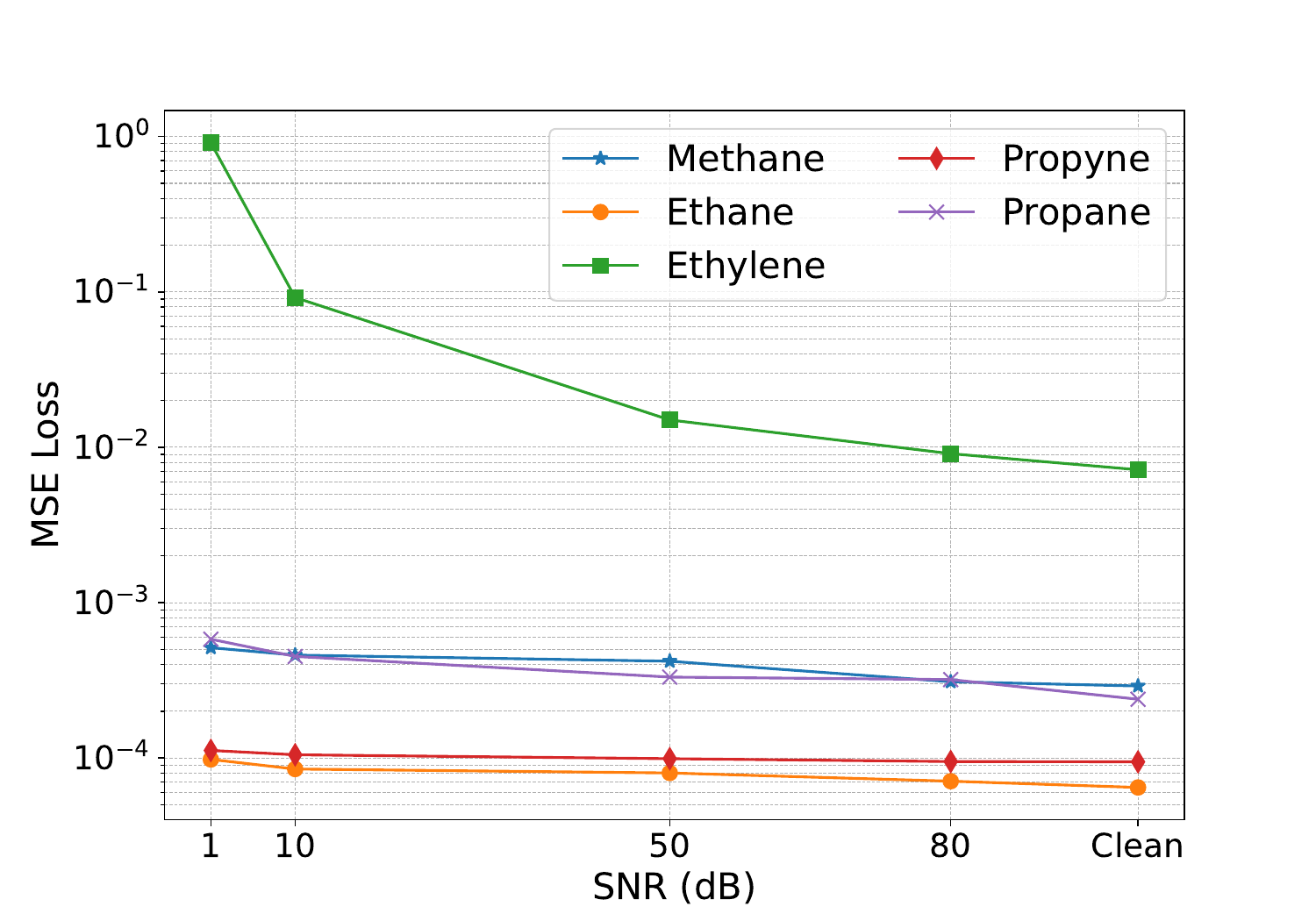}
    \caption{Effect of noise on species prediction.}
    \label{fig:sensitivity}
\end{figure}

Figure \ref{fig:sensitivity} presents the results of the noise sensitivity analysis, showing that as SNR decreases, MSE loss increases, with ethylene displaying the highest sensitivity to noise due to its lower absorbance. At SNR levels of 50 and above, the MSE for ethylene remains around \(10^{-2}\), allowing accurate predictions. Below this threshold, the model’s accuracy for ethylene declines. For the other species, the feature-engineered model consistently exhibits good MSE loss across all SNR levels.

    \chapter{Models Certification and Stability via Randomized Smoothing}\label{chapter7}
This chapter is adapted from the accepted manuscript version of an article by Sy et al., accepted for publication in \textit{ACS Omega}~\cite{sy2024voc}. 
\section{Introduction}
\label{sec:ch7:introduction}
Gas sensing is crucial in various sectors such as environment\cite{dhall2021review,mhanna2023selective}, energy\cite{Elkhazraji:23,sy2024laser}, and healthcare\cite{owen2014calibration}, significantly enhancing safety and operational efficiency\cite{feng2019review}. Accurate and selective gas sensors are vital for the detection of hazardous substances, enabling effective decision-making and risk management.  Researchers have explored various gas sensing techniques like photo-acoustics\cite{tomberg2018sub}, electro-chemical methods\cite{bakker2002electrochemical}, and gas chromatography\cite{eckenrode2001environmental}. However, laser-based spectroscopy stands out because it can identify gases by their unique spectral "fingerprints" \cite{farooq2022laser}. This method is non-intrusive, cost-effective, and efficient for measuring the composition, pressure, velocity, and temperature of gas mixtures. However, identifying these spectral fingerprints can be tricky, especially when different gases have overlapping features in complex mixtures, and this issue gets more pronounced with the effect of pressure and/or Doppler broadening in real-world settings.

Recent advancements in gas sensing have incorporated machine learning (ML) models for both classification and regression  \cite{nicolle2024mixtures,feng2019review,mhanna2022laser,javed2022quantification,mhanna2022deep,schwarm2023real,sy2023multi,peng2018gas,mozaffari2021convolutional,wang2021interpreting,chowdhury2022voc}. These models aim to automate spectral identification by learning unique absorption features that serve as fingerprints for each molecule. However, existing ML models, typically trained on i.i.d. (independent and identically distributed) datasets with known conditions, can encounter substantial accuracy when subjected to small perturbations or when tested on unseen conditions \cite{roelofs2019meta}. The limitation of generalizing to unknown conditions poses a significant challenge. Data augmentation techniques emerge as compelling solutions to enhance the robustness of ML models against potential perturbations\cite{perez2017effectiveness}. Widely employed in image classification\cite{shorten2019survey}, augmentation strategies involve mirroring, flipping, rotating, and zooming the original images to augment the dataset\cite{shorten2019survey,cubuk2019autoaugment}.

In spectroscopy, augmentations typically involve introduction of noise \cite{sy2023multi,gan2019multi,barton2021convolution,zhang2020noise} and/or baseline shifts\cite{wahl2020single} to the spectra. Other methods involve the application of extended multiplicative signal augmentation, addressing physical distortions arising from scattering and instrumental effects\cite{blazhko2021comparison}. A notable example of the effectiveness of data augmentation in gas sensing is shown by Al Ibrahim et al.\cite{al2023augmentations}. Their study introduces perturbations to composite spectra by adding fictitious spectra to the composite spectra through flipping, dilating, and mirroring the reference spectra, enhancing the model's generalization to unknown interferences. Despite their benefits, these augmentation techniques still require significant amounts of data to effectively train the ML models. This reliance on extensive datasets can be a limiting factor when such resources are not readily available \cite{brigato2021close}. In situations where a substantial amount of data is lacking, few-shot learning techniques can be employed.

In few-shot learning, the ML model is exposed to and trained on a limited dataset (few examples), with the expectation that it can accurately predict unseen instances \cite{wang2020generalizing}. To put it simply, in the context of gas sensing, this entails training the machine learning model under a limited set of conditions of pressure and temperature (P,T), after which it is challenged to predict outcomes under different conditions (P, T).  However, employing this approach poses challenges and may result in inaccurate predictions \cite{wang2020generalizing,seo2021self}, primarily due to the substantial variations in species spectra in response to the varying pressure and temperature conditions\cite{farooq2022laser,goldenstein2017infrared}. Factors such as Doppler-, self-, and collisional-broadening, along with line mixing, contribute significantly to these spectral variations \cite{farooq2022laser,goldenstein2017infrared,hanson2016spectroscopy}. Therefore, an augmentation technique that addresses these spectral variations is essential when adopting few-shot learning for precise predictions \cite{altae2017low}. Existing spectroscopic augmentation methods fall short of capturing the dynamic changes in spectra caused by changes in pressure and temperature. To tackle these challenges, our study presents a new and straightforward augmentation method that effectively tackles spectral changes caused by variations in pressure and temperature.

While augmentations contribute to improving the robustness of machine learning models, they do not provide complete defense against unseen perturbations
/attacks\cite{carlini2017towards,goodfellow2014explaining,szegedy2013intriguing}. In response to this limitation, various approaches have been developed to enhance a model's ability to defend itself against adversarial attacks. Many heuristic defenses have been proposed to create models resistant to adversarial perturbations; however, some of these defenses have proven vulnerable to more sophisticated adversaries\cite{carlini2017towards,athalye2018obfuscated}. Consequently, researchers have focused on strengthening empirical defenses\cite{madry2017towards} and developing certified defenses that offer robustness guarantees. Certified defenses ensure that classifiers deliver consistent predictions within a specified neighborhood of their inputs\cite{raghunathan2018certified,wong2018provable,cohen2019certified,salman2019provably,salman2020denoised}.

In critical domains such as gas sensing, the importance of model robustness, repeatability, and accuracy cannot be overstated. Consequently, effective and trustworthy approaches become crucial for the precise detection of species, particularly toxic ones. Thus employing provably certifiable and robust classifiers such as those seen in \cite{cohen2019certified, salman2020denoised} are necessary. The certification ensures that ML models can deliver reliable predictions within a predefined confidence radius, a crucial aspect in scenarios where precise and confident predictions are essential for decision-making and risk mitigation.

In this study, we focus on three key aspects: (1) introducing a novel augmentation technique which utilizes Voigt \cite{olivero1977empirical} convolutions with varying FWHM (full-width at half-maximum) to capture spectral changes due to pressure and temperature variations, in addition to noise and baseline shifts; (2) mitigating the need for extensive data by employing these augmentations to create a one-shot learning model for species classifications; and (3) providing a provable certification for predictions made by the newly developed one-shot classification model through randomized smoothing\cite{cohen2019certified}. We conduct a comprehensive comparison by classifying 12 volatile organic compounds (VOCs), replicating the VOC-net architecture from Chowdhury et al.\cite{chowdhury2022voc}; our model is trained under a single condition and, through augmentations, demonstrates comparable accuracy. To further understand the impact of augmentations and randomized smoothing, we compare four models: VOC-net, VOC-lite, VOC-plus, and VOC-certifire. VOC-net, requiring a substantial amount of high-quality data, is trained with a stratified split under all pressure conditions, rendering it susceptible to unknown conditions. In contrast, VOC-lite, which shares the architecture with VOC-net, is trained solely on one pressure condition and is thus vulnerable to unknown conditions. VOC-plus undergoes training under a single pressure condition, leveraging Voigt augmentations for improved performance. Subsequently, VOC-certifire employs VOC-plus as its pretrained classifier. During the testing phase, randomized smoothing is implemented. This process entails subjecting each test spectrum to multiple perturbations through augmentations. The final prediction for VOC-certifire is determined by aggregating the majority vote from VOC-plus predictions applied to the perturbed spectra. A detailed analysis of the certification process is conducted, exploring how the relation between radius and certified accuracy varies based on the number of perturbations, noise level, and confidence level. VOC-certifire ensures robust and reliable predictions within predefined confidence bounds, thus providing a useful tool for decision-making in gas sensing applications.

\section{Methodology}
\subsection{Dataset}
\begin{itemize}
    \item \textbf{Simulated data}: To benchmark our one-shot learning models, a fair comparison with existing classification models is essential. Chowdhury et al.\cite{chowdhury2022voc} introduced VOC-net, a convolutional neural network (CNN) model capable of classifying 12 VOCs, and evaluated its performance on both simulated and experimental datasets. In alignment with their approach, we chose to utilize a similar dataset to facilitate a meaningful comparison and used their experimental dataset for testing and validation. A concise overview of the VOC-net dataset generation is provided here, with additional details on the experimental setup available in\cite{chowdhury2022voc}.
    
The training data for VOC-net was generated using spectral simulations of twelve VOCs. These simulations were carried out based on spectroscopic parameters extracted from the HITRAN \cite{gordon2017hitran2016} and JPL\cite{pickett1998submillimeter} databases, utilizing the HAPI tool \cite{kochanov2016hitran} for spectral synthesis. Fig. \ref{fig:all_vocs} illustrates representative normalized simulated spectra for each molecule at 0.5 and 16 Torr. The frequency range considered spans from 220 to 330 GHz (7.33 – 11 \(cm^{-1}\)). The dataset encompasses spectra for the twelve VOC molecules, spanning a pressure range from 0.1 to 16.5 Torr (13.3 to 2200 Pa). Each spectrum corresponds to a single molecule and comprises 229 absorbance values, with a frequency resolution of 0.016 \(cm^{-1}\).
\item\textbf{Experimental Data:} Six VOCs were selected for experimental demonstration. Their spectra were obtained using an experimental setup, involving a THz microelectronics spectrometer operating in the 220–330 GHz range with a resolution of 0.5–15 MHz.  Measurements were conducted in a gas cell with a 21.6 cm absorption path at room temperature. Absorbance was calculated with incident and transmitted intensities and applying the Beer-Lambert relation. The resulting experimental dataset comprises 36 observations, with six measurements conducted for each of the six VOCs. Detailed information on the experimental setup can be found here\cite{chowdhury2022voc}.

\begin{figure}[!h] 
    \centering
    \includegraphics[width=0.85\textwidth]{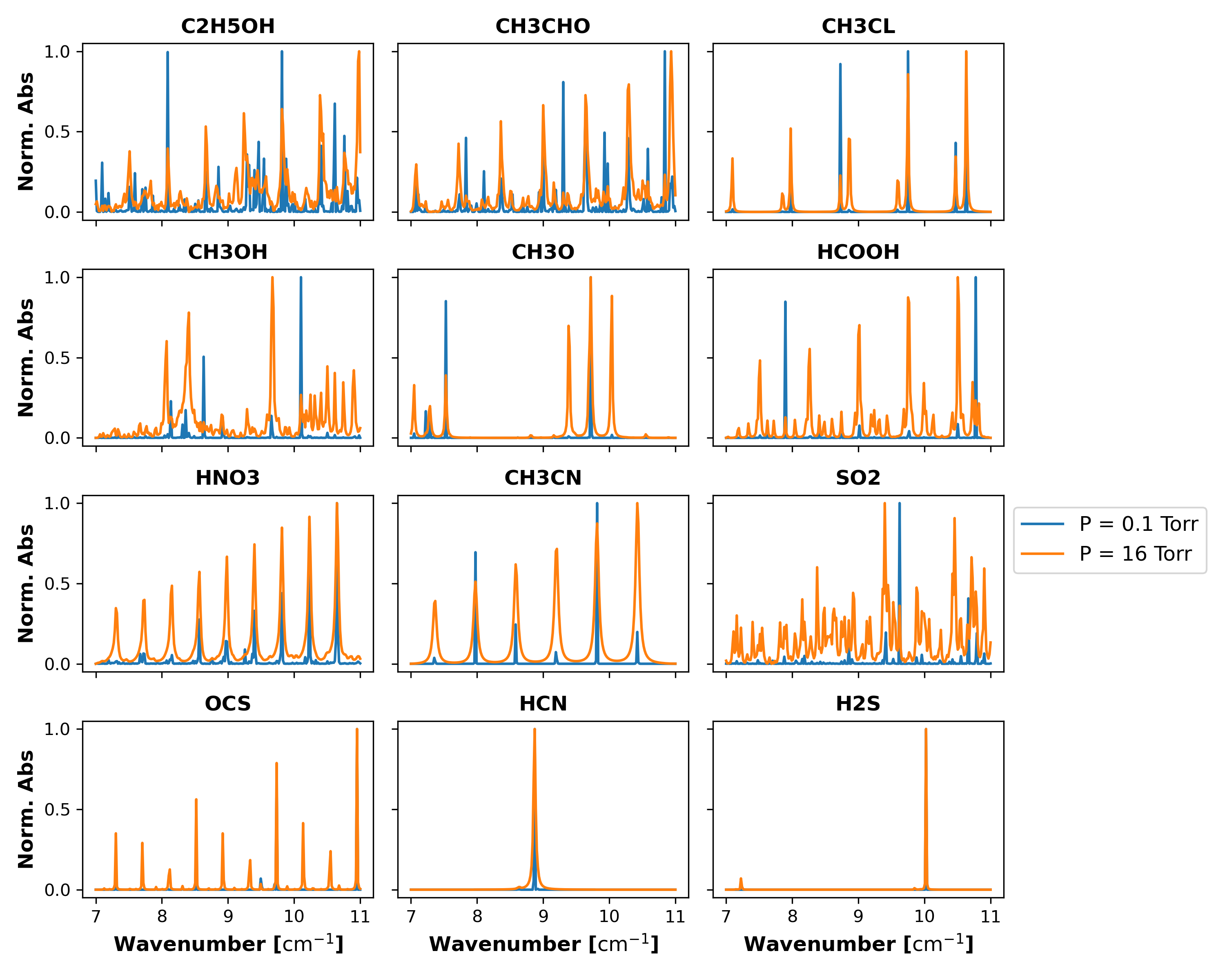}
    \caption{Normalized spectra of twelve VOCs at P = 0.5 Torr and P = 16 Torr}
    \label{fig:all_vocs}
\end{figure}
\item \textbf{Augmentations}: In one-shot learning models, where training exclusively occurs under a singular pressure condition, the incorporation of augmentations becomes paramount. This necessity arises from the intrinsic data hunger of machine learning models and the need to address pressure-induced spectral variations, as depicted in Fig.\ref{fig:all_vocs}. Our augmentations extend beyond addressing noise and baseline shifts. Notably, they involve the convolution of Voigt profiles with varying full-width at half-maximum (FWHM). Voigt profiles, fundamental in spectroscopic modelling\cite{whiting1968empirical}, result from the convolution of Lorentzian and Gaussian profiles. Gaussian width reflects Doppler broadening due to temperature-induced particle motion, while Lorentzian width indicates broadening resulting from collisions or non-thermal effects\cite{goldenstein2017infrared,hanson2016spectroscopy,whiting1968empirical}. The FWHM values, influenced by both Lorentzian (\(\theta\)) and Gaussian (\(\gamma\)) widths, serve as indicators of pressure and temperature effects. Higher FWHM values signify elevated pressures and temperatures, leading to increased spectral line broadening. Conversely, lower values denote tighter lines, indicative of conditions at lower pressures and temperatures\cite{goldenstein2017infrared,hanson2016spectroscopy}. The dynamic interplay of both \(\theta\) and \(\gamma\), spanning from 0.001 to 0.05, facilitates the representation of spectral variations induced by changes in pressure and temperature. This comprehensive approach ensures that the augmentations effectively encapsulate the diverse spectral variations arising from varying conditions. Fig.\ref{fig:augment_all} illustrates augmentations for ethnol spectra at P = 1 Torr, in three cases, where \(\theta\) and \(\gamma\) are small (0.001), moderate (0.01) and high (0.05). For higher pressures, the Voigt augmentation parameters can be adjusted to account for increased broadening, thereby capturing the spectral changes induced by elevated pressures, as detailed in the Supporting Information.
\end{itemize}
\begin{figure}[!h]
\renewcommand{\thesubfigure}{\arabic{subfigure}}
    \centering
    \begin{subfigure}{1.0\textwidth}
        \centering
        \includegraphics[width=0.85\linewidth]{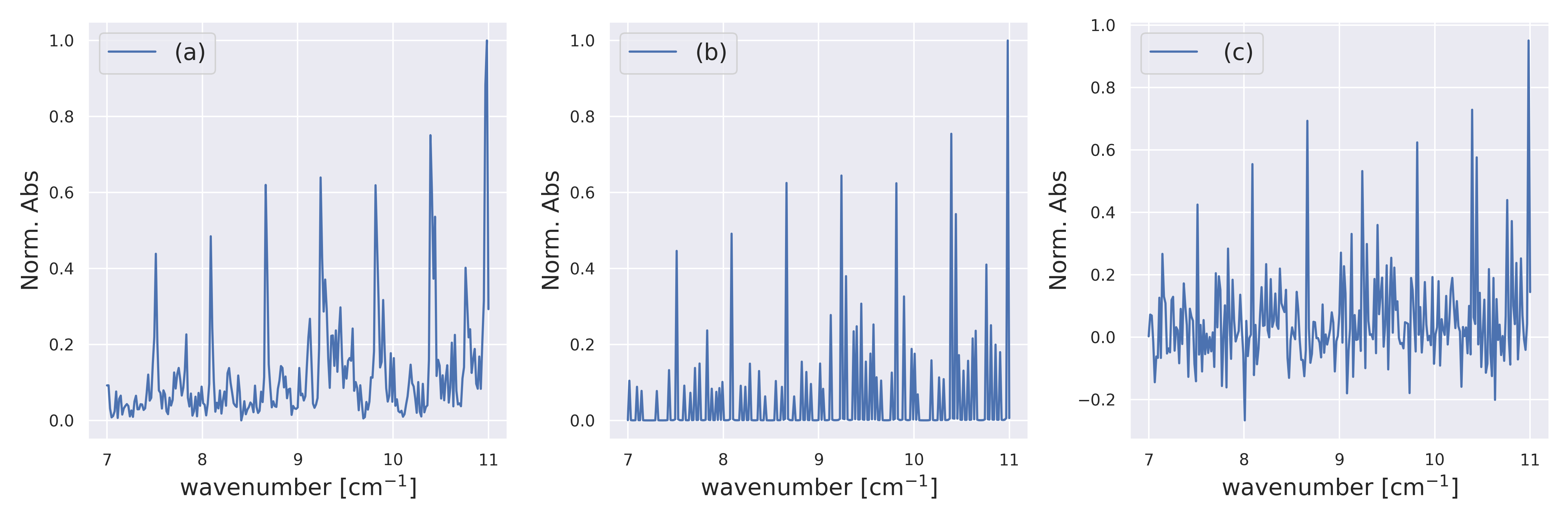}
        \caption{Low FWHM Voigt augmentations for ethanol spectrum at P = 1 Torr}
        \label{fig:augment1}
    \end{subfigure}


    \begin{subfigure}{1\textwidth}
        \centering
        \includegraphics[width=0.85\linewidth]{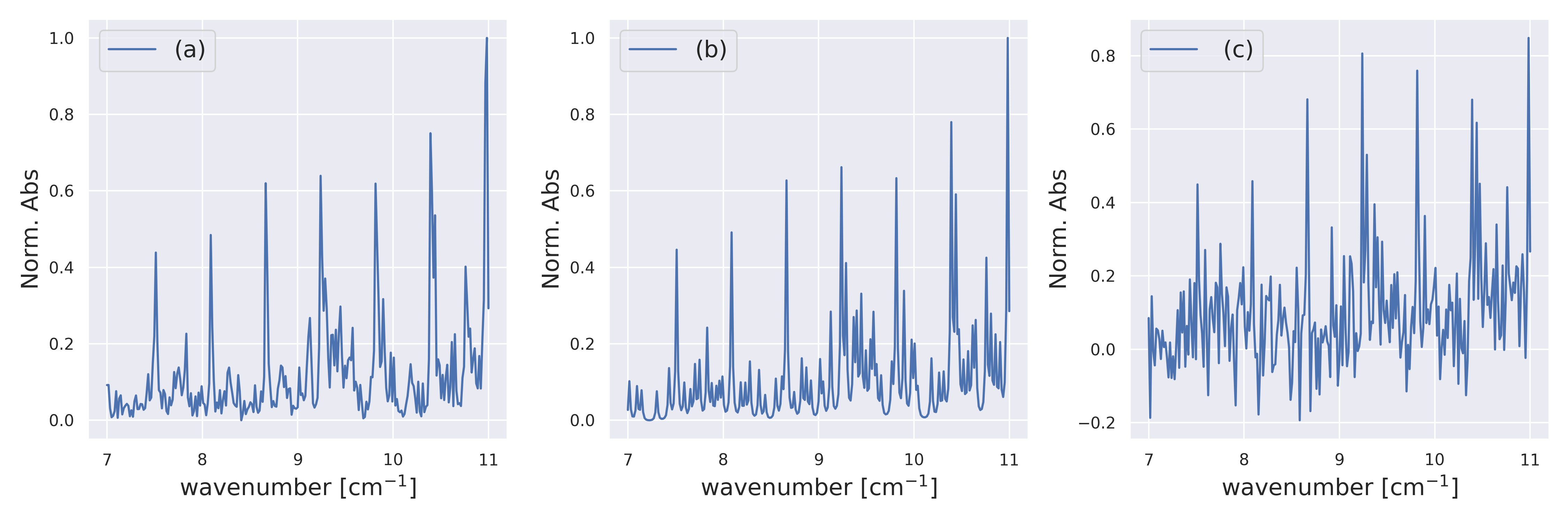}
        \caption{Moderate FWHM Voigt augmentations for ethanol spectrum at P = 1 Torr}
        \label{fig:augment2}
    \end{subfigure}


    \begin{subfigure}{1\textwidth}
        \centering
        \includegraphics[width=0.85\linewidth]{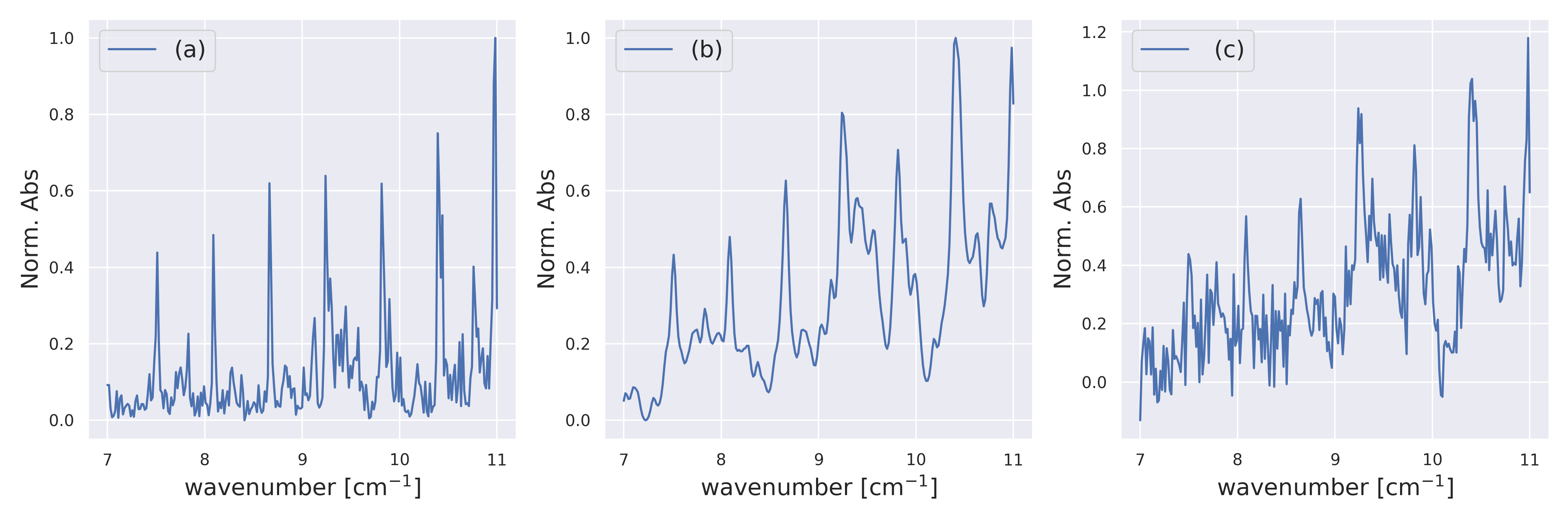}
        \caption{High FWHM Voigt augmentations for ethanol spectrum at P = 1 Torr}
        \label{fig:augment3}
    \end{subfigure}
    \caption{Effect of augmentations across varied FWHMs. Common to all figures: (a) illustrates the normalized spectrum at P = 1 Torr, (b) showcases the augmented spectrum with Voigt convolutions, and (c) presents the same spectrum augmented with Voigt convolutions, noise, and baseline shifts.}

    \label{fig:augment_all}
\end{figure}

In the upcoming sections, we will explore each model, offering insights into their architectures, training methodologies, and test datasets. For enhanced clarity and convenient reference, unique names have been assigned to each model. Table \ref{tab:models} succinctly summarizes the developed models along with their respective characteristics.

\begin{table}
\centering
  \caption{VOC models and their characteristics}
  \label{tab:models}
  \begin{tabular}{llll}
    \hline
    Models & Training data & Augmentations & Certified \\
    \hline
     VOC-net & all pressures  &  NO & NO  \\
    VOC-lite & one pressure &  NO   & NO   \\
    VOC-plus & one pressure   &  YES&  NO  \\ 
    VOC-certifire & one pressure  & YES & YES   \\
    \hline
  \end{tabular}
\end{table}
\subsubsection{VOC-net and VOC-lite}
An identical 1-D CNN VOC classifier developed in \cite{chowdhury2022voc} was replicated for both VOC-net and VOC-lite. Only a brief description of the model architecture is given below. Further details on hyperparameter tuning and model optimizations can be found here \cite{chowdhury2022voc}. VOC-net relies on a 1-D CNN, comprising two convolutional layers, each with three filters of kernel size three. A subsampling (pooling) layer is strategically positioned between the convolutional layers. This is followed by a flattened dense layer, a hidden layer housing 48 neurons, and an output layer featuring 12 neurons, each corresponding to a distinct VOC species.

\begin{figure}[!h]
    \centering
    \includegraphics[width=1\linewidth]{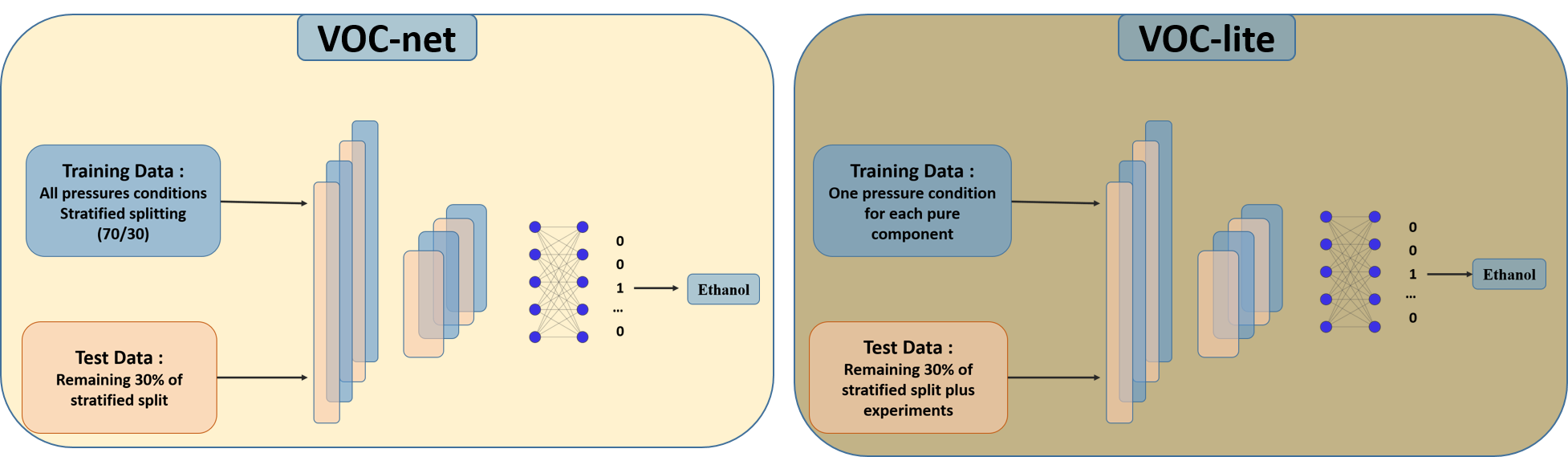}
    \caption{Schematic of VOC-net and VOC-lite training and testing procedures}
    \label{fig:VOC-net}
\end{figure}

VOC-net serves as the baseline model, and it undergoes training across all pressure conditions (0.1–16.5 Torr) using a (70/30) stratified splitting approach to ensure a balanced distribution between training and test data. The training dataset consists of 1377 simulated spectra of 12 VOCs at different pressure conditions, while the test dataset comprises 591 simulated spectra of the same VOCs across various pressures. Additionally, 36 experimental spectra of six VOCs are included in the test dataset; details of the experimental setup can be found in \cite{chowdhury2022voc}. Conversely, VOC-lite shares the same architecture as VOC-net but is trained under a singular pressure condition, specifically at P=1 Torr. Accordingly, the training data for VOC-lite consists of 12 observations, aiming for one-shot learning without augmentations. Schematic of the training and testing procedures for both VOC-net and VOC-lite is shown in Fig.\ref{fig:VOC-net}. To ensure a fair comparison, the test data are kept consistent across all models.

\subsubsection{VOC-plus and VOC-certifire}
The VOC-plus 1-D CNN architecture shares a similar structure with VOC-net, featuring a series of convolutional layers with ReLU activation and max-pooling to extract hierarchical features from one-dimensional molecular data.
VOC-plus and VOC-net differ primarily in their training datasets.
The training dataset for VOC-plus comprises of each VOC at a single pressure, followed by augmentations. Augmentations involve various transformations on each VOC spectrum, including convolutions with Voigt profiles featuring varying Gaussian and Lorentzian widths (ranging from 0.0001 to 0.05), baseline shifts, and different noise levels. Each spectrum is augmented 1000 times, resulting in a total of 12,000 augmented components for model training. The test data are the same as used for VOC-net and VOC-lite. Details of the training costs for all the VOC models are provided in the Supporting Information.

\begin{figure}[!h]
    \centering
    \includegraphics[width=0.9\linewidth]{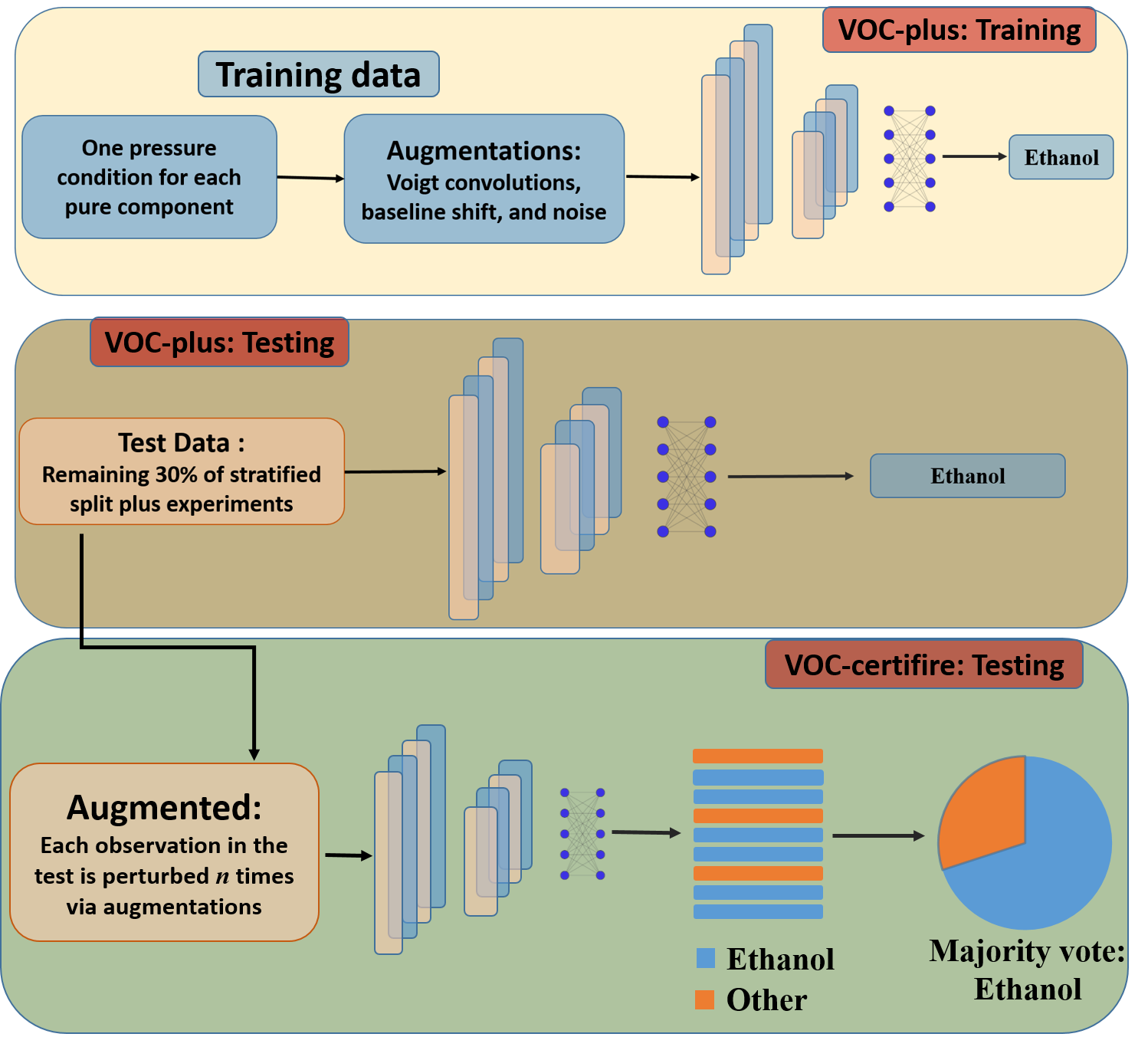}
    \caption{Schematics of training and testing procedure of VOC-plus and VOC-certifire}
    \label{fig:VOC-certif}
\end{figure}

As illustrated in Fig.\ref{fig:VOC-certif}, VOC-certifire employs the pre-trained VOC-plus model as its base classifier during training but takes a different path during testing. Notably, VOC-certifire incorporates randomized smoothing during its testing. This process involves perturbing each observation in the test data multiple times using the aforementioned augmentations before feeding it to the pre-trained VOC-plus model. The final prediction of VOC-certifire is determined by the majority vote of the pretained classifier predictions for these perturbed instances. In simpler terms, if 100 perturbations are applied through augmentations to a test spectrum, and the model predicts 80 instances as ethanol and 20 instances as another substance, the majority vote rule designates the underlying test spectrum to be ethanol.

\subsection{Randomized smoothing}

\label{sec:randomized smoothing}
Let a classifier $(f)$ map inputs from $(R^d)$ to classes in $(Y)$. The randomized smoothing procedure transforms the base classifier $(f)$ into a smoothed classifier $(g)$. Specifically, for a given input $(x)$, $(g)$ identifies the class most likely to be predicted by $(f)$ under isotropic Gaussian noise perturbations of $(x)$. Mathematically, this is expressed as:
\begin{equation}
\centering
g(x) = \arg\max_{c\in Y} \mathbb{P}[f(x + \delta) = c], \text{ where } \delta \sim \mathcal{N}(0, \sigma^2I)
\label{eq:eq1}
\end{equation}
Here, the noise/perturbation level \(\sigma\) governs the trade-off between robustness and accuracy. An increase in \(\sigma\) enhances the robustness of the smoothed classifier but reduces its standard accuracy.

\begin{figure}[h]
    \centering
    \begin{minipage}{0.45\textwidth}
        \centering
        \textbf{(a)}\\[0.5ex]
        \includegraphics[scale=0.9]{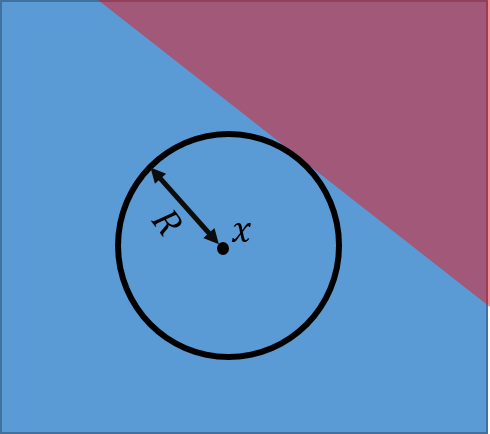}
        \label{fig:a_randomized}
    \end{minipage}\hfill
    \begin{minipage}{0.45\textwidth}
        \centering
        \textbf{(b)}\\[0.5ex]
        \includegraphics[scale=0.9]{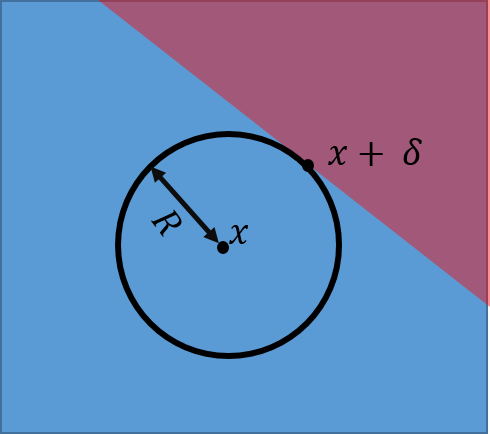}
        \label{fig:b_randomized}
    \end{minipage}
    \caption{Visual explanation of robustness radius \(R\) in smoothed classifiers.}
    \label{fig:randomized}
\end{figure}

A detailed robustness guarantee for the smoothed classifier \(g\) was initially introduced by \cite{cohen2019certified}. They proposed an efficient algorithm rooted in Monte Carlo sampling to facilitate both prediction and certification. The robustness guarantee relies on the Neyman-Pearson lemma \cite{neyman1933ix}. The procedure involves classifying \(\mathcal{N}(x, \sigma^2I)\) with the base classifier \(f\), where class \(c_A\) is returned with probability \(p_A = \mathbb{P}(f(x + \delta) = c_A)\), and the runner-up class \(c_B\) is returned with probability \(p_B = \max_{c \neq c_A} \mathbb{P}(f(x + \delta) = c)\). The smoothed classifier \(g\) is deemed robust around \(x\) within a radius \(R\), defined as:
\begin{equation}
R = \sigma^2 \left(\Phi^{-1}(p_A) - \Phi^{-1}(p_B)\right)
\label{eq:eq2}
\end{equation}

where \(\Phi^{-1}\) is the inverse of the standard Gaussian cumulative distribution function.
In Fig. \ref{fig:randomized}, a binary classifier certification is depicted. The red/blue half-spaces represent the decision regions in the smoothed classifier \(g\) obtained by majority voting of the base classifier \(f\). The smoothed classifier \(g\) ensures consistent predictions within an \(l_2\) circle with radius \(R\), as indicated by the black circle, showing the certified robustness radius \(R\) of Eq. \ref{eq:eq2}, as guaranteed by Eq. \ref{eq:eq1}. On the right, it illustrates that for any \(r > R\), there exists a perturbation \(\delta\) with \(\|\delta\|_2 = r\) such that \(g(x + \delta) \neq g(x)\). This implies that robustness is only guaranteed within the circle of radius \(R\).

\section{Results and discussion}
To evaluate the performance of our models, we employed three key metrics: accuracy, F1-score, and precision. Detailed equations for these metrics are provided in the Supporting Information accompanying this paper. A concise overview of our findings is provided in Table \ref{tab:table2} which presents a comparison of the aforementioned metrics for various models evaluated on both simulated and experimental datasets.
\begin{table}[!ht]
    \caption{Comparison of different VOC classification models }
    \label{tab:table2}
    \centering
    \begin{tabularx}{\textwidth}{||l|X|X|X|X|X|X||}
    \hline
        Models & \multicolumn{2}{|c|}{Accuracy} & \multicolumn{2}{c|}{F1-score} & \multicolumn{2}{c||}{Precision} \\ 
        \hline
        \hline
         & Simul & Exp & Simul & Exp & Simul & Exp \\ \hline
        VOC-net &  97.4\%& 88.8\% & 96.6\% & 89.2\% & 97.3\% & 91.2\% \\ \hline
        VOC-lite & 47.5\% & 50.2\% & 35.6\% & 40.5\%  & 31.7\% &  38.8\%\\ \hline
        VOC-plus & 81.7\% & 65\% & 83\% & 61\% & 93.7\% & 63\%\\ \hline
        VOC-certifire & 99\% & 94\%& 99\% & 96\% & 100\% & 100\%\\ \hline
    \end{tabularx}
\end{table}

\subsection{VOC-net and VOC-lite results}
\begin{figure}[!h]
    \centering
    \includegraphics[width=1\linewidth]{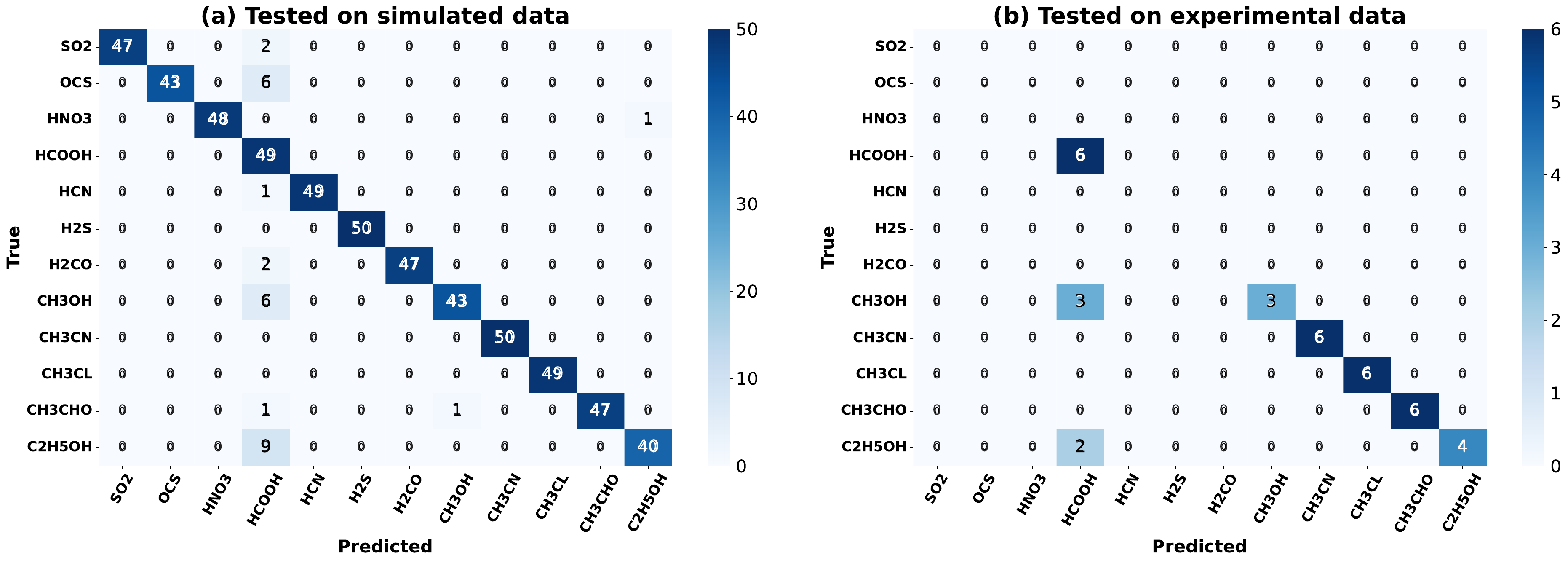}
    \caption{VOC-net confusion matrix. (a) Simulated data; (b) Experimental data}
    \label{fig:baseline}
\end{figure}

\begin{figure}[!h]
    \centering
    \includegraphics[width=1\linewidth]{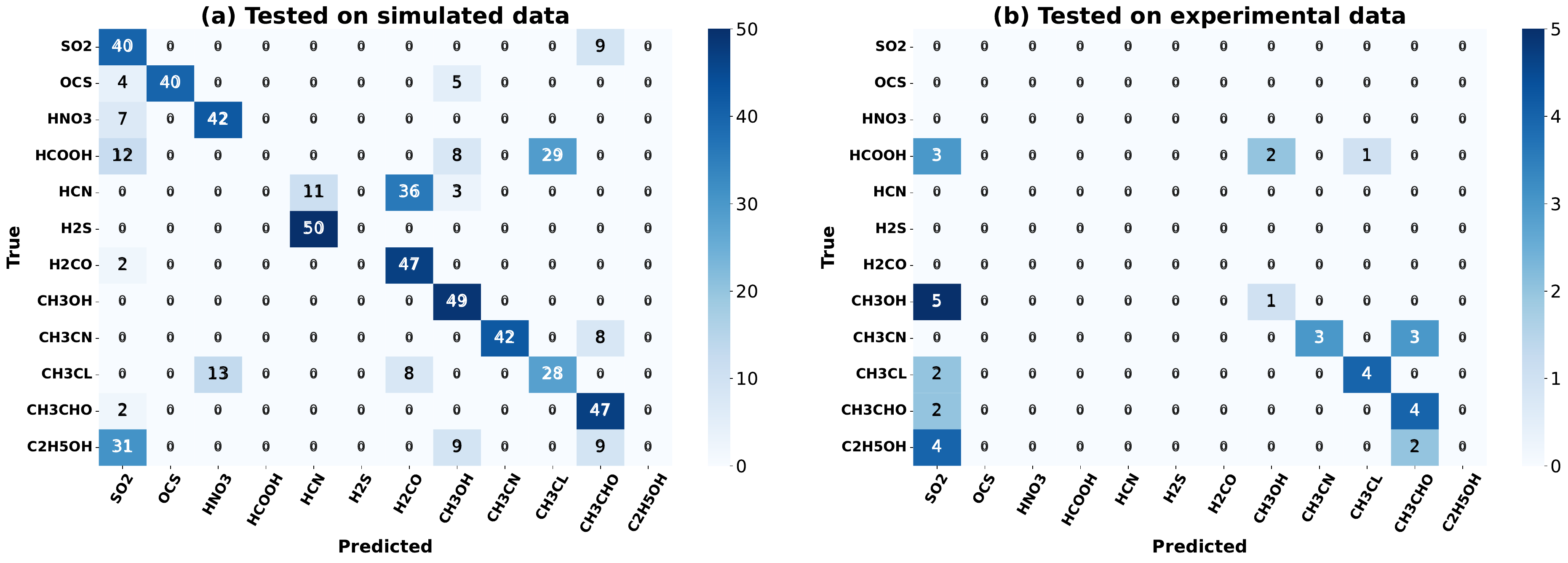}
    \caption{VOC-lite confusion matrix. (a) Simulated data; (b) Experimental data}
    \label{fig:lite}
\end{figure}

Figs.\ref{fig:baseline} and \ref{fig:lite} showcase the outcomes of the baseline model, VOC-net, and its counterpart, VOC-lite. While both models share the architecture of VOC-net, VOC-lite is trained under a single pressure condition (P = 1 Torr). The accuracy experiences a significant drop for both experimental and simulated data for VOC-lite. This decline highlights the limited ability of the VOC-lite model to generalize to unseen pressure conditions.

The critical factor contributing to VOC-net's successful generalization and high accuracy is its comprehensive training across all pressure conditions (0 to 16.5 Torr). In the experimental data, VOC-net exhibits only three misclassifications among 36 observations, whereas VOC-lite exhibits 30 misclassifications. This stark contrast emphasizes VOC-lite's inability to generalize across different pressure and temperature conditions, indicating a failure to capture the spectral variations induced by changes in operating conditions.

\subsection{VOC-plus and VOC-certifire results}
\begin{figure}[!h]
    \centering
    \includegraphics[width=1\linewidth]{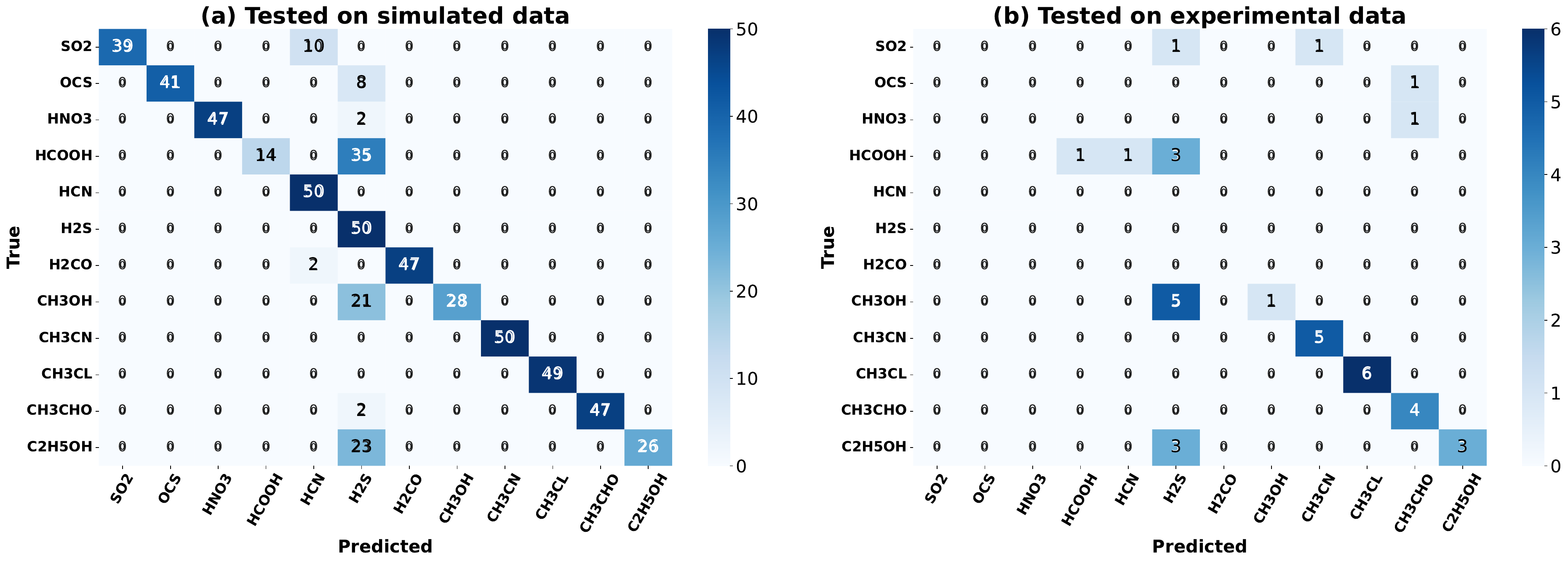}
    \caption{VOC-plus confusion matrix. (a) Simulated data; (b) Experimental data}
    \label{fig:plus}
\end{figure}

\begin{figure}[!h]
    \centering
    \includegraphics[width=1\linewidth]{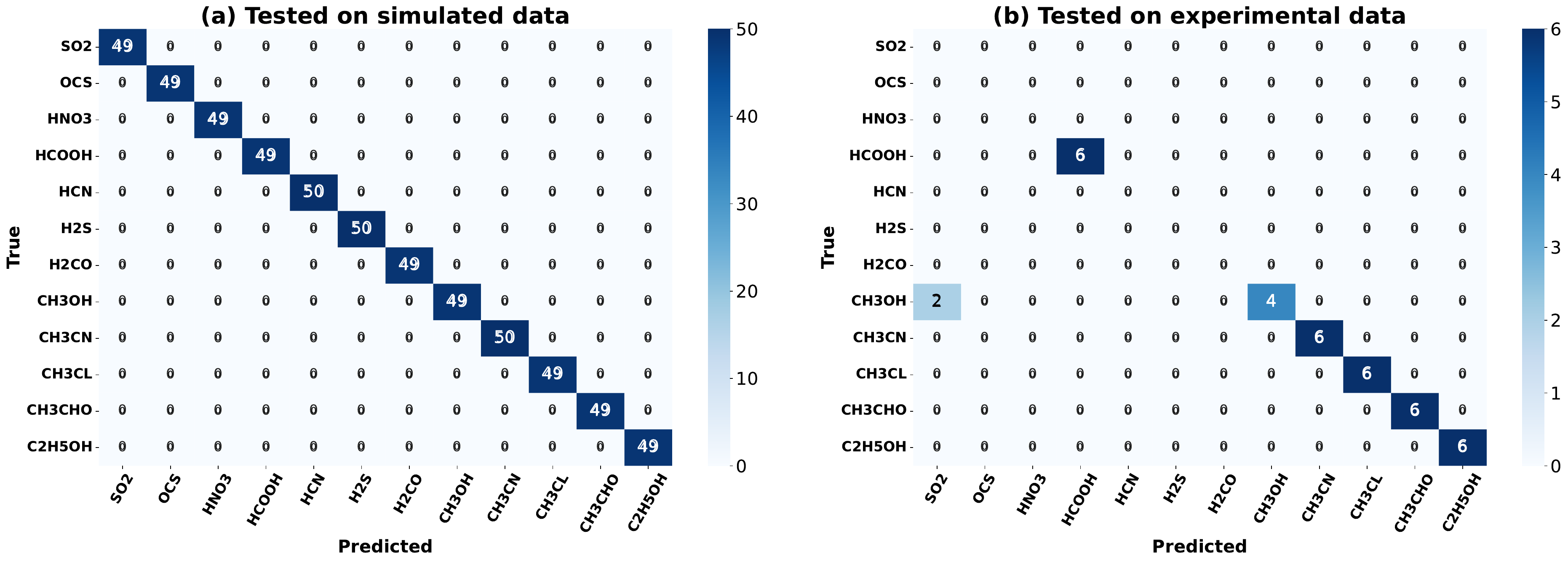}
    \caption{VOC-certifire confusion matrix. (a) Simulated data; (b) Experimental data}
    \label{fig:certified}
\end{figure}

Figs. \ref{fig:plus} and \ref{fig:certified} depict confusion matrices for VOC-plus and VOC-certifire models, evaluated on simulated and experimental data. VOC-plus is trained under a single pressure condition (P = 1 Torr) with augmentations (Voigt convolutions, noise, and baseline shifts). Noticeable improvements are observed, with VOC-plus achieving a significant increase in accuracy compared to VOC-lite, rising from 47.5\% to 81.7\% in simulated data. Additionally, VOC-plus exhibits 18 misclassifications compared to VOC-lite's 30 in experimental data. This underscores the efficacy of augmentations. VOC-certifire showcases remarkable accuracy, surpassing VOC-plus and achieving accuracy level comparable to the baseline model, VOC-net. In experimental data, VOC-certifire incurs one misclassification out of 36 observations. Despite utilizing VOC-plus as its pretrained classifier, VOC-certifire distinguishes itself by incorporating randomized smoothing during testing. This involves subjecting each observation in the test dataset to multiple perturbations via augmentations, including Voigt convolutions, noise, and baseline shifts. The final prediction for VOC-certifire is determined by the majority vote of the pretrained classifier predictions on the perturbed instances. A notable strength of VOC-certifire lies in its certifiability, assuring robustness under perturbations and ensuring consistent predictions even in the face of uncertainties. 
\subsection{Model certification}
\begin{figure}[H]
    \centering
    \begin{minipage}{0.48\textwidth}
        \centering
        \textbf{(a)}\\[0.5ex]
        \includegraphics[width=\linewidth]{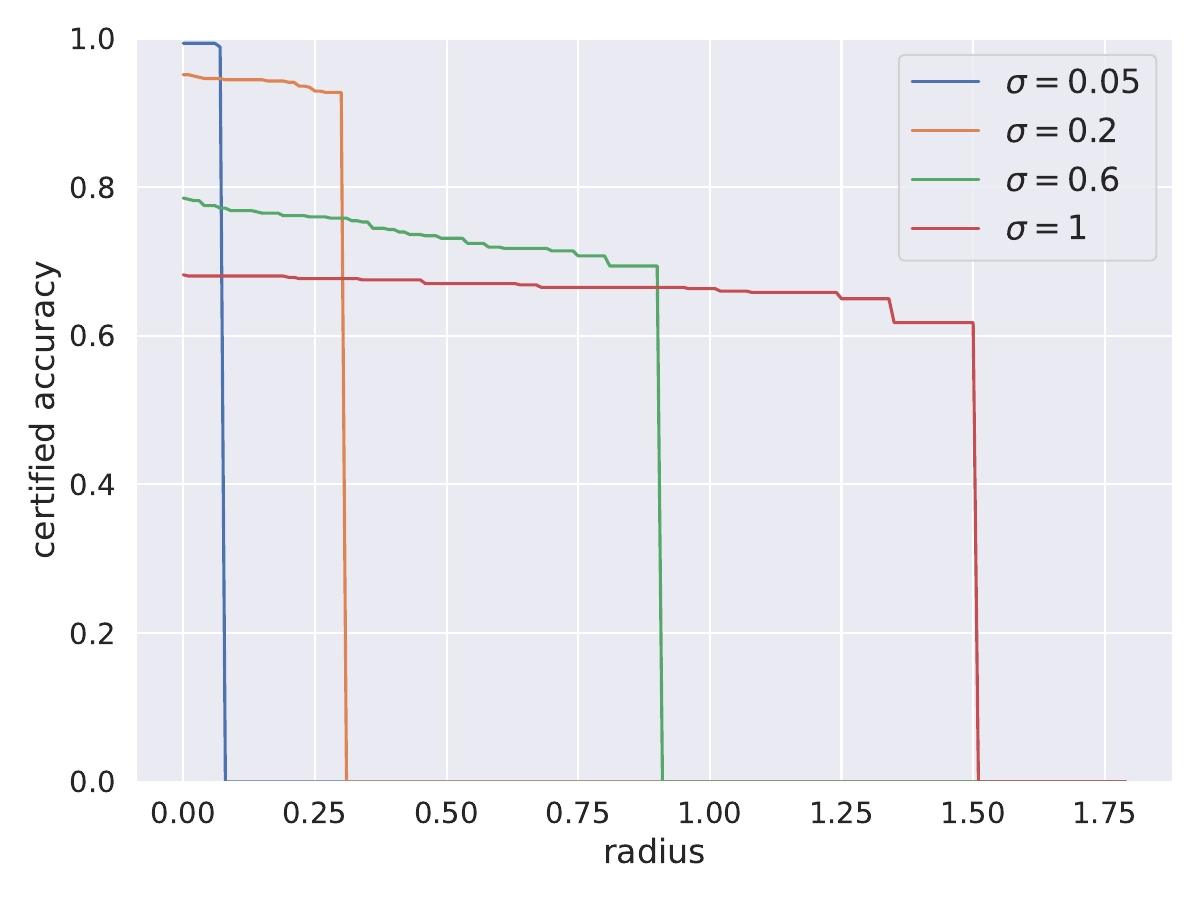}
        \label{fig:certified_sigma}
    \end{minipage}\hfill
    \begin{minipage}{0.48\textwidth}
        \centering
        \textbf{(b)}\\[0.5ex]
        \includegraphics[width=\linewidth]{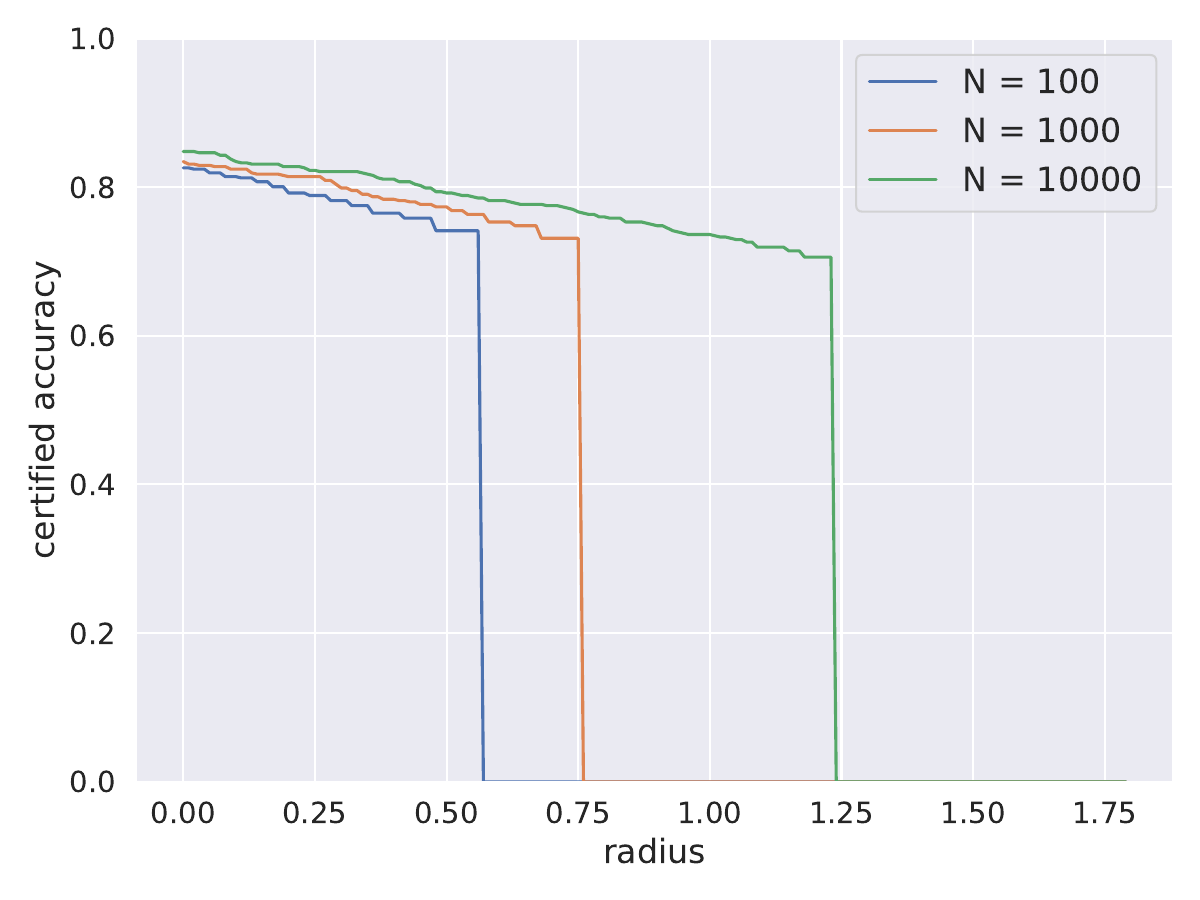}
        \label{fig:N3}
    \end{minipage}
    
    \vspace{2ex}
    
    \begin{minipage}{0.48\textwidth}
        \centering
        \textbf{(c)}\\[0.5ex]
        \includegraphics[width=\linewidth]{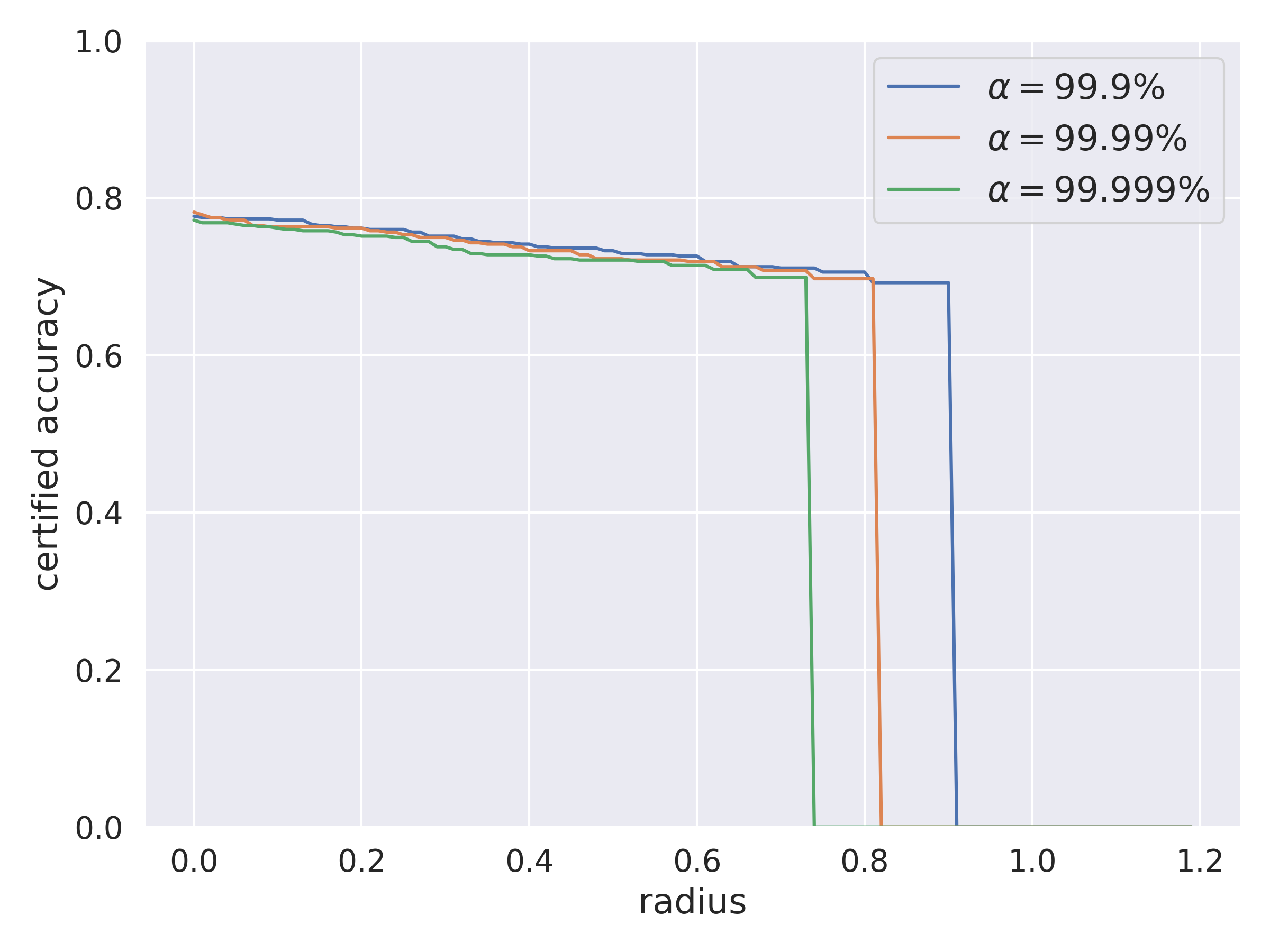}
        \label{fig:alphas2}
    \end{minipage}\hfill
    \begin{minipage}{0.48\textwidth}
        \centering
        \textbf{(d)}\\[0.5ex]
        \includegraphics[width=\linewidth]{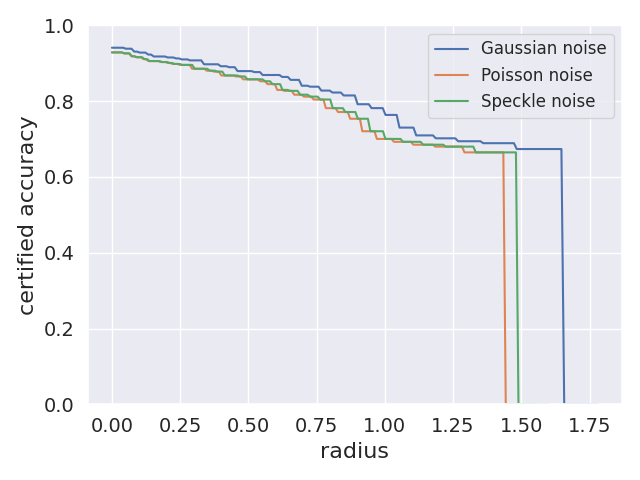}
        \label{fig:noisetype}
    \end{minipage}

    \caption{Certified accuracy analysis: (a) varying \(\sigma\), (b) varying \(N\), (c) varying \(\alpha\), and (d) different noise types.}
    \label{fig:m}
\end{figure}

The smoothed classifier \(g\) is guaranteed to produce consistent predictions within an \(\ell_2\) circle of radius \(R\), centered at observation (\(x\)), as illustrated by Eq.\ref{eq:eq2}. We assessed certified accuracy across different \(\ell_2\) radii by varying parameters (\(\alpha\) , \(\sigma\) and N). Fig.\ref{fig:m}(a) demonstrates the certified accuracy achieved through smoothing with varying \(\sigma\) for a fixed \(\alpha\) (99.9\%) and fixed N (1000), showcasing how \(\sigma\) plays a role in a robustness/accuracy tradeoff. Lower perturbation level \(\sigma\) values enable the certification of small radii with high accuracy, while larger radii exhibit lower certification accuracy. Conversely, higher \(\sigma\) values facilitate the certification of larger radii but result in reduced accuracy for smaller radii, aligning with the findings in \cite{cohen2019certified} regarding the tradeoff between adversarial attacks and standard accuracy. Fig.\ref{fig:m}(b) provides insights into how certified accuracy would change with varying randomized sample sizes \(N\) while \(\sigma\) and \(\alpha\) are fixed. Higher samples size results in higher certification radius. Fig. \ref{fig:m}(c) illustrates the impact of varying the confidence level parameter \(\alpha\) on certified accuracy while the sample size N and \(\sigma\) are fixed, highlighting its relatively low sensitivity to changes in \(\alpha\). Finally, Fig. \ref{fig:m}(d) assesses certified radius across different noise types, showing that the model, despite being trained only on Gaussian noise, maintains similar accuracy for unseen noise types.

    \chapter{Summary and Future Work}\label{chapter8}

\section{Summary of Results}
\subsection{Multi-Species Detection in Shock Tube Experiments Using a Single Laser and Deep Neural Networks}
In this chapter we developed a laser-based diagnostic capable of simultaneously measuring time-histories of methane, ethane, ethylene, propylene, and propyne. The diagnostic was used to study the evolution of major hydrocarbons during the pyrolysis of ethane and propane. We employed a machine learning approach that combined deep denoising autoencoders (DDAEs) with multiple linear regression (MLR). Our approach showed good agreement with model predictions (using the AramcoMech 3.0) and previous experimental work. The use of DDAEs allowed us to improve the signal-to-noise ratio (SNR) of the noisy signals and identify hidden features that would have otherwise been obscured. Our methodology represents a novel technique for reducing the cost and complexity of optical diagnostic setups for combustion applications by leveraging the power of data processing and machine learning. In future work, we aim to extend our study to a broader range of chemical systems, encompassing oxidation reactions as well. To enhance the temporal resolution, we plan to incorporate a bias-T configuration, as demonstrated by Nair, AP et al. \cite{nair2020mhz}. This approach will enable us to scan the laser at higher frequencies (in the MHz range) to achieve significantly improved temporal resolution while maintaining a similar scan depth. Additionally, we intend to develop more advanced and versatile denoising algorithms to further increase the accuracy of our measurements.
\subsection{Spectral Interference Mitigation: Technique to Tackle Noise and Unknown Interference }
In this chapter we present HT-SIMNet, an unsupervised deep learning framework for spectral interference mitigation in high-temperature combustion diagnostics. By leveraging the Noise2Noise paradigm and structured spectral augmentations—flipping, mirroring, and dilation—the model successfully isolates target species signals from overlapping 

interference without requiring clean reference spectra or prior knowledge of interferents. Evaluated on non-reactive mixtures of $C_1–C_3$ hydrocarbons and reactive n-butane pyrolysis experiments, HT-SIMNet demonstrated exceptional performance, achieving R² values of 0.95–0.99 for concentration predictions, a significant improvement over conventional methods ($R^2$ $<$ 0.35, with negative values indicating failure). The framework’s ability to suppress interference dynamically was further validated in reactive environments, where it accurately resolved methane and ethylene concentrations during pyrolysis, even as reactant (n-butane) interference diminished over time.  

The elimination of dependency on preexisting spectral databases and the reduction in experimental complexity position HT-SIMNet as a scalable, cost-effective solution for real-time diagnostics in shock tubes, rapid compression machines, and other high-temperature systems. Future work will extend this methodology to more complex combustion environments involving additional species, higher pressures, and broader temperature ranges. By advancing unsupervised machine learning applications in combustion science, HT-SIMNet addresses critical challenges in time-resolved multi-species detection, paving the way for robust, adaptive diagnostics in next-generation energy and propulsion systems.
\subsection{Unsupervised Spectral Source Separation: A reference-free technique for multi-species detection}
In this chapter we introduce UnblindMix, an unsupervised, reference-free diagnostic framework for multi-species detection in high-temperature combustion environments. UnblindMix infers species mole fractions and reconstructs high-temperature reference spectra directly from composite mixture absorbance data, eliminating the dependence on pre-tabulated reference spectra. The model’s performance was demonstrated on both non-reactive mixtures of \(C_1–C_3\) hydrocarbons and pyrolysis experiments of \(n\)-butane and \textit{iso}-butane at 923~K. The results showed good agreement between the predicted and measured absorbance spectra, as well as between the inferred and expected mole fraction profiles.

The integration of UnblindMix with infrared laser absorption spectroscopy (IR-LAS) provides a scalable and experimentally efficient solution for quantitative diagnostics under harsh combustion conditions. The framework addresses major challenges in the field, including the difficulty of obtaining accurate cross-sections at high temperatures and the presence of overlapping spectral features in complex mixtures. The physically constrained variant of UnblindMix, incorporating incorporating chemically-informed regularization such elemental carbon balance and monotonic species formation, further enhances consistency with chemical expectations. 
Future efforts will aim to extend the methodology to reacting systems with time-varying thermodynamic conditions, such as shock tubes and rapid compression machines, where temperature and pressure evolve dynamically. Additional developments will focus on integrating probabilistic inference for improved uncertainty quantification, and on adapting the model to recover individual spectral line shapes in high-resolution applications. The incorporation of spatial priors and tomographic extensions will also be explored to enable application to non-uniform flows.

\subsection{Spectral Feature Engineering : Feature Transformation for an Enhanced Multi-species Detection}
In this chapter, we demonstrate the effectiveness of a feature-engineered model for detecting low mole fractions of C\(_1\)-C\(_3\) hydrocarbons using laser-based vibrational spectroscopy. Utilizing a distributed feedback inter-band cascade laser (DFB-ICL) tuned over 2984.5 - 2989.5 cm\(^{-1}\), we measured methane, ethane, propane, propyne, and ethylene. Traditional models often struggle with overlapping and weak absorbance spectra. However, our feature-engineered model, which applies convolutions to the first derivatives and composite spectra, significantly improves detection accuracy. This work not only addresses critical limitations in traditional models but also paves the way for more reliable, sensitive, and accurate gas sensing technologies.
\subsection{Models Certification and Stability via Randomized Smoothing}
In this chapter, we explored VOC classification models, starting with VOC-net and VOC-lite, highlighting the significance of comprehensive training data for accurate generalization. To overcome the challenge of extensive data requirements, we introduced one-shot learning models coupled with straightforward yet impactful augmentations. These augmentations were designed to effectively capture pressure-induced spectral variations for both low and high pressures, faithfully representing real-world applications.  We introduced VOC-plus, a one-shot learning model utilizing these augmentations during training, which yielded a significant improvement in accuracy compared to VOC-lite. Furthermore, we introduced VOC-certifire, a certifiable model that leverages VOC-plus as its base classifier through randomized smoothing. VOC-certifire demonstrated high accuracy in comparison to the baseline model, highlighting its robust and certified nature, ensuring consistent predictions even under unforeseen adversarial attacks.

The randomized smoothing procedure played a crucial role in certifying the robustness of VOC-certifire model. We conducted a thorough evaluation of parameters such as perturbation level \(\sigma\) , sample size (\(N\)), and confidence level (\(\alpha\)), offering insights into the tradeoff between robustness and accuracy. Summarized in Table \ref{tab:table2}, our results underscore the significance of augmentations in enhancing one shot-learning model performance, and model's robustness against unforeseen perturbations. This study not only offers a comparative analysis of VOC classification models but also provides valuable insights into the balance between the robustness and accuracy of machine learning models. These findings contribute to the advancement of developing trustworthy machine learning models, particularly crucial in gas sensing applications where reliability and accuracy are crucial for well informed decision-making and effective risk mitigation strategies. Future work will involve exploring mixture classifications using one-shot learning and investigating the impact of augmentations on handling unknown interference and pressure-induced spectral changes within mixtures.

\section{Recommendations for Future Work}

The methodologies developed in this dissertation establish a robust foundation for advancing laser-based multi-species gas detection enhanced by machine learning (ML). These techniques address key challenges in combustion diagnostics, interference mitigation, unsupervised source separation, and model certification. Building on these contributions, the following directions are proposed to expand the impact of this work across a broader range of scientific and societal applications, including atmospheric sensing, environmental safety, and health diagnostics.

\subsection*{1. Application to Complex Chemical Systems and Biomarker Detection}
Future studies should extend the developed frameworks to chemically diverse environments, including biofuel combustion, oxidation chemistry, and polluted urban air. In parallel, adapting these tools for breath analysis—targeting VOC biomarkers such as acetone, isoprene, and ammonia—can support early diagnosis of metabolic, respiratory, and infectious diseases. This dual-purpose generalization requires robust model architectures capable of handling complex and overlapping spectra across a wide range of concentrations and molecular families.

\subsection*{2. Real-Time Diagnostics in Field and Clinical Environments}
Real-time capability is essential for both reactive systems and clinical decision-making. Integrating MHz-range laser scanning (via bias-T circuits) and deploying models on embedded hardware (e.g., FPGAs, edge devices) can enable real-time diagnostics in shock tubes, ambient air monitoring, and portable breath analyzers. This will also facilitate the development of compact, low-power sensor platforms suitable for continuous operation in the field or at the point-of-care.

\subsection*{3. Adaptation to Environmental Variability: Pressure, Temperature, and Humidity}
Spectroscopic signals are often affected by external conditions such as pressure, temperature, and humidity—whether in a combustion chamber, open-air environment, or the human respiratory tract. Future models should incorporate physics-informed augmentations and domain-adaptive training strategies to ensure robust performance across such variable conditions, enhancing model transferability to outdoor sensing and breath-based clinical diagnostics.

\subsection*{4. Hybrid Physics–ML Architectures for Atmospheric and Clinical Sensing}
A promising avenue involves embedding physical constraints (e.g., Beer–Lambert law, mass balance, or known environmental baselines) directly into ML architectures. Such hybrid models can improve interpretability and enhance trust in scenarios requiring regulatory compliance or medical diagnostics. These hybrid approaches are also expected to improve generalization in trace-level atmospheric sensing of species like CH$_4$, CO$_2$, O$_3$, or benzene.

\subsection*{5. Autonomous Feature Learning and Transferability Across Domains}
While handcrafted spectral features improved detection in this work, future approaches should explore self-attention mechanisms and transformer-based architectures for autonomous feature learning. Transfer learning between domains (e.g., from synthetic combustion spectra to real atmospheric or biomedical spectra) could reduce data requirements and broaden deployment scenarios with minimal retraining.

\subsection*{6. Continual Learning for Dynamic Mixtures}
The UnblindMix framework introduced here can be expanded to support online learning and adaptive mixture decomposition. This is particularly valuable for tracking evolving gas compositions in time-varying environments such as combustion exhausts, air pollution events, or metabolic transitions in patient breath. Future efforts should implement continual learning strategies that allow for real-time model updates and drift compensation.

\subsection*{7. Certifiable Predictions for Safety and Medical Risk Assessment}
To ensure deployment in critical scenarios—such as industrial leak detection or disease screening—models must offer certifiable predictions. Building on the VOC-certifire framework, future studies should extend robustness certification methods to multi-class gas mixtures, quantify uncertainties, and support risk-informed decision-making in health and environmental applications.

\subsection*{8. Deployment on UAVs, Fixed Stations, and Wearable Sensors}
Lastly, the developed ML-enhanced sensing strategies can be integrated into mobile and remote sensing platforms, including UAV-based leak detection systems, urban air quality stations, and wearable breath analyzers. These platforms will require compact instrumentation, real-time processing, and robustness to environmental variability, all of which are supported by the techniques presented in this thesis.

\refstepcounter{chapter}%
\chapter*{\thechapter \quad Publications List}
\addcontentsline{toc}{chapter}{Publications List}

\section*{Peer-Reviewed Journal Articles -- First Author}

\begin{itemize}
    \item \textbf{Mohamed Sy}; Dapeng Liu; Aamir Farooq, ``Laser-based speciation of 1,4-pentadiene thermal decomposition behind reflected shock waves,'' \textit{Proceedings of the Combustion Institute}, 2026. \cite{sy2026pentadiene}

    \item \textbf{Mohamed Sy}; Sk Aman Ali; Khalil Djebbi; Ali Elkhazraji; Pan Luo; Ibrahim Atwah; Aamir Farooq, ``Wide-range gas analysis of C\textsubscript{1}--C\textsubscript{5} alkanes using a machine learning--enhanced dual-path optical cell,'' \textit{Sensors and Actuators B: Chemical}, p.~139825, 2026. \cite{sy2026wide}

    \item \textbf{Mohamed Sy}; Emad Al Ibrahim; Aamir Farooq, ``UnblindMix: An unsupervised reference-free framework for multi-species detection in high-temperature combustion diagnostics,'' \textit{Combustion and Flame}, vol.~280, p.~114383, 2025. \cite{sy2025unblindmix}

    \item \textbf{Mohamed Sy}; Sarah Aamir; Aamir Farooq, ``Multi-component gas sensing via spectral feature engineering,'' \textit{Sensors and Actuators B: Chemical}, vol.~430, p.~137285, 2025. \cite{sy2025multiSAB}

    \item \textbf{Mohamed Sy}; Jiabiao Zou; Mohammad Adil; Ali Elkhazraji; Mhanna Mhanna; Aamir Farooq, ``Laser-based speciation of isoprene thermal decomposition behind reflected shock waves,'' \textit{Proceedings of the Combustion Institute}, vol.~40, no.~1--4, p.~105460, 2024. \cite{sy2024laser}

    \item \textbf{Mohamed Sy}; Emad Al Ibrahim; Aamir Farooq, ``VOC-Certifire: Certifiably robust one-shot spectroscopic classification via randomized smoothing,'' \textit{ACS Omega}, vol.~9, no.~37, pp.~39033--39042, 2024. \cite{sy2024voc}

    \item \textbf{Mohamed Sy}; Mhanna Mhanna; Aamir Farooq, ``Multi-speciation in shock tube experiments using a single laser and deep neural networks,'' \textit{Combustion and Flame}, vol.~255, p.~112929, 2023. \cite{sy2023multi}

    \item \textbf{Mohamed Sy}; Emad Al Ibrahim; Ali Elkhazraji; Aamir Farooq, ``Unsupervised Source Separation for Multi-Speciation: Quantifying Hydrocarbons Without Their Reference Spectra,'' 2024. \cite{sy2024unsupervisedJournal}
\end{itemize}

\section*{Peer-Reviewed Conference Papers -- First Author}

\begin{itemize}
    \item \textbf{Mohamed Sy}; Mohammed A.~Sait; Shams Musheer; Aamir Farooq, ``SIMNet: An Unsupervised Framework for Accurate Gas Species Quantification Under Unknown Interference,'' \textit{2025 Conference on Lasers and Electro-Optics (CLEO)}, 2025. \cite{sy2025simnet}

    \item \textbf{Mohamed Sy}; Mohammed S.~Khan; Ali Elkhazraji; Pan Luo; Ibrahim Atwah; Aamir Farooq, ``ML-Enhanced Wide Dynamic Range C\textsubscript{1}--C\textsubscript{5} Alkanes Analyzer,'' \textit{Optical Sensors}, 2025. \cite{sy2025ml}

    \item \textbf{Mohamed Sy}; Jiabiao Zou; Ali Elkhazraji; Mohammad Adil; Mhanna Mhanna; Aamir Farooq, ``Multi-species Time-History Measurements During 1,3-Butadiene and Isoprene Oxidation Behind Reflected Shock Waves,'' \textit{AIAA SCITECH 2025 Forum}, 2025. \cite{sy2025multiAIAA}

    \item \textbf{Mohamed Sy}; Emad Al Ibrahim; Ali Elkhazraji; Aamir Farooq, ``Reference-Free Multi-Species Gas Detection via Unsupervised Learning,'' \textit{2024 IEEE SENSORS}, 2024. \cite{sy2024reference}

    \item \textbf{Mohamed Sy}; Ali Elkhazraji; Mohammed S.~Khan; Pan Luo; Ibrahim Atwah; Aamir Farooq, ``ML-Enhanced Laser-Based Analyzer for Selective C\textsubscript{1}--C\textsubscript{5} Alkanes Detection,'' \textit{Laser Applications to Chemical, Security and Environmental Analysis}, 2024. \cite{sy2024ml}

    \item \textbf{Mohamed Sy}; Emad Al Ibrahim; Ali Elkhazraji; Aamir Farooq, ``Unsupervised Source Separation Technique for Multi-Speciation Using a Single Laser: Quantifying Hydrocarbons Without Their Reference Spectra,'' \textit{2024 Conference on Lasers and Electro-Optics (CLEO)}, 2024. \cite{sy2024unsupervisedCLEO}

    \item \textbf{Mohamed Sy}; Mhanna Mhanna; Aamir Farooq, ``Multi-Speciation Using a Tunable Laser and Deep Neural Networks,'' \textit{2023 Conference on Lasers and Electro-Optics (CLEO)}, 2023. \cite{sy2023multiCLEO}
\end{itemize}

\section*{Peer-Reviewed Journal Articles -- Co-Author}

\begin{itemize}
    \item Mohammed Almomtan; Janardhanraj Subburaj; Emad Al Ibrahim; \textbf{Mohamed Sy}; El Mehdi Kharkhache; Danyaal Alam; Aamir Farooq, ``{SAFprop}: A Dual Spectroscopic Approach for Predicting Properties of Sustainable Aviation Fuels,'' \textit{Sustainable Energy \& Fuels}, 2026. \cite{almomtan2026safprop}

    \item Borislav Hinkov; Johannes Kunsch; Werner M\"antele; Lukasz Sterczewski; \'Angel S\'anchez-Illana; Jaume B\'ejar-Grimalt; V\'ictor Navarro-Esteve; David Perez-Guaita; Alexander Mittelst\"adt; Philippa Clark; \textbf{Mohamed Sy}; \textit{et al.}, ``Infrared photonics for healthcare: A roadmap for proactive and predictive health management,'' \textit{arXiv preprint arXiv:2602.08248}, 2026. \cite{hinkov2026infrared}

    \item Ahmad F.~Alsewailem; Janardhanraj Subburaj; \textbf{Mohamed Sy}; Vinny Gupta; Matthew J.~Dunn; Assaad R.~Masri; Aamir Farooq, ``Near-field quantification of hydrogen fluoride emissions during Li-ion battery thermal runaway using laser absorption spectroscopy,'' \textit{Journal of Power Sources}, vol.~667, p.~239221, 2026. \cite{alsewailem2026near}

    \item Ali Elkhazraji; \textbf{Mohamed Sy}; Mohammad Khaled Shakfa; Aamir Farooq, ``A mid-infrared laser diagnostic for simultaneous detection of furan and nitric oxide,'' \textit{Proceedings of the Combustion Institute}, vol.~40, no.~1--4, p.~105366, 2024. \cite{elkhazraji2024mid}

    \item Mhanna Mhanna; \textbf{Mohamed Sy}; Ali Elkhazraji; Aamir Farooq, ``Multi-speciation in shock tube kinetics using deep neural networks and cavity-enhanced absorption spectroscopy,'' \textit{Proceedings of the Combustion Institute}, vol.~40, no.~1--4, p.~105733, 2024. \cite{mhanna2024multi}

    \item Mhanna Mhanna; \textbf{Mohamed Sy}; Ayman Arfaj; Jose Llamas; Aamir Farooq, ``Highly sensitive and selective laser-based BTEX sensor for occupational and environmental monitoring,'' \textit{Applied Optics}, vol.~63, no.~11, pp.~2892--2899, 2024. \cite{mhanna2024highlyAO}

    \item Ali Elkhazraji; \textbf{Mohamed Sy}; Mhanna Mhanna; Joury Aldhawyan; Mohammad Khaled Shakfa; Aamir Farooq, ``Selective BTEX detection using laser absorption spectroscopy in the CH bending mode region,'' \textit{Experimental Thermal and Fluid Science}, vol.~151, p.~111090, 2024. \cite{elkhazraji2024selective}

    \item Mhanna Mhanna; \textbf{Mohamed Sy}; Ali Elkhazraji; Aamir Farooq, ``A laser-based sensor for selective detection of benzene, acetylene, and carbon dioxide in the fingerprint region,'' \textit{Applied Physics B}, vol.~129, no.~9, p.~139, 2023. \cite{mhanna2023laserAPB}

    \item Mhanna Mhanna; \textbf{Mohamed Sy}; Aamir Farooq, ``A selective laser-based sensor for fugitive methane emissions,'' \textit{Scientific Reports}, vol.~13, no.~1, p.~1573, 2023. \cite{mhanna2023selectiveSciRep}

    \item Ali Elkhazraji; Mohammad Khaled Shakfa; Nawaf Abualsaud; Mhanna Mhanna; \textbf{Mohamed Sy}; Marco Marangoni; Aamir Farooq, ``Laser-based sensing in the long-wavelength mid-infrared: chemical kinetics and environmental monitoring applications,'' \textit{Applied Optics}, vol.~62, no.~6, pp.~A46--A58, 2023. \cite{elkhazraji2023laser}

    \item Mhanna Mhanna; \textbf{Mohamed Sy}; Ali Elkhazraji; Aamir Farooq, ``Deep neural networks for simultaneous BTEX sensing at high temperatures,'' \textit{Optics Express}, vol.~30, no.~21, pp.~38550--38563, 2022. \cite{mhanna2022deep}

    \item Mhanna Mhanna; \textbf{Mohamed Sy}; Ayman Arfaj; Jose Llamas; Aamir Farooq, ``Laser-based selective BTEX sensing using deep neural networks,'' \textit{Optics Letters}, vol.~47, no.~13, pp.~3247--3250, 2022. \cite{mhanna2022laserOL}
\end{itemize}

\section*{Peer-Reviewed Conference Papers -- Co-Author}

\begin{itemize}
    \item Mohammed Almomtan; Sirine Naguez; Janardhanraj Subburaj; \textbf{Mohamed Sy}; Emad Al Ibrahim; Aamir Farooq, ``Domain Adaptation for Fuel Property Prediction Using Vibrational Spectroscopy,'' \textit{AIAA SCITECH 2026 Forum}, 2026. \cite{almomtan2026domain}

    \item Zhenhai Wang; \textbf{Mohamed Sy}; Syed Tajammul Ahmad; Luca Moretti; Mathieu Walsh; Davide Gatti; Jerome Genest; Marco Marangoni; Aamir Farooq, ``ML-Enabled High-Speed Dual-Comb Spectrometer for N\textsubscript{2}O Detection in the Long-Wave Infrared Region,'' \textit{2025 Conference on Lasers and Electro-Optics (CLEO)}, 2025. \cite{wang2025ml}

    \item Mohammed Almomtan; Janardhanraj Subburaj; Emad Al Ibrahim; \textbf{Mohamed Sy}; El Mehdi Kharkhache; Danyaal Alam; Aamir Farooq, ``Predicting Sustainable Aviation Fuel Properties using ATR-FTIR and Confocal Raman Spectroscopy,'' \textit{AIAA AVIATION FORUM and ASCEND 2025}, 2025. \cite{almomtan2025predicting}

    \item Mhanna Mhanna; \textbf{Mohamed Sy}; Ali Elkhazraji; Aamir Farooq, ``A Selective Benzene, Acetylene, and Carbon Dioxide Sensor Near 14.84\,$\mu$m,'' \textit{CLEO: Applications and Technology}, 2023. \cite{mhanna2023benzeneCLEO}

    \item Mhanna Mhanna; \textbf{Mohamed Sy}; Ali Elkhazraji; Aamir Farooq, ``Multi-Species Sensing Using CEAS and DNN in Shock Tube Kinetics,'' \textit{Computational Optical Sensing and Imaging}, 2022. \cite{mhanna2022multiCEAS11, mhanna2022computational}

    \item Mhanna Mhanna; \textbf{Mohamed Sy}; Ali Elkhazraji; Aamir Farooq, ``Laser-Based Sensor for Multi-Species Detection Using CEAS and DNN,'' \textit{2022 Conference on Lasers and Electro-Optics (CLEO)}, 2022. \cite{mhanna2022laserCEAS}

    \item Mhanna Mhanna; \textbf{Mohamed Sy}; Aamir Farooq, ``Interference-Free Methane Laser Sensor Using Cepstral Analysis,'' \textit{CLEO: QELS Fundamental Science}, 2022. \cite{mhanna2022interference}

    \item Mhanna Mhanna; \textbf{Mohamed Sy}; Aamir Farooq, ``Selective BTEX Measurements Using Deep Neural Networks,'' \textit{CLEO: Science and Innovations}, 2021. \cite{mhanna2021selectiveCLEO1}

    \item Mhanna Mhanna; \textbf{Mohamed Sy}; Aamir Farooq, ``Selective BTEX Sensing with Laser Absorption and DNNs,'' \textit{Optical Sensors}, 2021. \cite{mhanna2021selectiveOS}
\end{itemize}

\section*{Preprints}

\begin{itemize}
    \item Mhanna Mhanna; \textbf{Mohamed Sy}; Ali Elkhazraji; Aamir Farooq, ``A Selective Benzene, Acetylene, and Carbon Dioxide Sensor in the Fingerprint Region,'' \textit{arXiv preprint arXiv:2209.11832}, 2022. \cite{mhanna2022benzeneArxiv}
\end{itemize}

\section*{Patents}

\begin{itemize}
    \item Mhanna Mhanna; Muhammad Arsalan; \textbf{Mohamed Sy}; Aamir Farooq, ``Method and System for Water Cut Sensing in an Oil--Water Flow,'' 2025. \cite{mhanna2025methodPatent}

    \item Mhanna Mhanna; Muhammad Arsalan; \textbf{Mohamed Sy}; Aamir Farooq, ``Method and System for Water Cut Sensing in an Oil--Water Flow,'' US Patent App.~18/535,964, 2024. \cite{mhanna2024methodPatent}

    \item Aamir Farooq; Mhanna Mhanna; \textbf{Mohamed Sy}, ``Laser-Based Selective BTEX Sensing with Deep Neural Network,'' US Patent App.~17/901,068, 2023. \cite{farooq2023patent}
\end{itemize}

    \begin{onehalfspacing}
    \renewcommand*\bibname{\centerline{REFERENCES}} 
    \addcontentsline{toc}{chapter}{References}
    \newcommand{\BIBdecl}{\setlength{\itemsep}{0pt}}
    \bibliographystyle{IEEEtran}
    \bibliography{References}
\end{onehalfspacing}

    \appendix
\newpage
\begin{center}
    \vspace*{2\baselineskip}
    { \textbf{{\large APPENDICES}}} 
    \addcontentsline{toc}{chapter}{Appendices}
\end{center}
\begingroup
    \let\clearpage\relax
\refstepcounter{chapter}%
\chapter*{Unsupervised Spectral Source Separation: A Reference-free Technique for Multi-species Detection} \label{appendixA}
\section{Effect of Input Spectral Resolution}
\setcounter{figure}{0} 
\label{sec:sample:NMF-4}
Lowering the resolution of input spectra can reduce computational costs but may affect model performance. We evaluate how UnblindMix behaves when trained with lower-dimensional input data compared to the default 600-point spectra.
\begin{figure}[H]
    \centering
    \includegraphics[width=\linewidth]{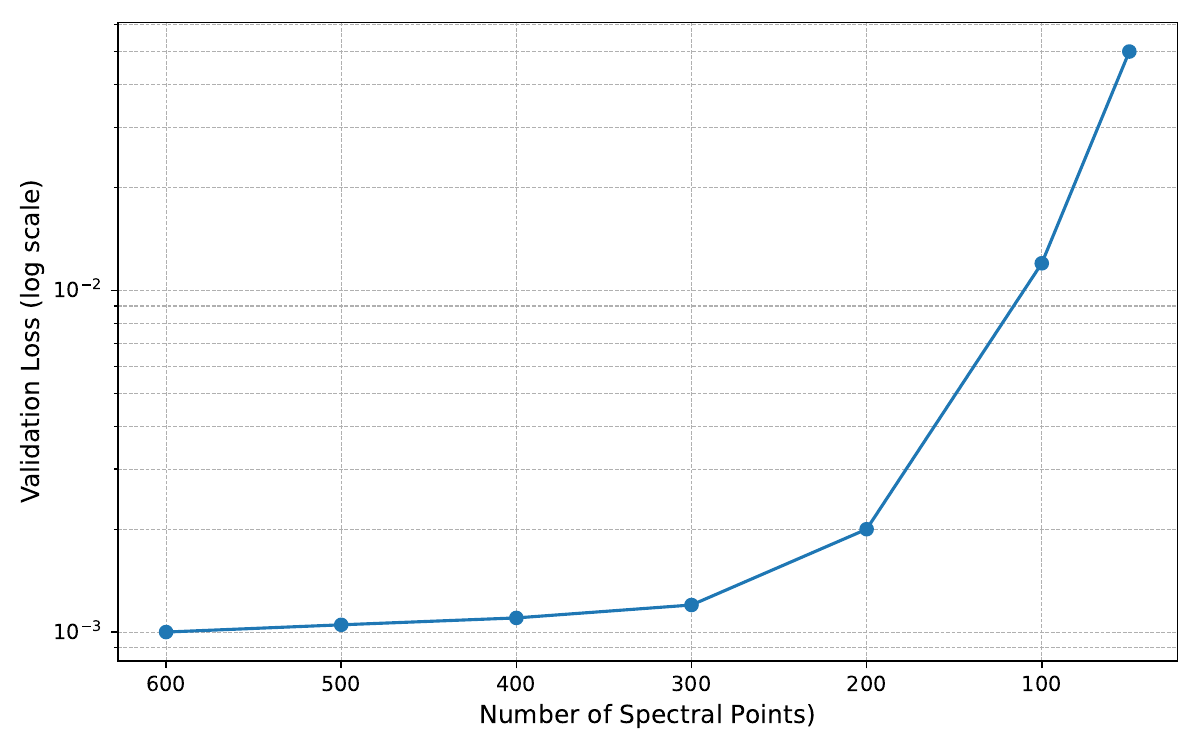}
    \caption{Effect of input spectral resolution on UnblindMix performance. Lower-resolution spectra lead to higher validation errors and reduced accuracy in spectral reconstruction.}
    \label{fig:figS1}
\end{figure}

\section{Effect of Physical Constraints}
\setcounter{figure}{0} 
\label{sec:sharppeaks}

Physical constraints can guide the model toward physically meaningful solutions. These constraints include enforcing carbon atom balance, monotonic species appearance profiles, and bounding product yields.

\begin{figure}[H]
    \centering
    \includegraphics[width=0.9\linewidth]{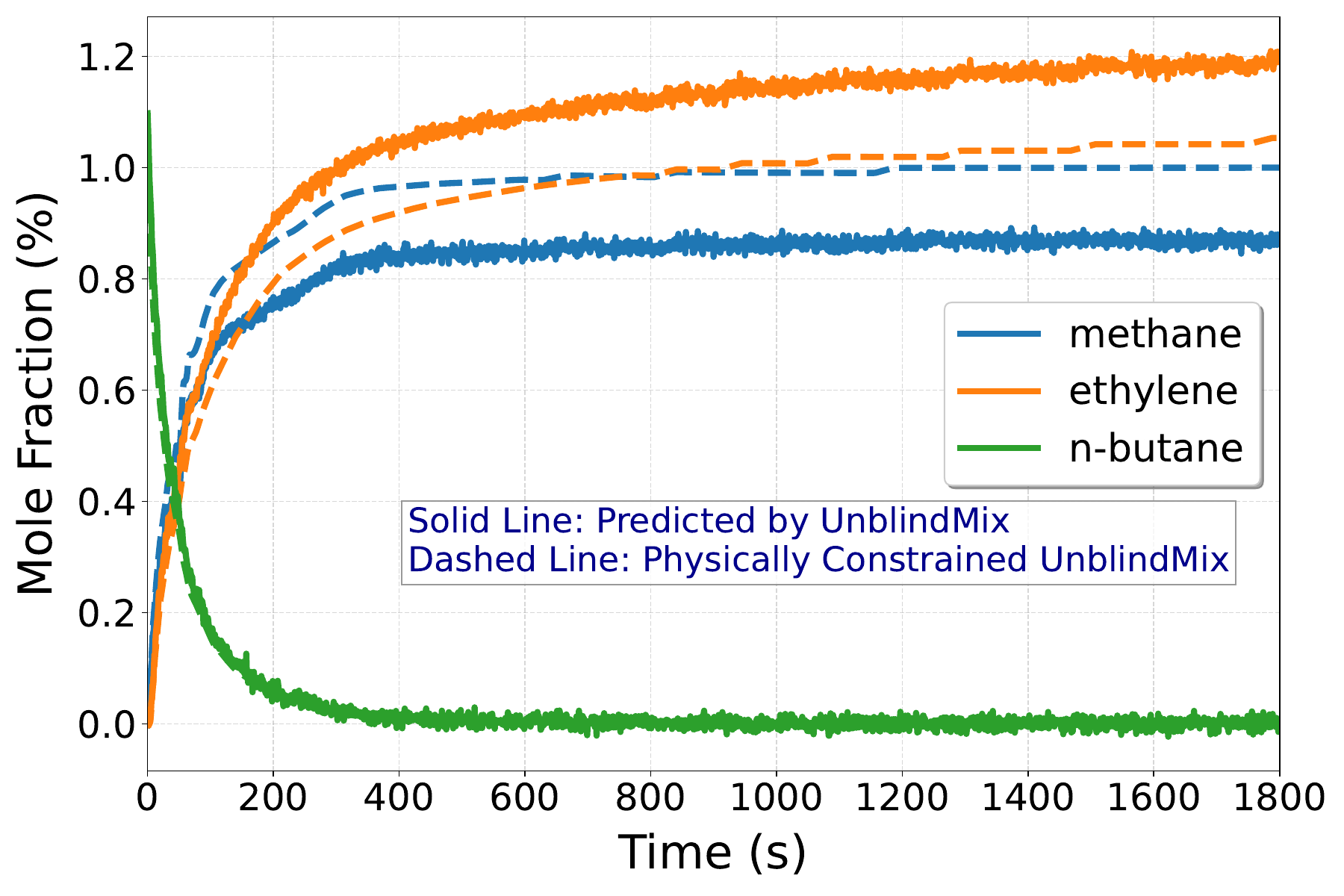}
    \caption{Effect of physical constraints in UnblindMix training. Constraints include carbon balance, monotonicity in species profiles, and bounded yields. The constrained model produces smoother concentration trends and better reflects physical behavior.}
    \label{fig:figS3}
\end{figure}
\newpage
\section{Mixture Compositions}
\begin{table}[htbp]
\centering
\caption{Gas mixtures (\% mole fractions)}
\begin{tabular}{ccccc}
\hline
\textbf{Methane} & \textbf{Ethane} & \textbf{Ethylene} & \textbf{Propyne} & \textbf{N\textsubscript{2}} \\
\hline
0.82 & 0.40 & 0.44 & 0.38 & 97.96 \\
0.67 & 0.00 & 0.72 & 0.55 & 98.06 \\
0.91 & 0.40 & 0.22 & 0.00 & 98.47 \\
0.83 & 0.40 & 0.37 & 0.00 & 98.40 \\
0.42 & 0.00 & 0.71 & 0.66 & 98.21 \\
0.97 & 0.32 & 0.00 & 0.51 & 98.20 \\
0.81 & 0.40 & 0.46 & 0.00 & 98.33 \\
0.22 & 0.33 & 0.28 & 0.27 & 98.90 \\
0.88 & 0.35 & 0.00 & 0.62 & 98.15 \\
0.69 & 0.00 & 0.21 & 0.84 & 97.26 \\
0.19 & 0.17 & 0.28 & 0.23 & 99.13 \\
0.54 & 0.26 & 0.61 & 0.31 & 98.28 \\
0.83 & 0.00 & 0.37 & 0.48 & 98.32 \\
0.41 & 0.40 & 0.00 & 0.67 & 98.52 \\
0.76 & 0.00 & 0.72 & 0.28 & 98.24 \\
0.91 & 0.33 & 0.00 & 0.56 & 98.20 \\
0.65 & 0.21 & 0.47 & 0.00 & 98.67 \\
0.34 & 0.40 & 0.45 & 0.00 & 98.81 \\
0.88 & 0.40 & 0.00 & 0.35 & 98.37 \\
0.52 & 0.00 & 0.81 & 0.48 & 98.19 \\
0.41 & 0.26 & 0.25 & 0.00 & 99.08 \\
0.93 & 0.00 & 0.36 & 0.61 & 98.10 \\
0.95 & 0.19 & 0.00 & 0.44 & 98.42 \\
0.99 & 0.00 & 0.45 & 0.53 & 98.03 \\
\hline
\end{tabular}
\end{table}



\refstepcounter{chapter}%
\chapter*{Models Certification and Stability via Randomized Smoothing} \label{appendixB}
\section{Performance metrics:}
\begin{itemize}
    \item \textbf{Accuracy:} \[ \frac{\text{True Positives (TP)} + \text{True Negatives (TN)}}{\text{Total Predictions}}\]
    Accuracy measures the overall correctness of the model by considering both true positive and true negative predictions. A higher accuracy indicates a more reliable model.
    
    \item \textbf{Precision:}\[\frac{\text{True Positives (TP)}}{\text{True Positives (TP)} + \text{False Positives (FP)}}\]
    Precision focuses on the accuracy of positive predictions. A higher precision implies fewer false positives among the positive predictions.
    
    \item \textbf{F1 Score:}\[\frac{2 \times \text{True Positives (TP)}}{2 \times \text{True Positives (TP)} + \text{False Positives (FP)} + \text{False Negatives (FN)}}\]
    F1 Score is a balanced metric that considers both precision and recall. It ranges from 0 to 1, with higher values indicating a better balance between precision and recall.
\end{itemize}
\section{Numerical cost}
\subsection{Data generation}
Using the tic() and toc() timer in Python, generating a spectrum with Voigt convolutions takes 0.15 ms, while generating a spectrum with HAPI \textit{(HITRAN Application Programming Interface)} in the same range and resolution takes 0.17 ms. Assuming a linear scaling for both, the time required for data generation is just slightly favored for augmentations as shown in Fig. \ref{fig:augs}.
\begin{figure}[!h] 
    \centering
    \includegraphics[width=0.55\textwidth]{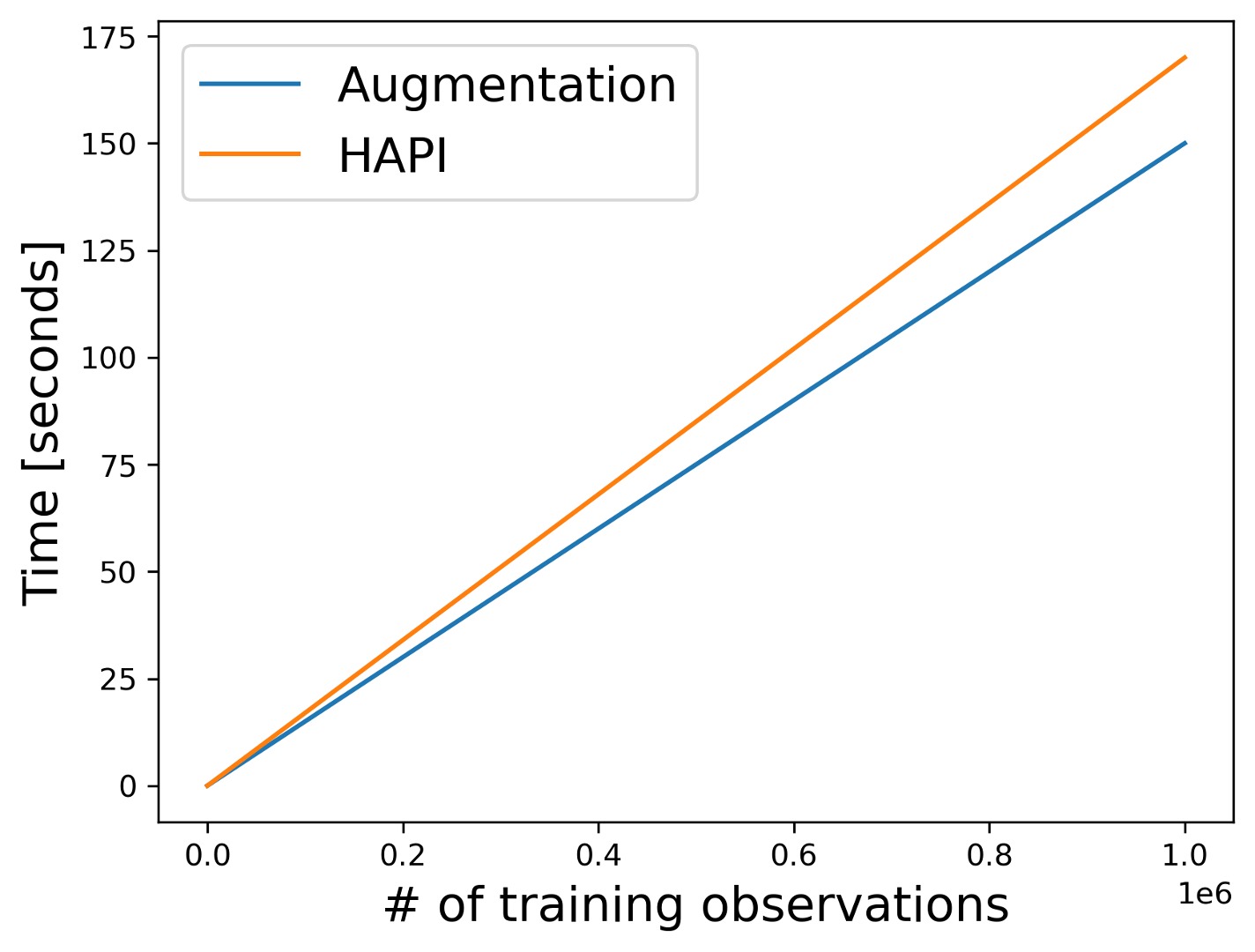}
    \caption{Numerical cost of data generation cost}
    \label{fig:augs}
\end{figure}
\subsection{Training cost}
Table \ref{tab:model_params} provides a detailed overview of the model architecture and its trainable parameters. The model comprises a total of 18,246 trainable parameters. Training was conducted on a GPU with 128 GB of memory over 10 epochs. The total training time for the VOC-net model was approximately 200 ms, for VOC-lite approximately 23 ms, and for VOC-plus approximately 14 minutes.
\begin{table}[h!]
\centering
\caption{Layer Architecture and Parameters of the Model}
\begin{tabular}{l c c}
\hline
\textbf{Layer (type)} & \textbf{Output Shape} & \textbf{Param \#} \\
\hline
C1 (Conv1D)           & (None, 249, 3)        & 12          \\
S2 (AveragePooling1D) & (None, 124, 3)        & 0           \\
C3 (Conv1D)           & (None, 122, 3)        & 30          \\
Flatten              & (None, 366)           & 0           \\
Dense (Dense)         & (None, 48)            & 17,616      \\
Dropout              & (None, 48)            & 0           \\
Dense (Dense)         & (None, 12)            & 588         \\
\hline
\textbf{Total params:} & \multicolumn{2}{c}{\textbf{18,246}} \\
\textbf{Trainable params:} & \multicolumn{2}{c}{\textbf{18,246}} \\
\textbf{Non-trainable params:} & \multicolumn{2}{c}{\textbf{0}} \\
\hline
\end{tabular}
\label{tab:model_params}
\end{table}

\section{Extension to higher pressures conditions}
In the VOC-certifire model, training was focused on handling unseen pressure conditions, achieving high accuracy in predicting and classifying VOCs across various pressures, even though the model was trained on a single pressure condition. To extend the model's applicability beyond 16 Torr, additional data is not required. Instead, adjusting the broadening coefficients in the voigt convolutions can account for new pressure conditions. As illustrated in the figure below, Voigt convolution augmentations effectively capture spectral variations under high-pressure environmental conditions. Fig. \ref{fig:all_vocs}a) displays the methanol spectrum at P = 0.001 atm simulated using HAPI. Fig. \ref{fig:all_vocs}b) on the right panel compares the spectrum simulated using HAPI at P = 1 atm with the augmented spectrum generated using Voigt convolutions.

\begin{figure}[!h] 
    \centering
    \includegraphics[width=0.65\textwidth]{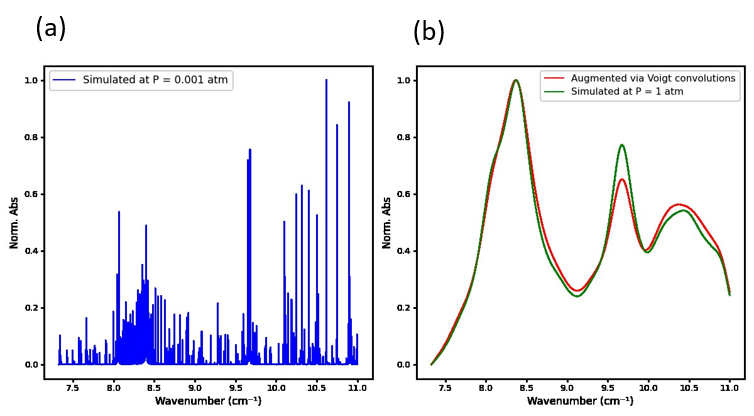}
    \caption{a) Methanol spectrum at 0.001 atm simulated with HAPI, b) Methanol spectrum at 1 atm spectrum simulated with HAPI and the augmented spectrum using Voigt convolutions.}
    \label{fig:all_vocs}
\end{figure}
\endgroup

\end{document}